\documentclass[aps,prx,twocolumn,superscriptaddress,nofootinbib,nolongbibliography,nobalancelastpage,10pt,floatfix]{revtex4-2}

\usepackage{graphicx}
\usepackage{dcolumn}
\usepackage{bm}
\usepackage{braket}
\usepackage{booktabs}

\usepackage{multirow}
\usepackage{amsmath,amssymb,amsfonts}
\usepackage{amsthm}
\usepackage{mathrsfs}
\usepackage{xcolor}
\usepackage{textcomp}
\usepackage[ruled,vlined]{algorithm2e}
\usepackage{listings}
\usepackage{tikz}
\usepackage{fontawesome}
\usetikzlibrary{arrows.meta,positioning,fit,calc,shapes.symbols,shapes.misc}
\usepackage{array}
\usepackage{makecell}
\usepackage{xurl}
\usepackage[colorlinks,citecolor=blue,linkcolor=blue,urlcolor=blue]{hyperref}

\usepackage[margin=0.65in]{geometry}
\usepackage[T1]{fontenc}
\usepackage[utf8]{inputenc}
\usepackage{lmodern}
\usepackage{xcolor}
\usetikzlibrary{arrows.meta,positioning,calc}
\newcolumntype{R}{>{$}r<{$}}
\newcolumntype{L}{>{$}l<{$}}

\newcolumntype{:}{|}
\newcommand{\tup}[1]{\left(#1\right)}

\newcommand{\entryhead}[1]{\par\smallskip\noindent\textbf{#1}\par}

\theoremstyle{plain}
\newtheorem{theorem}{Theorem}
\newtheorem{proposition}[theorem]{Proposition}
\theoremstyle{definition}
\newtheorem{definition}{Definition}

\theoremstyle{remark}

\newcommand{\criticize}[3]{\ifmmode\textcolor{\ifnum#2=5 red\else\ifnum#2=4 orange\else\ifnum#2=3 yellow!80!red\else\ifnum#2=2 yellow!60!black\else\ifnum#2=0 green!60!black\else gray\fi\fi\fi\fi\fi!\ifnum#3<3 30\else\ifnum#3<5 70\else 100\fi\fi!black}{\text{[#1 (S#2/C#3)]}}\else\textcolor{\ifnum#2=5 red\else\ifnum#2=4 orange\else\ifnum#2=3 yellow!80!red\else\ifnum#2=2 yellow!60!black\else\ifnum#2=0 green!60!black\else gray\fi\fi\fi\fi\fi!\ifnum#3<3 30\else\ifnum#3<5 70\else 100\fi\fi!black}{[#1 (S#2/C#3)]}\fi}

\newif\ifshowcomments
    \showcommentstrue 

\begin{document}

\title{Co-Designing Quantum Codes with Transversal Diagonal Gates\\ via Multi-Agent Systems}

\author{Xi He}
\affiliation{Department of Physics, The University of Texas at Dallas, Richardson, Texas 75080, USA}

\author{Sirui Lu}
\affiliation{Max-Planck-Institut f\"ur Quantenoptik, 85748 Garching bei M\"unchen, Germany}

\author{Bei Zeng}
\email{bei.zeng@utdallas.edu}
\affiliation{Department of Physics, The University of Texas at Dallas, Richardson, Texas 75080, USA}

\date{\today}

\begin{abstract}
Exact scientific discovery requires more than heuristic search: candidate constructions must be turned into exact objects and checked independently. We address this gap by extending TeXRA with an independent Lean 4 verification layer, turning it into a human-guided multi-agent platform for exact scientific discovery. The platform couples symbolic synthesis, combinatorial and linear-programming search, exact reconstruction of numerical candidates, and formal verification in Lean. We apply this platform to nonadditive quantum error-correcting codes with prescribed transversal diagonal gates within the subset-sum linear-programming (SSLP) framework. In the distance-2 regime where logical states occupy distinct residue classes, the platform yields a Lean-certified catalogue of 14,116 codes for $K\in\{2,3,4\}$ and up to six physical qubits, realizing cyclic logical orders 2 through 18, from which we extract closed-form infinite families. We also construct a residue-degenerate $((6,4,2))$ code implementing the logical controlled-phase gate $\mathrm{diag}(1,1,1,i)$. At distance 3, we resolve the transversal-$T$ problem for $((7,2,3))$ codes within the complementary binary-dihedral $\mathrm{BD}_{16}$ setting: among the 12 candidates surviving the SSLP filters, 10 admit exact realizations and 2 are excluded by no-go proofs. All accepted constructions, families, and no-go results are formalized and checked in Lean, illustrating how AI-assisted workflows can bridge search, exact reconstruction, and formal proof in the physical sciences.
\end{abstract}

\maketitle

\section{Introduction}
\label{sec:intro}

AI systems are increasingly taking on research tasks~\cite{Wang2023Scientific,Lu2024AI,Binz2025How,Glazer2024FrontierMath,Lu2025Can}, but their scientific impact depends strongly on problem structure. The most favorable settings are those in which candidate solutions are difficult to invent, the search space is large but structured, and correctness can be checked exactly once a promising candidate is written down. Classical AI has long excelled at search and optimization, from game playing~\cite{Silver2016Mastering,Silver2017Mastering} and protein structure prediction~\cite{Jumper2021Highly} to automated theorem proving~\cite{Romera-Paredes2024Mathematical,Trinh2024Solving}. Large language models~\cite{Brown2020Language} add complementary capabilities in tool use~\cite{Schick2023Toolformer,Anthropic2024Model,Patil2023Gorilla}, code generation~\cite{Tian2024SciCode,Jimenez2024SWEbench}, and symbolic reasoning from examples~\cite{OpenAI2024OpenAI,Guo2025DeepSeekR1}. Coupled to formal backends that machine-check the resulting claims, these capabilities make it possible to build systems for exact scientific discovery rather than heuristic assistance alone. Quantum code discovery with prescribed transversal gates is a particularly natural testbed: one must traverse large combinatorial spaces of candidate supports, extract regularities from sparse successful examples, convert numerical instances into exact algebraic constructions, and verify the final claims rigorously. The challenge is therefore not a single calculation but a coordinated loop of formulation, search, synthesis, and proof.

\begin{figure*}[t]
    \centering
    \includegraphics[width=0.95\linewidth]{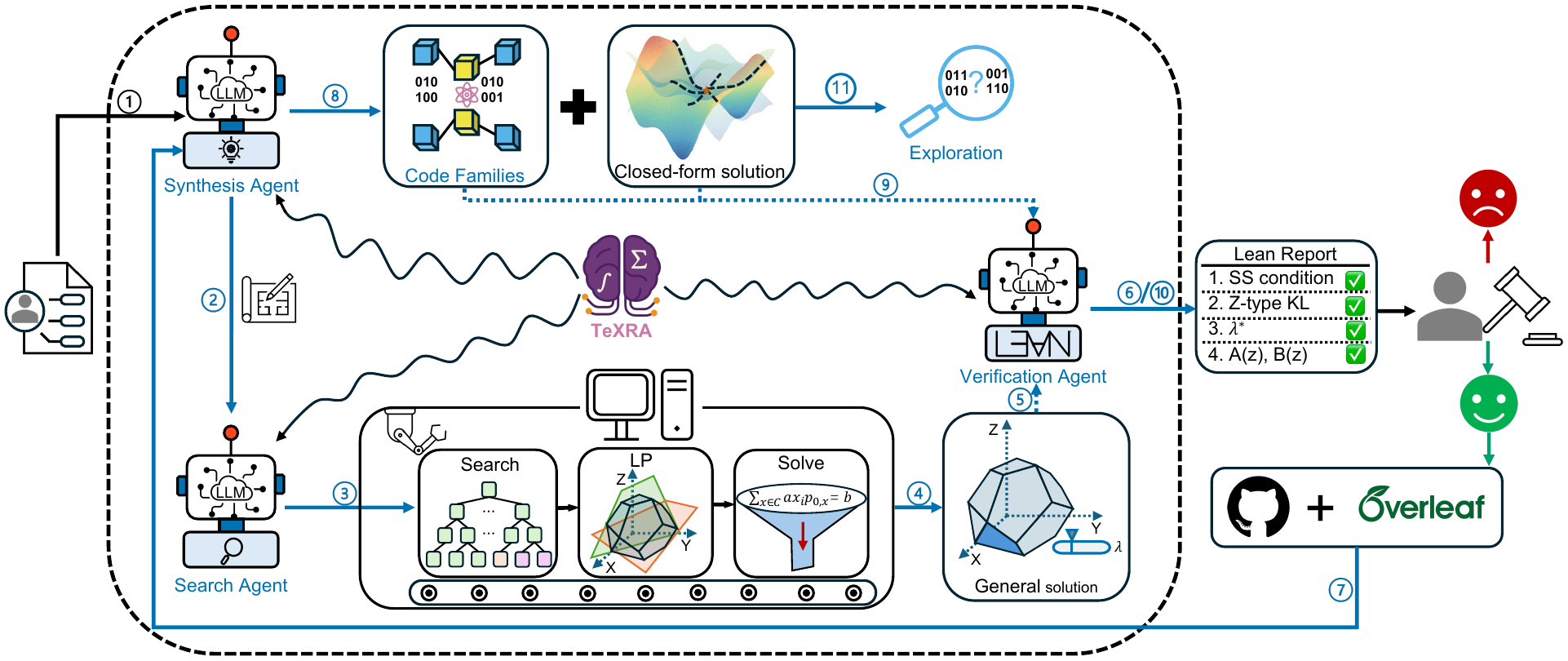}
    \caption{Human-guided multi-agent workflow for quantum code discovery. A problem specification initializes the Synthesis Agent (1), which derives symbolic reformulations, parameter templates, and proof goals and dispatches executable search programs to the Search Agent (2). The Search Agent performs combinatorial enumeration, linear-program construction, and numerical solving (3) to generate candidate code constructions (4). These are passed to an independent Verification Agent, implemented in TeXRA with Lean, which checks subset-sum conditions, $Z$-type Knill-Laflamme equalities, induced logical phases, and associated polynomial identities and returns formal reports (5-6). Lean-verified instances are then fed back through the shared TeXRA workspace to the Synthesis Agent (7), which abstracts general code families, closed-form solutions, and no-go arguments (8); these higher-level results are again verified in Lean (9-10) and can guide further exploration (11). The dashed boundary marks the independent Lean verification layer. The core search-synthesis-verification loop is automated, while human researchers initialize the problem, steer targets and priorities, and accept verified outputs.}
    \label{fig1:sche}
\end{figure*}

To address this problem, we extend TeXRA~\cite{Texra2025TeXRA} with an independent Lean 4 verification layer and develop it into a human-guided multi-agent platform~\cite{Wu2023AutoGen,Du2023Improving,Sumers2023Cognitive,Novikov2025AlphaEvolve} and apply it to the discovery of nonadditive quantum error-correcting codes. TeXRA, driven here by GPT-5~\cite{Openai2025GPT}, provides a shared research workspace in which agents can create and edit \LaTeX{}, Python, Lean, and data files. Within this workspace, agents combine tool-use loops~\cite{Yao2023ReAct} with derivation-then-edit workflows, execute searches and verification scripts, and exchange intermediate artifacts through a common working directory. The architecture used here, summarized in Fig.~\ref{fig1:sche}, has three specialized components. A Synthesis Agent derives symbolic reformulations, parameter templates, exact ans\"atze, and proof goals from the problem specification; a Search Agent turns these into executable combinatorial and linear-programming sweeps; and an independent Verification Agent, implemented in TeXRA with Lean, formalizes and checks constructions, logical actions, and no-go results. Researchers initialize the problem and steer targets and priorities, while the core search-synthesis-verification loop is automated. The AI contribution of this work is therefore not merely the application of a frontier model to a physics problem, but the implementation of a platform that integrates executable search, symbolic synthesis, and a separate formal proof layer. All constructions, logical actions, and no-go results reported in this paper are verified in Lean.

Quantum error correction encodes information into a $K$-dimensional subspace of an $n$-qubit system, enabling the detection or correction of physical errors~\cite{Nielsen2010Quantum,Calderbank1998Quantum,Steane1996Error,Knill1997Theory}. Logical operations must be implemented without spreading errors uncontrollably; transversal gates, which act independently on each physical qubit, are attractive for this reason but are sharply constrained by no-go theorems forbidding universal transversal gate sets~\cite{Zeng2011Transversality,Eastin2009Restrictions} and by strong group-theoretic restrictions on the logical operations they can realize~\cite{Liu2023Approximate,Anderson2016Classification}. 
Recent systematic enumerations of small stabilizer subsystem codes~\cite{Cross2025Small} emphasize the value of structured searches in the low-qubit regime.
Beyond stabilizer codes~\cite{Calderbank1998Quantum,Gottesman1997Stabilizer}, nonadditive constructions~\cite{Rains1997Nonadditive} substantially enlarge the design space, including codeword-stabilized (CWS) codes~\cite{Cross2009Codeword,Chuang2009Codeword,Yu2007Graphical,Yu2008Nonadditive,Grassl2008Nonadditive}, permutation-invariant (PI) codes~\cite{Pollatsek2004Permutationally,Ouyang2014Permutation,Ouyang2021Permutation,Ouyang2024Measurement}, and broader nonadditive families~\cite{Du2024Characterizing}.  Across these families, the transversal gate structure is rich, from early nonadditive phenomena~\cite{Rains2002Quantum} to recent permutation-invariant codes realizing the binary icosahedral group $2\mathrm{I}$~\cite{Kubischta2023Family} and higher-order diagonal phases~\cite{Kubischta2024Permutation}. In this paper we focus on diagonal transversal gates, which yield abelian, typically cyclic logical groups. The Subset-Sum Linear Programming (SSLP) framework~\cite{Zhang2025Transversal} captures a particularly promising portion of this design space by rewriting the diagonal-transversal problem in terms of congruence structure plus linear conditions on $Z$-marginals.

In the diagonal setting considered here, SSLP partitions computational-basis strings according to modular data associated with the target transversal phases.
Each logical basis state is assigned to one such class, so a transversal diagonal gate induces a predictable logical phase determined by the class label. The KL conditions~\cite{Knill1997Theory} then split into two layers: structural separation conditions that prevent bit-flip errors from mixing logical states, and constraints on how amplitudes are distributed within the classes so that single-qubit $Z$ statistics agree across logical states. The latter are linear in the squared amplitudes. This decomposition makes large regions of code space accessible to exhaustive or near-exhaustive search, but it does not by itself produce explicit constructions or classifications. One must still navigate exponentially many support choices, recognize general families from isolated successful instances, reconstruct exact amplitudes from numerical data, and resolve the coupled polynomial constraints that reappear in exact full-KL checks. We therefore ask: for given $((n,K,d))$, which diagonal transversal groups can arise, and how can the corresponding codes be constructed exactly?

Using the TeXRA-based workflow above across multiple $((n,K,d))$ regimes, we obtain a certified catalogue of 14,116 previously unreported nonadditive codes after deduplication in the tractable distance-$2$ nondegenerate-residue regime, where the logical states occupy distinct residue classes. For code dimensions $K\in\{2,3,4\}$ and up to $n=6$ physical qubits, we find new codes realizing cyclic logical gate orders from 2 to 18, with explicit exact constructions specifying amplitudes, parameters, and phases (Sec.~\ref{subsec:catalogue}). From these data we extract closed-form infinite families that recover and generalize many of the individual instances (Sec.~\ref{subsec:families}). Relaxing the distinct-residue assumption, we also construct a $((6,4,2))$ code realizing the controlled-phase gate $\mathrm{diag}(1,1,1,i)$ in a residue-degenerate setting where three logical states share a residue class (Sec.~\ref{subsec:extension}). Most notably, we resolve the small-code transversal-$T$ problem within the SSLP framework for $((7,2,3))$ codes in the binary-dihedral $\mathrm{BD}_{16}$ specialization. In earlier SSLP work~\cite{Zhang2025Transversal}, this case was reduced to 12 surviving candidates, but deciding them requires solving or excluding highly coupled exact polynomial constraints rather than merely passing subset-sum and LP filters. Our platform converts these surviving numerical candidates into exact algebraic constructions, formalizes the corresponding proof obligations, and resolves all 12 cases, proving that 10 admit exact transversal-$T$ realizations and that the remaining 2 are impossible (Sec.~\ref{sec:bd16-distance3}). Together, these results show how a TeXRA-based, Lean-verified multi-agent platform can turn a rich but difficult-to-navigate nonadditive code space into a rigorous pipeline for discovery, classification, and proof, while illustrating a broader model for exact AI-assisted research in the physical sciences.

\section{The multi-agent system}
\label{sec:methods}

This section turns the schematic workflow of Fig.~\ref{fig1:sche} into a concrete research protocol.
We first formulate the search for nonadditive quantum codes with prescribed transversal diagonal gates in the SSLP language, separating modular support data, linear $Z$-type Knill-Laflamme constraints, and the remaining exact full-KL conditions. We then explain how this decomposition is realized in a shared TeXRA workspace: the Synthesis Agent derives reformulations and exact ans\"atze, the Search Agent executes finite combinatorial and linear-programming sweeps, and the Verification Agent independently certifies all accepted constructions and no-go results in Lean. The point of the multi-agent design is that these three stages require different computational strengths, but can exchange exact intermediate artifacts through a common project directory.

\subsection{Multi-agent workspace setup and agent workflows}
\label{subsec:workspace}

\begin{figure*}[t]
    \centering
    \includegraphics[width=0.95\linewidth]{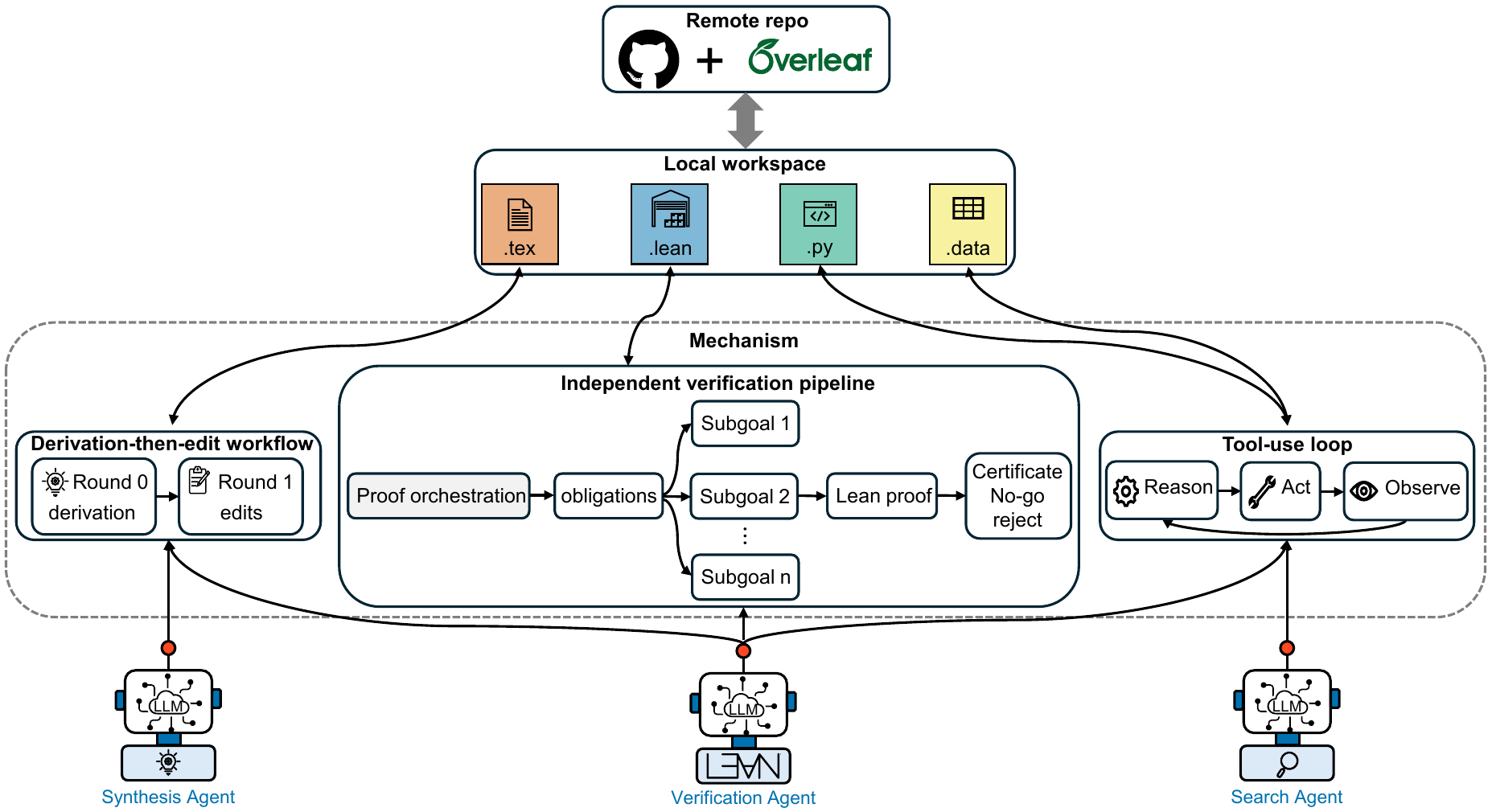}
    \caption{TeXRA-enabled workspace linking agents and tools.
        All agents operate on a shared local project directory containing \LaTeX{} sources, Python scripts, Lean~4 files, and data, synchronized with a remote repository and accessing a common LLM backend via API. (a)~Derivation-then-edit workflow: the Synthesis Agent produces derivations in successive rounds; the researcher reviews, instructs, and accepts or rejects the suggested edits. (b)~Tool-use loop: the Search and Verification Agents follow iterative reason-act-observe cycles, writing and executing code or Lean proofs to update the workspace. The Verification Agent operates independently: it receives only exported exact data (supports, probabilities, amplitudes) and does not see the search transcripts or intermediate reasoning of other agents.}
    \label{fig:agents-workspace}
\end{figure*}

The workspace is built on TeXRA~\cite{Texra2025TeXRA}, a VS~Code~\cite{VSCode} extension that integrates large language models into a local development environment. TeXRA provides two operational modes (see Fig.~\ref{fig:agents-workspace}): a \emph{tool-use loop}~\cite{Yao2023ReAct,Schick2023Toolformer} in which the agent iteratively reasons, calls functions (creating or editing files, executing scripts, reading outputs), and observes results; and a \emph{derivation-then-edit workflow}~\cite{Wei2022Chainofthought,Shinn2023Reflexion} in which the agent first expands its reasoning in an internal scratchpad, then generates structured \LaTeX{} edits that the researcher reviews via \texttt{latexdiff}~\cite{Tilmann2024latexdiff}. We used OpenAI's GPT-5~\cite{Openai2025GPT} as the model provider. The shared project directory contains \LaTeX{} source files, Python scripts, Lean~4 source files, data files, and chat logs, connected as a git repository to Overleaf so that all collaborators can review agent-generated content and track changes.

TeXRA allows flexibility in user-defined multi-agent architecture and agent roles; we customized its modes into three specialized roles by choosing different prompts, contexts, and tool permissions for each. The Synthesis Agent ran in derivation-then-edit mode with the SSLP literature~\cite{Zhang2025Transversal} and the \LaTeX{} draft as context. The Search Agent ran in tool-use mode with permission to write and execute Python scripts. The Verification Agent ran in a separate tool-use session with access to the Lean~4 compiler and the existing proof library, but not to the search transcripts.

The shared project directory serves as the common memory between agents. Each agent can read files written by the others: the Verification Agent reads the Search Agent's output files to obtain exact supports and probabilities, and reads the existing Lean modules in the \texttt{SS/} directory to know what infrastructure is available to import. Within a single TeXRA session, the agent retains its conversation history and can build on earlier work; across sessions, the human researcher bridges the gap by selecting which outputs to include in the next agent's prompt. The researcher also decides when to move between stages-for example, reviewing the Search Agent's candidate list before passing selected entries to the Verification Agent, or feeding Lean-verified instances back to the Synthesis Agent for pattern analysis. Git version control tracks all changes, so collaborators can review agent-generated content through Overleaf.

We organize the work into three specialized agents~\cite{Wu2023AutoGen,Du2023Improving,Sumers2023Cognitive} under human orchestration (Fig.~\ref{fig1:sche}). The \emph{Synthesis Agent} operates primarily in derivation-then-edit mode. Given the SSLP problem specification and the relevant literature as context, it reformulates the mathematical problem into code-ready terms: deriving the combinatorial support conditions, setting up the linear-programming constraints, proposing exact amplitude templates, and, after search results are available, analyzing Lean-verified instances to identify recurring patterns that can be lifted into closed-form families or no-go arguments. The \emph{Search Agent} works in the tool-use loop. It writes and executes Python scripts that enumerate canonical parameter sets, construct residue classes, solve the $Z$-type linear programs, perform rational reconstruction of numerical solutions, and record the outputs. When a search branch becomes unproductive, the human researcher can steer the agent toward different parameter ranges or tighter filters.

The \emph{Verification Agent} produces formal proofs in Lean~4, a proof assistant in which the compiler itself checks that every logical step is valid (see Sec.~\ref{subsec:audit} for details on Lean and the proof library). It operates in a separate TeXRA session and receives only the exported exact data (supports, rational probabilities, algebraic amplitudes) together with a proof goal. It does not see the search transcripts or the reasoning of the other agents, so its output cannot be contaminated by earlier mistakes.

In practice, the researcher composes a natural-language specification for the agent (e.g., ``verify that these 14~strings form the residue-$0$ support for weight vector $\mathbf w$ modulo~$8$, that the rational probabilities satisfy the $Z$-type KL equalities, and that all weight-$\le 2$ Pauli matrix elements vanish or match''). The agent then generates Lean source code implementing these checks. The code is compiled against the local Lean~4 toolchain and the project's reusable proof library. If the compiler returns errors, the agent reads the error message and revises the code. The same applies to the Search Agent's Python scripts: if a script crashes or produces unexpected output, the agent reads the traceback and edits the code. For the Verification Agent specifically, a common failure is that the agent calls a Mathlib lemma by a name that does not exist or that was renamed in a recent update; the compiler rejects the call, and the agent greps the Mathlib~\cite{mathlib} source tree to find the correct name and signature.

This loop works because the Lean compiler provides exact, actionable error messages; the agent does not need to produce correct code on the first attempt, only to converge.

The reusable proof library (\texttt{SS/}) was itself built through this same loop. The human researcher set each proof goal (e.g., ``prove that support inside a subset-sum class implies the diagonal action''), and the agent wrote, compiled, and debugged the Lean code over multiple iterations. Once a module was complete, subsequent proofs imported it rather than re-deriving the lemmas. The cost of each new certificate decreased as the library grew.

For a routine distance-$2$ certificate, the compile-diagnose-revise loop runs several iterations. Distance-$3$ cases are harder: exact number-field arithmetic must be set up and the surviving KL constraints evaluated term by term. The most difficult proof, the no-go for $\mathbf w=(0,1,1,2,3,3,5)$ with its $19$-equation contradiction argument, took roughly $10$ hours in the agent loop. A proof is accepted only when it compiles with no errors, no warnings, and no \texttt{sorry} placeholders.

This separation provides three advantages: specialization (each agent is prompted for a single function), verification independence (the Verification Agent cannot be influenced by earlier reasoning that may contain errors or hallucinations~\cite{Lightman2023Lets}), and natural checkpoints for human oversight (researchers review proposals before large computations and validate outputs before they propagate).

\subsection{Problem formulation}
\label{subsec:formulation}

The distance-$2$ catalogue, the residue-degenerate $((6,4,2))$ construction, and the distance-$3$ $\mathrm{BD}_{16}$ analysis all address the same basic question. For fixed code parameters $((n,K,d))$, can one find orthonormal logical states $\{\ket{j_L}\}_{j=0}^{K-1}$ such that a diagonal transversal operator realizes a nontrivial logical gate, while the code still detects all Pauli errors of weight $<d$? The search is therefore constrained simultaneously by fault-tolerant gate structure and by quantum error-detection requirements.

A transversal gate acts independently on the physical qubits, which is why it is attractive for fault tolerance: a fault on one qubit is not spread across the block by the gate itself. In the diagonal setting considered here, we write
\[
    U(\mathbf w,m):=
    \bigotimes_{i=1}^n Z\!\left(\tfrac{2\pi w_i}{m}\right),
    \qquad
    \omega_m=e^{2\pi i/m},
\]
with modulus $m$ and weight vector $\mathbf w=(w_1,\dots,w_n)\in(\mathbb Z_m)^n$. Because this operator is diagonal in the computational basis,
\[
    U(\mathbf w,m)\ket{x} = \omega_m^{\langle \mathbf w,x\rangle}\ket{x},
\]
where $\langle \mathbf w,x\rangle \equiv \sum_i w_i x_i \pmod{m}$ is the modular inner product. The modular inner product partitions bit strings into subset-sum classes $C_s(\mathbf w)$, defined by the residue condition
$x\in C_s(\mathbf w)$ iff $\langle \mathbf w,x\rangle\equiv s \pmod m$.
If a logical basis state is supported inside one such class,
\[
    \mathrm{supp}(\ket{j_L})\subseteq C_{S_j}(\mathbf w),
\]
then every basis component of $\ket{j_L}$ acquires the same phase, and hence
\[
    U(\mathbf w,m)\ket{j_L} = \omega_m^{S_j}\ket{j_L}.
\]
Thus the modular data $(m,\mathbf w,\mathbf S)$, with
$\mathbf S=(S_0,\dots,S_{K-1})$, directly determine the candidate logical diagonal action.

The gate condition alone is far from sufficient. The states must also satisfy the Knill-Laflamme (KL) conditions, which are the standard algebraic form of quantum error detection. Physically, they require that every detectable error act identically on all logical states: the error may occur, but it must not reveal which logical state was encoded. Writing $E_{\mathrm D}$ for the detectable Pauli set, the conditions are
\begin{align*}
    \bra{j_L}P\ket{k_L} & =0 \qquad (j\neq k), \\
    \bra{j_L}P\ket{j_L} & =\lambda_P
    \qquad (P\in E_{\mathrm D}),
\end{align*}
with $\lambda_P$ independent of $j$. For distance $2$, $E_{\mathrm D}$ contains all single-qubit Paulis. For distance $3$, it contains all Pauli operators of weight at most $2$. In the rest of the paper, ``full-KL'' means this complete set of constraints, not merely the linear subset used for screening.

The Subset-Sum Linear Programming (SSLP) framework~\cite{Zhang2025Transversal} is useful because it exposes a partial separation between the easy and hard parts of the problem. If
\[
    \ket{j_L}=\sum_x a_{j,x}\ket{x},
    \qquad
    p_{j,x}=|a_{j,x}|^2,
\]
then the diagonal $Z$-type KL equalities depend only on the probabilities $p_{j,x}$, not on the phases of the amplitudes. In the distance-$2$ setting they reduce to linear conditions of the form
\[
    \sum_{x\in C_{S_j}(\mathbf w)}
    (1-2x_i)\,p_{j,x} =    t_i,
    \qquad
    i\in[n],
\]
with the same $t_i$ for every logical state $j$. This linear subset of the KL conditions can therefore be checked by a small linear program-solving for non-negative probabilities that satisfy the marginal equalities. SSLP also uses the residue classes themselves to screen out many support patterns for which bit-flip errors would obviously connect different logical states.

The $Z$-type LP handles roughly half the KL constraints-those involving diagonal Pauli operators. The remaining constraints, generated by bit-flip errors ($X$ and $Y$ operators), couple amplitudes across different basis strings and cannot be linearized. After the supports are fixed, these off-diagonal conditions become nonlinear polynomial equations in the complex amplitudes. At the same time, the underlying search space is large: one must scan over the modulus $m$, the weight vector $\mathbf w$, the residue pattern $\mathbf S$, and the possible supports inside the subset-sum classes. Distinct residues simplify part of the problem because different logical states then occupy disjoint classes; repeated residues are harder because orthogonality and error cancellation must be created by interference among amplitudes, as in Sec.~\ref{subsec:extension}. For the distance-$2$ catalogue, the subset-sum and LP stages are strong enough to make exhaustive or near-exhaustive search practical. For the distance-$3$ $\mathrm{BD}_{16}$ case, however, they only reduce the problem to a small set of candidates, and the main difficulty lies in resolving the remaining nonlinear full-KL equations exactly. This separation between broad combinatorial screening and hard exact completion is the central structure exploited by the multi-agent workflow below.

\subsection{Synthesis agent for full KL solutions}
\label{subsec:comb}

The Synthesis Agent operates at three distinct stages of the pipeline: \emph{before search}, reformulating the problem into combinatorial and linear-programming terms; \emph{during search}, providing feasibility certificates and no-go filters; and \emph{after search}, extracting analytical families and no-go proofs from Lean-verified candidates.
In the distance-$2$ nondegenerate-residue regime, the pre-search role dominates: the agent reformulates code existence into a combinatorial support-selection problem together with a small linear-feasibility problem, which the Search Agent then executes at scale. Beyond this search stage, the post-search role takes over: the agent analyzes Lean-verified candidates to identify structured amplitude patterns that satisfy the remaining full KL constraints, with the goal of either lifting isolated solutions into analytical families or deriving explicit obstructions. The first role underlies the systematic catalogue; the second underlies the analytical families, the residue-degenerate $((6,4,2))$ controlled-phase code, and the distance-$3$ $((7,2,3))$ transversal-$T$ analysis.

\paragraph{Combinatorial reformulation for distance-$2$ search.}
A key simplification is that subset-sum residue classes can enforce the difficult parts of the distance-$2$ KL conditions combinatorially. A single-bit flip at site $i$ maps a string $x$ to $x\oplus e_i$, changing its residue by $\pm w_i$ modulo $m$:
\[
    \langle \mathbf w, x\oplus e_i\rangle \equiv \langle \mathbf w, x\rangle \pm w_i \pmod m.
\]
Thus, if distinct logical states are supported on residue classes $C_{S_j}(\mathbf w)$, we can forbid all Hamming-$1$ adjacencies between different logical supports by imposing
\begin{equation}
    \label{eq:shift-screen-synth}
    S_j - S_k \ \not\equiv\ \pm w_i \pmod m
    \qquad \text{for every } i\in[n] \text{ and } j\neq k.
\end{equation}
When Eq.~\eqref{eq:shift-screen-synth} holds, every single-qubit $X_i$ or $Y_i$ error either leaves a logical support block or lands outside the union of occupied residue classes, so all off-diagonal weight-$1$ KL terms vanish without phase engineering. In practice this condition serves as a fast sufficient screen for the distance-$2$ search; when some coordinates are pinned within a residue class, we may replace it by an explicit computation of the classical union distance $d(C)$.

The remaining weight-$1$ KL constraints come from the diagonal $Z_i$ operators. For a binary string $x$, define its sign vector
\[
    v(x):=((-1)^{x_1},\dots,(-1)^{x_n})\in\{\pm1\}^n,
\]
and for each residue class set
\[
    V_j:=\{v(x):\ x\in C_{S_j}(\mathbf w)\}.
\]
The set of all single-site $Z$-expectation vectors realizable by a state supported on class $j$ is exactly the convex hull $\operatorname{conv}(V_j)\subset[-1,1]^n$. Therefore the distance-$2$ $Z$-type KL equalities hold if and only if these convex sets share a common point. Equivalently, there must exist probabilities $\{p_{j,x}\}$ and a common expectation vector $q=(t_1,\dots,t_n)$ such that
\begin{align}
    \label{eq:Zmatch-comb}
     &\sum_{x\in C_{S_j}(\mathbf w)} (1-2x_i)\,p_{j,x}\ =\ t_i,
    \quad \forall i\in[n],\ \forall j, \notag\\
     &\sum_{x\in C_{S_j}} p_{j,x}=1,\ \ p_{j,x}\ge 0.
\end{align}
The convex-hull formulation is the core of the SSLP framework: a nonlinear amplitude problem is reduced to support selection plus a linear-feasibility problem on probabilities. The concrete large-scale implementation of this screen is described in Sec.~\ref{subsec:search}.

\paragraph{Sparse representatives and fast no-go filters.}
Assume $\bigcap_j \operatorname{conv}(V_j)\neq \varnothing$, and choose a rational common point $q$ in this intersection; such a point exists because each $\operatorname{conv}(V_j)$ is a rational polytope. For each class $j$, Carath\'eodory's theorem~\cite{Caratheodory1907Uber} implies that $q$ can be written as a convex combination of at most $n+1$ points of $V_j$. Since the corresponding linear system has rational data, these coefficients may be taken rational. Clearing denominators then yields a positive integer $L$, an integer vector $t:=Lq\in\mathbb Z^n$, and non-negative integer vectors $u_j$, each supported on at most $n+1$ entries, such that
\begin{equation}
    \label{eq:int-feas-comb}
    A_j u_j\ =\ t,\qquad
    \mathbf 1^\top u_j\ =\ L,\qquad
    u_j\ \in\ \mathbb Z_{\ge 0}^{|C_{S_j}|}
    \quad (\forall j),
\end{equation}
where the columns of $A_j$ are the sign vectors in $V_j$. Normalizing by $L$ recovers rational probabilities. The exact consequence of Carath\'eodory is therefore a bound on the support size of each witness, not a bound of the form $L\le n+1$ on the common denominator; this sparsity bound helps explain why sparse exact solutions are common in the catalogue.

The same convex viewpoint also yields fast no-go certificates for the search stage. The emptiness of the common intersection $\bigcap_j \operatorname{conv}(V_j)$ is exactly the infeasibility of Eq.~\eqref{eq:Zmatch-comb}; a particularly cheap sufficient certificate is a pairwise linear separator. Concretely, if there exist $\alpha\in\mathbb Z^n$ and $\beta\in\mathbb R$ such that
\[
\max_{x\in C_{S_j}} \alpha\cdot v(x)\ <\ \beta\ <\ \min_{y\in C_{S_k}} \alpha\cdot v(y)
\quad\text{for some } j\neq k,
\]
then $\operatorname{conv}(V_j)$ and $\operatorname{conv}(V_k)$ are disjoint, hence
\[
    \bigcap_{\ell} \operatorname{conv}(V_{\ell})=\varnothing,
\]
and Eq.~\eqref{eq:Zmatch-comb} is infeasible. For more than two residue classes, not every failure of common intersection is witnessed by such a pairwise separator, so this criterion is sufficient rather than complete. Nevertheless, these separators are inexpensive to evaluate and are especially useful for homogeneous-weight patterns and affine-slice residue classes.

\paragraph{Synthesis for the remaining full KL constraints.}
Passing the residue screen and the $Z$-only feasibility test is only a necessary condition for code existence. The remaining full KL equations involve amplitude cross-terms generated by bit flips, and at this stage the Synthesis Agent switches from search reformulation to ansatz discovery. It inspects Lean-verified numerical or exact candidates for recurring support decompositions, parity constraints, complement symmetries, and character-like sign patterns, then proposes low-dimensional exact templates for amplitudes and phases. For distance $2$, this process promotes isolated search hits into analytical families such as the $C_0=\{0^n,1^n\}$ family and the even-parity family, and it is also how the residue-degenerate $((6,4,2))$ controlled-phase construction was organized into a human-readable proof.

The same synthesis loop applies in the distance-$3$ setting, where the subset-sum and LP stages remain intact but the detectable error set expands to all Pauli errors of weight at most $2$. In the binary-dihedral $\mathrm{BD}_{16}$ specialization for transversal $T$, focusing on small $((7,2,3))$ codes and the complementary convention $\ket{1_L}=X^{\otimes 7}\ket{0_L}$, the agent reduces the remaining full KL conditions to explicit finite constraints on the amplitudes of $\ket{0_L}$. It then proposes exact support/sign ans\"atze for the realizable weight vectors and, for the LP-feasible but unrealizable cases, isolates incompatible subsets of the weight-$2$ KL equations to obtain no-go proofs. These obstructions are qualitatively different from the convex no-go certificates above: the subset-sum and LP filters succeed, but no exact state can satisfy all remaining full KL conditions simultaneously.

All agent-proposed families, exact solutions, and no-go arguments are accepted only after independent Lean-based verification of the exact KL equalities and induced logical actions.

\subsection{Search space and enumeration pipeline for $d=2$}
\label{subsec:search}

We now describe how the SSLP formulation is instantiated at scale to enumerate codes with transversal diagonals and distance $2$.

\subsubsection{Canonical search space and guards}

To avoid redundant enumeration, we restrict to canonical representatives. The Search Agent enumerates weight vectors $\mathbf w$ in non-decreasing order $(1\le w_1\le\cdots\le w_n\le m-1)$, and residue tuples $\mathbf S$ with $S_0=0$ and $1\le S_1<\cdots<S_{K-1}\le m-1$. Equivalence under qubit permutations is tracked so that each discovered code is genuinely distinct: two parameter sets $(\mathbf w,\mathbf S)$ are identified whenever one is obtained from the other by permuting qubit indices, and the deduplication is performed within each fixed $(n,K,m)$ triple. Two lightweight guards are applied before any heavy computation:
(i) an optional coprime filter on the residues $S_j$, and
(ii) the residue-shift screen condition
\begin{equation}
    \label{eq:shift-screen-comb}
    S_j - S_k \ \not\equiv\ \pm w_i \pmod m \quad \text{for every } i\in[n] \text{ and } j\neq k,
\end{equation}
which forbids all Hamming-1 adjacencies between supports of different logical states.

\subsubsection{Supports, union distance, and Z-only feasibility}

For each parameter set $(\mathbf w,m,\mathbf S)$ that passes the initial guards, we construct the residue classes $C_{S_j}(\mathbf w)$ by evaluating
$\langle \mathbf w,x\rangle \bmod m$ for all $x\in\{0,1\}^n$. Any set for which a class $C_{S_j}(\mathbf w)$ is empty is discarded.

We then compute the minimum distance of the union support,
\begin{equation}
    C :=\bigcup_{j=0}^{K-1} C_{S_{j}}(\mathbf w),\qquad
    d(C)=\min_{\substack{x\ne y\\x,y\in C}} d_{\mathrm H}(x,y).
    \label{eq:distance2}
\end{equation}
For the catalogue reported in the main text we impose $d(C)=2$, consistent with the distance-$2$ setting; in large sweeps we often rely on Eq.~\eqref{eq:shift-screen-comb} as a fast sufficient condition and only compute $d(C)$ for promising candidates.

Next, we test $Z$-only feasibility by solving the linear program implied by Eq.~\eqref{eq:Zmatch-comb}. Let $p_{j,x}$ denote non-negative, block-normalized probabilities on $C_{S_j}(\mathbf w)$ such that $\langle j_{L} |Z_i| j_{L} \rangle=\sum_{x\in C_{S_j}(\mathbf{w})} p_{j,x}(1-2x_i)$. The KL equalities for all $Z_i$ are satisfied if there exists a set $\{p_{j,x}\}$ obeying
\begin{equation}
    \begin{split}
         &\sum_{x\in C_{S_0}(\mathbf{w})} p_{0,x} (1-2x_i)
        = \sum_{y\in C_{S_j}(\mathbf{w})} p_{j,y} (1-2y_i),\\
         &\text{for} \ \forall i\in[n],\ \forall j=1,\dots,K-1;\\
         &\sum_{x\in C_{S_j}(\mathbf{w})} p_{j,x} = 1,\qquad
        p_{j,x}\ge 0 \quad \  \forall j,\ \forall x\in C_{S_j}(\mathbf{w}).
    \end{split}
    \label{eq:Zmatch}
\end{equation}
We assemble these into a block-structured linear system $A_{\mathrm{eq}} \mathbf{p}=b_{\mathrm{eq}}$ with $\mathbf{p}\ge 0$, where $\mathbf{p}$ concatenates all $p_{j,x}$. For $K=2$, this is an $(n+2)\times(|C_{S_0}|+|C_{S_1}|)$ system. We solve Eq.~\eqref{eq:Zmatch} numerically using standard LP solvers and then convert numerical solutions to exact rationals as described below.

\subsubsection{Rational reconstruction}
\label{subsubsec:rational-main}

Although $A_{\mathrm{eq}}$ and $b_{\mathrm{eq}}$ are integer-valued, the LP solver operates in floating-point arithmetic, and naive rounding of the resulting $\mathbf p^{(\mathrm{num})}$ can violate the constraints. To obtain exact rational solutions $\mathbf p\in\mathbb Q^N$ we exploit the underlying integer structure in two complementary ways.

First, when the solution is close to a basic feasible solution (BFS), at most $(K-1)n+K$ entries are nonzero. We then identify a full-rank basis $B$ of that size, form the integer submatrix $A_B\in\mathbb{Z}^{((K-1)n+K)\times((K-1)n+K)}$, and solve $A_B \mathbf{p}_B=b_{\mathrm{eq}}$ exactly over $\mathbb Q$, with $p_i=0$ for $i\notin B$. The resulting vector $\mathbf{p}$ is accepted only if $A_{\mathrm{eq}}\mathbf{p}=b_{\mathrm{eq}}$ and $\mathbf{p}\ge 0$ hold exactly.

Second, when the numerical solution is not clearly a BFS, we first rationalize each coordinate by continued fractions (imposing a denominator bound) and then project back onto the affine constraint space $A_{\mathrm{eq}}\mathbf p=b_{\mathrm{eq}}$ using exact rational linear algebra. Small negative entries caused by rounding are clipped to $0$ and the vector is re-projected and block-normalized. Detailed pseudocode for these two procedures is given in Appendix~\ref{alg:exactbfs-supp}-\ref{alg:ratproj-supp}~\cite{Bareiss1968Sylvester,Bertsimas1997Introduction,Schrijver1986Theory,Vonzurgathen2003Modern,Lenstra1982Factoring,Ferguson1999Analysis}.

\subsubsection{Search skeleton and recorded outputs}

The full search procedure over $(m,\mathbf w,\mathbf S)$ proceeds as follows:
(i) enumerate canonical $(\mathbf w,\mathbf S)$ subject to initial guards;
(ii) construct residue classes and enforce non-emptiness;
(iii) check the distance-$2$ screen (either via Eq.~\eqref{eq:shift-screen-comb} or by computing $d(C)$);
(iv) solve the $Z$-only LP Eq.~\eqref{eq:Zmatch};
(v) perform rational reconstruction;
(vi) assemble logical states and verify all distance-$2$ KL conditions and the transversal logical action in Lean.

For each successful hit we record $(n,m,K,\mathbf w,\mathbf S)$, the rational probabilities $\{p_{j,x}\}$, the explicit logical states $\{\ket{j_L}\}$, the per-site expectations $\langle Z_i\rangle$, and the logical order $\mathcal O$.

\subsection{Formal verification in Lean}
\label{subsec:audit}

The search pipeline outputs exact rational data for each candidate code: supports, probabilities, and when needed algebraic amplitudes. Independent Python scripts verify the full distance-$2$ catalogue by re-checking every KL equality and logical action in exact rational arithmetic. For the analytical families, the distance-$3$ constructions, and the no-go proofs, we go further and formalize the verification in Lean~4~\cite{Moura2021Lean4}.

Lean~4 is a proof assistant: a programming language in which mathematical statements and their proofs are expressed as code. When Lean compiles a file successfully, every claim in that file has been checked by Lean's kernel, a small trusted core that verifies each logical step. A proof that compiles is not merely a program that ran without errors; it is a certificate that the stated theorem follows from the axioms. The marker \texttt{sorry} in Lean denotes an unproved assumption; a codebase with no \texttt{sorry} has no gaps in its logical chain. Our proofs build on Mathlib~\cite{mathlib}, a community-maintained library of formalized mathematics that provides foundations we rely on, including finite-set operations, modular arithmetic, and rational-number fields.

The formal development is organized as the library \texttt{lean-qec}. It accepts only exported exact data and produces kernel-checked proof objects; it certifies representative distance-$2$ instances, proves the analytical families as universally quantified theorems over admissible parameters, and verifies all~$12$ distance-$3$ $\mathrm{BD}_{16}$ cases including the two impossibility proofs. The complete build ($3{,}092$ jobs, ${\sim}4{,}300$ lines across ${\sim}30$ modules) compiles with zero errors, zero warnings, and no \texttt{sorry}. Figure~\ref{fig:lean-verification-architecture} summarizes the architecture. 

\begin{figure*}[t]
    \centering
    \resizebox{0.9\textwidth}{!}{%
    \begin{tikzpicture}[
        >=Latex,
        font=\footnotesize,
        line join=round,
        line cap=round,
        x=1cm,y=1cm,
        head/.style={
            draw=black!80,
            rounded corners=3pt,
            line width=0.9pt,
            fill=gray!20,
            align=left,
            inner sep=5pt
        },
        main/.style={
            draw=black!80,
            rounded corners=3pt,
            line width=0.85pt,
            fill=white,
            align=left,
            inner sep=5pt
        },
        outcome/.style={
            draw=black!80,
            rounded corners=3pt,
            line width=0.85pt,
            fill=gray!6,
            align=left,
            inner sep=5pt
        },
        nogo/.style={
            draw=black!70,
            rounded corners=3pt,
            line width=0.85pt,
            dashed,
            fill=gray!3,
            align=left,
            inner sep=5pt
        },
        flow/.style={
            -Latex,
            draw=black!80,
            line width=0.95pt
        },
        sideflow/.style={
            -Latex,
            draw=black!75,
            line width=0.9pt,
            dashed
        },
        branchtitle/.style={
            font=\bfseries\footnotesize,
            align=left,
            text=black!85
        }
    ]

    \node[head, text width=3.6cm] (input) at (0,5.3) {%
        \textbf{Exact data exported to Lean}\\
        supports, rational probabilities,\\
        and, when needed, exact amplitudes
    };

    \node[head, text width=3.6cm] (core) at (0,2.7) {%
        \textbf{Reusable SSLP core}\\
        \texttt{BitString}, \texttt{Basic},\\
        \texttt{ResidueShiftScreen}, \texttt{DiagonalAction},\\
        \texttt{ZTypeKL}
    };

    \node[main, text width=3.6cm] (verify) at (0,0) {%
        \textbf{\texttt{Verify.lean}}\\
        exact finite certificate predicates\\
        discharged by \texttt{native\_decide}
    };

    \draw[flow] (input) -- (core);
    \draw[flow] (core) -- (verify);

    \node[branchtitle, anchor=west] at (4,6.55) {Distance-$2$: finite decidable checks and families};

    \node[main, text width=3.7cm] (d2a) at (6,5.3) {%
        \textbf{Distance-$2$ certificate}\\
        residue support, no-Hamming-$1$ screen,\\
        normalization, rational $Z$-type KL
    };

    \node[outcome, text width=4.2cm] (d2b) at (10.9,5.3) {%
        \textbf{Representative D2 examples and analytical families}\\
        \texttt{Examples/D2/*}\\
        \texttt{SS/FamilyI*}, \texttt{SS/LambdaV2*}
    };

    \node[branchtitle, anchor=west] at (4,1.6) {Distance-$3$: two-layer exact verification};

    \node[main, text width=3.6cm] (d3a) at (6,0) {%
        \textbf{Layer 1: finite rational check}\\
        \texttt{Distance3Verify}, \texttt{BDSymmetry}\\
        \texttt{Distance3Data}, \texttt{Layer1OK}\\
        support, BD relation, normalization, balances
    };

    \node[main, text width=3.9cm] (d3b) at (10.9,0) {%
        \textbf{Support-intersection screen}\\
        \texttt{Pauli}, \texttt{FullKL}\\
        identifies trivially vanishing KL constraints
    };

    \node[outcome, text width=4.1cm] (d3c) at (16.0,0) {%
        \textbf{Exact Layer 2 and certified feasible $\mathrm{BD}_{16}$ instances}\\
        \texttt{QSqrt23i} $\rightarrow$ \texttt{QSqrt235i} $\rightarrow$ \texttt{FullKLEval}\\
        \texttt{BD16Defs}\\
        \texttt{Examples/D3/BD16v*}
    };

    \node[nogo, text width=4.0cm] (nogo) at (10.9,2.7) {%
        \textbf{Machine-checked no-go proofs}\\
        reduced KL subsystems and\\
        \texttt{Examples/D3/NoGo/*}
    };

    \draw[flow] ([yshift=4pt]verify.east) -- ++(1,0) |- (d2a.west);
    \draw[flow] ([yshift=0pt]verify.east) -- (d3a.west);

    \draw[flow] (d2a.east) -- (d2b.west);
    \draw[flow] (d3a.east) -- (d3b.west);
    \draw[flow] (d3b.east) -- (d3c.west);
    \draw[sideflow] (d3b.north) -- (nogo.south);

    \end{tikzpicture}%
    }
    \caption{Architecture of the Lean~4 verification library \texttt{lean-qec}, which provides the authoritative proof layer of the pipeline. Exact exported supports, rational probabilities, and, when needed, exact algebraic amplitudes are passed to the Lean verification layer, whose shared SSLP core is organized around \texttt{Verify.lean}. The upper branch certifies distance-$2$ constructions and universally quantified analytical families through exact finite decidable checks. The lower branch specializes to the two-layer verification of the \texorpdfstring{$((7,2,3))$}{((7,2,3))} \texorpdfstring{$\mathrm{BD}_{16}$}{BD16} classification: Layer~1 packages the finite rational support and balance checks in \texttt{Distance3Verify} together with the complementary-symmetry module \texttt{BDSymmetry}; the middle stage uses \texttt{Pauli} and \texttt{FullKL} to identify which KL constraints vanish trivially by the support-intersection test; the final stage evaluates the surviving equations in exact number-field arithmetic and packages the shared \texorpdfstring{$\mathrm{BD}_{16}$}{BD16} bundle used by the certified feasible instances. The dashed side branch denotes the corresponding machine-checked no-go proofs. Representative modules are shown.}
    \label{fig:lean-verification-architecture}
\end{figure*}

An early version of the library was a single large file, but we found that the Verification Agent produced more reliable proofs when each module was short enough to fit within its working context. We therefore refactored into focused modules matching the decomposition visible in Fig.~\ref{fig:lean-verification-architecture}: a reusable SSLP core (\texttt{Basic.lean} through \texttt{Verify.lean}) handling support membership, the residue-shift screen, diagonal action, and $Z$-type KL equalities; family directories (\texttt{FamilyI/}, \texttt{LambdaV2/}) for the universally quantified analytical proofs; and a distance-$3$ extension adding Pauli enumeration, the support-intersection screen, exact number-field arithmetic, and $\mathrm{BD}_{16}$ certificate bundling. Adding a new code instance then requires only a new data file that imports the existing modules, not new proof infrastructure. The next two subsections describe the two main verification paths-finite decidable checks for distance~$2$, and two-layer exact verification for distance~$3$-in concrete terms.

\subsubsection{Distance-$2$ certificates from finite decidable checks}

For the nondegenerate-residue distance-$2$ regime, Lean operates on exact finite data
\[
    D=(n,m,K,\mathbf w,\mathbf S,\{\operatorname{supp}_j\}_{j=0}^{K-1},\{p_{j,x}\}),
\]
where $\operatorname{supp}_j\subseteq C_{S_j}(\mathbf w)$ and $p_{j,x}\in \mathbb Q_{\ge 0}$ satisfy $\sum_{x\in \operatorname{supp}_j} p_{j,x}=1$. In the library this is packaged as an \texttt{ExampleData} record. The ambient basis type is \texttt{BitString n := Fin n -> Bool}, so $\{0,1\}^n$ becomes a finite decidable type of size $2^n$. The bundled predicate \texttt{ExampleData.OK} asserts four exact facts: (i) each claimed support element lies in the advertised subset-sum residue class; (ii) the residue-shift screen Eq.~\eqref{eq:shift-screen-comb}, or an equivalent explicit no-Hamming-$1$ condition, holds between different logical classes; (iii) the rational probabilities are block-normalized exactly; and (iv) the $Z$-type KL equalities Eq.~\eqref{eq:Zmatch} hold exactly over $\mathbb Q$. Because all quantifiers range over finite types, these checks can be decided by direct computation. Lean's \texttt{native\_decide} tactic compiles the predicate into native machine code, evaluates it (which takes seconds for $n\le 7$), and, if it returns \texttt{true}, accepts the result as a kernel-checked proof. No manual proof steps are needed; the computation itself is the proof. The final certificate is a single-line theorem of the form \texttt{theorem verified : myCode.OK := by native\_decide}. For the representative files \texttt{Examples/D2/Ex522.lean} and \texttt{Examples/D2/Ex622.lean}, this one-line theorem is the final certificate.

The reusable lemmas then propagate \texttt{ExampleData.OK} to the actual coding-theoretic statement. In particular, \texttt{Dact\_eq\_global\_of\_SS} proves that support inside a subset-sum class implies the advertised logical action
\[
    U(\mathbf w,m)\ket{j_L}=\omega_m^{S_j}\ket{j_L},
\]
and \texttt{no\_hamming1\_neighbor\_of\_screen} turns Eq.~\eqref{eq:shift-screen-comb} into exact inter-class Hamming separation, which kills every single-qubit $X_i$ or $Y_i$ Knill-Laflamme term automatically. Together with the already-certified $Z$-type equalities, these lemmas yield the full weight-$1$ KL conditions for distance~$2$ without any floating-point tolerances. The same mechanism also certifies the analytical families: files in \texttt{FamilyI/} and \texttt{LambdaV2/} formalize these constructions as universally quantified theorems over admissible parameters, so one Lean proof covers infinitely many instantiations.

\subsubsection{Two-layer exact verification for the \texorpdfstring{$((7,2,3))$}{((7,2,3))} \texorpdfstring{$\mathrm{BD}_{16}$}{BD16} problem}

The distance-$3$ $\mathrm{BD}_{16}$ classification requires a stronger certificate because the surviving full-KL equations are quadratic in the amplitudes rather than linear in the probabilities. The distance-$3$ modules therefore use a two-layer architecture. Layer~1 is still purely finite and rational: a \texttt{Distance3Data} record specifies the weight vector, the distinguished residue support, the complementary convention $\ket{1_L}=X^{\otimes 7}\ket{0_L}$, and the rational probabilities $p_x=|c_x|^2$. The predicate \texttt{Layer1OK} checks the subset-sum support data, the $\mathrm{BD}_{16}$ residue relation $S_1\equiv \sum_i w_i-S_0 \pmod m$, exact normalization, and the linear balance equations
\[
    \sum_{x\in S_0}(-1)^{x_i}p_x=0,\qquad i=1,\dots,7.
\]
In particular, the two-body diagonal constraints $Z_iZ_j$ are automatic in this complementary setting once the single-site balances hold, so Layer~1 remains an LP-style rational check and is again discharged by \texttt{native\_decide}.

Layer~2 handles the genuinely nonlinear part: the weight-$\le 2$ Pauli matrix elements involving bit flips. Writing $P=i^{N_Y(P)}X^{x(P)}Z^{z(P)}$, the verifier evaluates
\[
    \bra{0_L}P\ket{0_L} =    i^{N_Y(P)}
    \sum_{\substack{s\in S_0 \\ s\oplus x(P)\in S_0}}
    (-1)^{z(P)\cdot s} \overline{c_{s\oplus x(P)}}\,c_s.
\]
The first simplification is combinatorial: if $S_0\cap(S_0\oplus x(P))=\varnothing$, then the sum is empty and the corresponding KL matrix element vanishes identically. The verifier checks this intersection test for every weight-$\le 2$ Pauli before attempting any algebraic computation. The test eliminates the large majority of constraints: in the representative file \texttt{BD16v1.lean}, only $25$ of the $210$ weight-$\le 2$ Pauli constraints have nonempty intersections and require explicit evaluation.

For the remaining $25$ constraints, the verifier must compute the matrix-element sum and confirm it is zero. The amplitudes in the $\mathrm{BD}_{16}$ solutions involve $\sqrt{2}$, $\sqrt{3}$, $\sqrt{5}$, and~$i$. Neither Lean nor Mathlib provides arithmetic over this number field, so the Verification Agent built it from scratch: each amplitude is stored as a tuple of $16$ rational coefficients in the basis $\{1,\sqrt{2},\sqrt{3},\sqrt{6},\sqrt{5},\sqrt{10},\sqrt{15},\sqrt{30}\}\times\{1,i\}$, with multiplication rules encoding the identities $\sqrt{2}^2=2$, $\sqrt{2}\cdot\sqrt{3}=\sqrt{6}$, and so on. Lean also requires a proof that addition in this representation is commutative and associative before it will accept sums over finite sets; these properties are derived from the corresponding properties of the rationals. The resulting arithmetic is exact: equality reduces to comparing $16$ rational numbers with no rounding. For each surviving Pauli $P(x,z)$, the evaluator computes the matrix-element sum
\[
    \sum_{s\,\in\, S_0\,\cap\,(S_0\oplus x)}(-1)^{z\cdot s}\,\overline{c_{s\oplus x}}\,c_s
\]
term by term in this exact arithmetic-conjugating, multiplying, and accumulating-then checks that all $16$ coefficients of the result are zero. The final certificate is again a single \texttt{native\_decide} theorem. The files \texttt{Examples/D3/BD16v1.lean} through \texttt{BD16v10.lean} implement this two-layer verification for $12$ exact instances, covering all $10$ feasible $\mathrm{BD}_{16}$ weight vectors reported in Sec.~\ref{sec:bd16-distance3}.

Proving that a candidate code \emph{cannot} exist is something numerical methods cannot do here-they can only report failure to find a solution. The Lean formalization handles this as well. For the two excluded $\mathrm{BD}_{16}$ vectors, the proof extracts a small subsystem of the KL equations-$17$ equations for $\mathbf w=(1,1,1,2,2,2,6)$ and $19$ for $\mathbf w=(0,1,1,2,3,3,5)$-and shows it has no solution over~$\mathbb C$. The proof strategy has five steps: (i)~solve the linear diagonal equations for the probabilities $|c_k|^2$; (ii)~combine pairs of off-diagonal equations into bilinear relations among amplitudes (e.g., $\bar b\, a = -\bar d\, c'$); (iii)~use the probability constraints to show that all relevant amplitudes are nonzero; (iv)~exploit global phase freedom to make all amplitudes real; and (v)~multiply the three bilinear relations together and cancel common nonzero factors, arriving at $af = -af$ with $a,f\neq 0$, a contradiction. The full argument is formalized and machine-checked in Lean. Throughout, the final outputs of the pipeline are exact kernel-checked certificates, with no unresolved \texttt{sorry} placeholders.

\section{Results for $d=2$}
\label{sec:results}

\subsection{Catalogue of distance-$2$ diagonal-transversal codes ($K \leq 4$, $n \leq 6$)}
\label{subsec:catalogue}

\begin{figure*}[t]
    \centering
    \begin{minipage}[t]{0.67\linewidth}
        \centering
        \includegraphics[width=\linewidth]{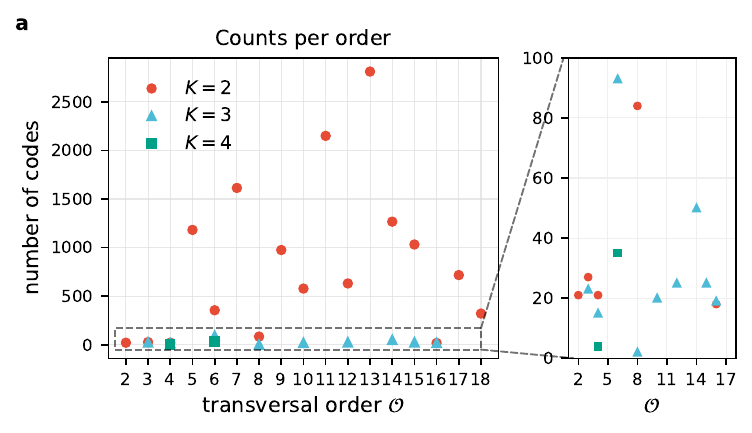}\par\vspace{3pt}
        \includegraphics[width=\linewidth]{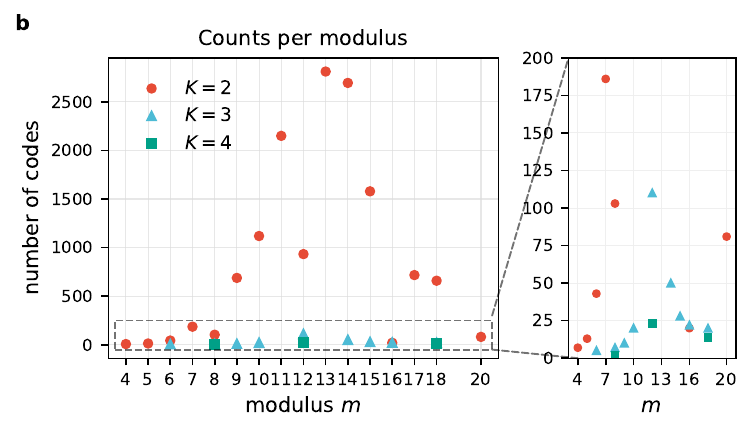}
    \end{minipage}\hfill
    \raisebox{-0.4\linewidth}{%
    \begin{minipage}[t]{0.33\linewidth}
        \centering
        \includegraphics[width=\linewidth]{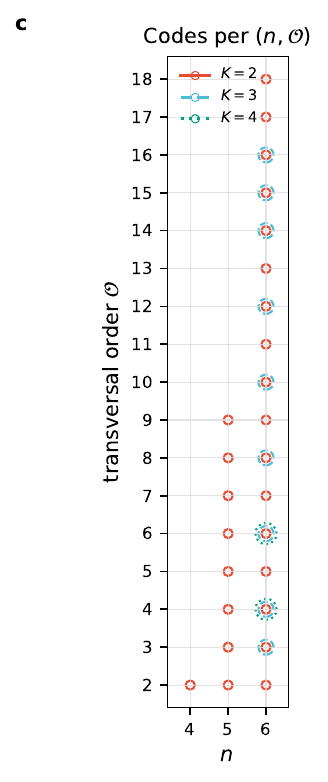}
    \end{minipage}}
    \caption{Global statistics of the diagonal-transversal code catalogue. We aggregate all nondegenerate distance-$2$ codes with $K\leq 4$ and $n\leq 6$ obtained from the multi-agent SSLP search. (a) Number of codes found at each transversal order $\mathcal{O}$, separated by logical dimension $K=2,3,4$ (colours and markers as indicated in the legend). (b) Number of codes as a function of the modulus $m$, resolved by $K$. (c) Existence landscape in the $(n,\mathcal{O})$ plane: for each pair $(n,\mathcal{O})$ where at least one code is found, we draw concentric rings centred at $(\mathcal{O},n)$, with the inner, middle and outer rings indicating the presence of codes with $K=2$, $K=3$ and $K=4$, respectively. Panels~(a) and~(b) encode absolute counts, while panel~(c) highlights which regions of the $(n,\mathcal{O})$ grid are populated by different logical dimensions.}
    \label{fig:search_codes}
\end{figure*}

We next characterise the landscape of distance-2 diagonal-transversal codes uncovered by the multi-agent SSLP search in the regime $K\leq 4$ and $n\leq 6$. After solving all subset-sum instances, performing rational reconstruction, and re-checking the Knill-Laflamme and logical transversality conditions, we obtain a certified catalogue of nondegenerate codes. In total, the catalogue contains $14{,}116$ distinct canonical parameter sets $(m,\mathbf w,\mathbf S)$, where two parameter sets are identified if one is obtained from the other by permuting qubit indices within a fixed $(n,K,m)$ triple. Fig.~\ref{fig:search_codes} summarises global statistics over the full search space, while Appendix~C collects representative parameter tables and explicit examples.

The representative tables in Appendix~C list one canonical parameter set $(m,\mathbf{w},\mathbf{S})$ for each realised triple $(K,n,\mathcal{O})$ and exhibit a clear hierarchy across logical dimension. For $K=2$, the catalogue is already dense: all orders $\mathcal{O}=2,\dots,18$ occur at $n=6$, and several lower orders also admit shorter realisations at $n=4$ or $5$. For $K=3$, the currently known representatives all occur at $n=6$ with $\mathcal{O}\in\{3,4,6,8,10,12,14,15,16\}$, while the present $K=4$ catalogue consists of two $n=6$ constructions with $\mathcal{O}\in\{4,6\}$. Low-order representatives often use nearly homogeneous weight vectors, whereas larger orders typically require more heterogeneous patterns.

Figure~\ref{fig:search_codes}a aggregates all nondegenerate codes by transversal order. Several intermediate orders, such as $\mathcal{O}=11$ and $13$, support noticeably more codes than their neighbours, suggesting particularly favourable congruence structure in the underlying subset-sum constraints. The curves for $K=3$ and $K=4$ are shifted down by several orders of magnitude, consistent with the reduced combinatorial volume for higher-dimensional logical spaces at fixed $(n,m)$ and with the additional convex-hull intersection constraint. Nevertheless, most admissible orders host at least one $K=3$ code, and a subset also supports a $K=4$ code, showing that diagonal-transversal resources are not confined to very low logical order.

Figure~\ref{fig:search_codes}b reorganizes the same data as a function of the modulus. Certain moduli support markedly richer families of codes, with $m=13$ and $14$ particularly prolific for $K=2$. Only a subset of these moduli admit higher-$K$ codes, but overlap is common: several values of $m$ carry simultaneous solutions for $K=2$ and $K=3$, and some also host $K=4$ codes. This arithmetic clustering indicates that once a modulus supports one feasible residue pattern, nearby patterns often generate additional solutions at different orders and with different weight profiles, turning each productive modulus into a local family of diagonal-transversal constructions.

The existence map in the $(n,\mathcal{O})$ plane is disentangled from absolute counts in Fig.~\ref{fig:search_codes}c. For each pair where at least one code exists, concentric rings mark the presence of $K=2$, $K=3$, and $K=4$ codes. The inner ring almost saturates the accessible grid for $n=6$, consistent with the dense $K=2$ statistics in panels~(a) and~(b), while $K=3$ and $K=4$ appear exclusively at $n=6$ but over many orders. Notably, no point in the grid supports a $K=3$ or $K=4$ code without a companion $K=2$ code at the same $(n,\mathcal{O})$, suggesting that $K=2$ solutions may form a combinatorial backbone from which higher-dimensional constructions can be built.

Taken together, the statistics in Fig.~\ref{fig:search_codes} provide a compact global picture of the small-distance diagonal-transversal landscape. They highlight where the parameter space is dense, which orders and moduli are especially productive, and where gaps remain open. These observations guide our extensions to larger $n$ and higher distances, and suggest concrete analytical questions about when a given diagonal-transversal group can be realised by a nonadditive code.

\subsection{Analytical families and scaling patterns}
\label{subsec:families}

The catalogue of certified codes serves not only as a list of isolated instances, but also as a dataset from which to extract structure. After completing sweeps over $n\le 6$ and $K\in\{2,3,4\}$, the Synthesis Agent analyzed Lean-verified instances selected by the human researcher and proposed parameterizations that explain recurring residue patterns, amplitude assignments, and logical phase structures. From these proposals we identified several infinite families of distance-$2$ codes that can be constructed analytically within the SSLP framework. We highlight two representative families that capture the main scaling patterns. In both, the distance-$2$ verification is transparent: residue or parity bookkeeping removes the single-qubit $X/Y$ terms, while symmetry reduces the remaining $Z$-type KL conditions to simple marginal equalities. Full derivations and explicit small-$n$ instances are given in Appendix~\ref{supp:families-examples}.

\subsubsection{Family I: $C_0=\{0^n,1^n\}$}
\label{subsubsec:family1}

The first family is seeded by the extremal codewords $\{0^n,1^n\}$. For each $n\ge 2$ we choose a modulus $m\ge n$ and weight vector
\[
    \mathbf w=(1,\dots,1,m-(n-1))\in(\mathbb Z_m)^n,
\]
and write
\[
    C_s=\{x\in\{0,1\}^n:\ \langle \mathbf w,x\rangle \equiv s \pmod m\}.
\]
Writing $y=(u,b)$ with $u\in\{0,1\}^{n-1}$, $b\in\{0,1\}$, and $t=\mathrm{wt}(u)$, we have
\[
    \langle \mathbf w,y\rangle \equiv t+b(m-(n-1)) \pmod m.
\]
Hence the residue-$0$ class contains exactly the two extremal strings,
\[
    C_0=\{0^n,1^n\}.
\]
In the two-slice window $m-(n-1)\le s\le n-1$, the class $C_s$ decomposes as
\begin{align*}
     & A_s=\{(u,0):u\in\{0,1\}^{n-1},\ \mathrm{wt}(u)=s\}, \\
     & B_t=\{(u,1):u\in\{0,1\}^{n-1},\ \mathrm{wt}(u)=t\}, \\
     & t=n-1+s-m,
\end{align*}
so $C_s=A_s\sqcup B_t$. Choosing $\ket{0_L}$ on $C_0$ and $\ket{1_L}$ on $C_s$, the SSLP $Z$-marginal constraints admit the site-symmetric solution
\begin{align*}
     & \ket{0_L}= \sqrt{1-\frac{s}{m}} \ket{0^n}+\sqrt{\frac{s}{m}} \ket{1^n}, \\
     & \ket{1_L}
    =\sum_{y\in A_s}\sqrt{\frac{m-s}{m\binom{n-1}{s}}} \ket{y} +\sum_{y\in B_t}\sqrt{\frac{s}{m\binom{n-1}{t}}} \ket{y}.
\end{align*}
By symmetry these states have the same single-site expectation value on every qubit,
\[
    \langle Z_i\rangle = 1-\frac{2s}{m},\qquad i=1,\dots,n.
\]
The associated diagonal transversal gate
$U(\mathbf w,m)=\bigotimes_j Z(2\pi w_j/m)$ acts on the code space as
\[
    \overline U=\mathrm{diag}(1,\omega_m^{s}),
\]
so the generated logical group has order
\[
    \mathcal O=\frac{m}{\gcd(m,s)}.
\]
Moreover, if
\[
    s\not\equiv \pm 1,\ \pm(m-(n-1)) \pmod m,
\]
then a single bit flip changes the residue by $\pm w_i$ and cannot carry a basis state between the occupied classes $C_0$ and $C_s$. All weight-1 $X_i/Y_i$ off-diagonal KL terms therefore vanish combinatorially, so the code has distance~2.

This family explains many of the high-order entries in the catalogue. For example, at $((n,K))=((5,2))$ taking $m=5$ and $s=2$ reproduces a $((5,2,2))$ code with $\overline U=\mathrm{diag}(1,\omega_5^2)$, while at $((n,K))=((6,2))$ taking $m=7$ and $s=3$ produces a $((6,2,2))$ code with $\overline U=\mathrm{diag}(1,\omega_7^3)$ and order $\mathcal O=7$, matching the independently discovered instances.

\subsubsection{Family II: even-parity subset-sum codes}
\label{subsubsec:family2}

A second family arises by restricting supports to the even-parity subcode
\[
    \mathsf E=\{\sigma\in\{0,1\}^n:\mathrm{wt}(\sigma)\equiv 0\ (\bmod\ 2)\}
\]
for even $n$. Given a modulus $m$, a weight vector $\mathbf w\in(\mathbb Z_m)^n$, and distinct residues $\mathbf S=\{S_0,\dots,S_{K-1}\}\subset\mathbb Z_m$ with $S_0=0$, we consider the even-parity subset-sum classes
\[
    C^{(+)}_{S_k}(\mathbf w)=\bigl\{\sigma\in\mathsf E:\textstyle\sum_{j=1}^{n} w_j\sigma_j\equiv S_k \pmod m\bigr\},
\]
and define logical states as uniform superpositions over these supports:
\[
    \ket{k_L} = \frac{1}{\sqrt{|C^{(+)}_{S_k}|}}\sum_{\sigma\in C^{(+)}_{S_k}}\ket{\sigma},\qquad k=0,\dots,K-1.
\]
Because every support lies in $\mathsf E$, any single-qubit $X_i$ or $Y_i$ error maps basis states to odd-parity strings outside the support of every logical state, so all $X/Y$ matrix elements vanish identically. For the $Z$-error constraints it suffices that each support $C^{(+)}_{S_k}$ be column-balanced: for every qubit index $i$, exactly half of the strings in $C^{(+)}_{S_k}$ have a $1$ at position $i$. Then
\[
    \langle k_L|Z_i|k_L\rangle=0\qquad \text{for all }k\text{ and }i,
\]
and, because the supports are disjoint, also
\[
    \langle k_L|Z_i|\ell_L\rangle=0\qquad \text{for }k\neq \ell.
\]
Thus the weight-1 KL conditions hold and the code has distance~2. The diagonal transversal gate $U(\mathbf w,m)$ induces
\[
    \overline U=\mathrm{diag}(\omega_m^{S_0},\dots,\omega_m^{S_{K-1}}),
\]
so the logical group generated by $U$ has order
\[
    \mathcal O = \frac{m}{\gcd(m,S_1,\dots,S_{K-1})}.
\]
In the catalogue instances, column balance is enforced by involutive symmetries of each support (for example bitwise complement or structured pairings), which is why simple uniform amplitudes already solve the KL equations; concrete realizations and sufficient symmetry conditions are given in Appendix (Section~\ref{subsubsec:supp:family2}).

Within this family we obtain examples with different dimensions and logical orders. Choosing suitable $(m,\mathbf w,\mathbf S)$ for $n=4$ and $n=6$ yields $((n,K,d))=((4,2,2))$ and $((6,2,2))$ codes with logical order $\mathcal O=2$, while a construction at $n=6$ with $K=3$ realizes $\overline U=\mathrm{diag}(1,\omega_9^3,\omega_9^6)$ of order $\mathcal O=3$. Together with Family~I, these constructions account for a substantial fraction of the catalogue entries with small $n$ and demonstrate how multi-agent-guided analysis can lift isolated search hits into scalable code constructions.

\subsection{Beyond nondegenerate residues}
\label{subsec:extension}

All constructions discussed so far impose two simplifying guards on the SSLP framework: (i) residue nondegeneracy, where each logical state $\ket{j_L}$ is supported on a distinct residue class $C_{S_j}(\mathbf w)$, and (ii) the classical union-distance condition $d(C)=2$. Together these reduce the design problem to combinatorial residue screens plus small $Z$-only programs: all single-qubit $X/Y$ matrix elements vanish by bookkeeping, while the remaining $Z$-marginal equalities become linear feasibility tests. The catalogue in the previous sections was obtained entirely within this regime.

To access a richer design space, we now relax both guards. If several logical states share the same residue class, the diagonal transversal $U(\mathbf w,m)$ is degenerate on that block, and the Knill-Laflamme conditions must be enforced by structured sign or phase patterns rather than by residue separation alone. Allowing Hamming-$1$ neighbours inside the union support similarly shifts the burden from combinatorial screening to amplitude design. The following construction shows that the multi-agent system can still find exact distance-$2$ codes in this harder regime.

As a test case, we ask the agents to construct a distance-$2$ code implementing a non-trivial two-qubit logical gate, namely a controlled phase with eigenvalues $(1,1,1,i)$. The Synthesis and Search Agents jointly identify a residue-degenerate $((6,4,2))$ code with modulus $m=4$ and weight vector $\mathbf w=(1,3,2,2,2,2)\in\mathbb Z_4^6$, for which the residue of $x=(x_1,\dots,x_6)$ is
\begin{align*}
    \operatorname{res}(x) & \equiv \mathbf w\cdot x \pmod 4                              \\
                          & = x_1 - x_2 + 2\,(x_3\oplus x_4\oplus x_5\oplus x_6)\pmod 4.
\end{align*}
Thus the residue is determined by the first two bits and the parity of the last four. We organize the last four qubits into even- and odd-parity subsets indexed by $t\in\mathbb F_2^3$ via
\[
    \phi(t)=(t_1,t_2,t_3,t_1\oplus t_2\oplus t_3),
    \psi(t)=(t_1,t_2,t_3,1\oplus t_1\oplus t_2\oplus t_3),
\]
and introduce characters
\[
    \chi_3(t)=(-1)^{t_1},\qquad \chi_4(t)=(-1)^{t_2},\qquad \chi_5(t)=(-1)^{t_3},
\]
together with sign patterns
\[
    s_0(t)=1,\qquad s_1(t)=\chi_3(t)\chi_4(t),\qquad s_2(t)=\chi_3(t)\chi_5(t).
\]
The four logical states are then
\[
    \ket{j_L}
    =\frac{1}{4}\sum_{t\in\mathbb F_2^3} s_j(t)(\ket{00\,\phi(t)}+\ket{11\,\phi(t)}),
    \qquad j=0,1,2,
\]
and
\[
    \ket{3_L}
    =\frac{1}{4}\sum_{t\in\mathbb F_2^3} \chi_5(t)(\ket{10\,\phi(t)}+\ket{01\,\psi(t)}).
\]
Three logical states therefore share residue value $0$, while the fourth occupies residue value $1$. All four states are normalized flat-amplitude superpositions; character orthogonality on $\mathbb F_2^3$ gives orthogonality within the residue-$0$ block, and residue separation makes $\ket{3_L}$ orthogonal to the other three.

The distance-$2$ KL equations are enforced partly by residue screening and partly by sign cancellation. For qubits $3$-$6$, any bit flip toggles the parity of the last four bits and shifts the residue by $2$, mapping occupied strings to residue values $2$ or $3$ that are unoccupied by the code. Hence all $X_i$ and $Y_i$ matrix elements vanish for $i\in\{3,4,5,6\}$. For the corresponding $Z_i$, each logical state has a $50/50$ split between $0$ and $1$ in coordinate $i$, so the diagonals vanish, while the off-diagonals reduce to averages of nontrivial characters and therefore cancel.

For qubits $1$ and $2$, the construction lies genuinely beyond the classical union-distance condition: the union support contains Hamming-$1$ neighbours, for example $00\,\phi(t)$ and $10\,\phi(t)$. The potentially nonzero $X_1,X_2,Y_1,Y_2$ overlaps between the residue-$0$ block and $\ket{3_L}$ are nevertheless proportional (up to phases) to character sums of the form
\[
    \sum_{t\in\mathbb F_2^3} s_j(t)\chi_5(t),
\]
which vanish by orthogonality; within the residue-$0$ block, a single flip of qubit $1$ or $2$ leaves the support. The $Z_1$ and $Z_2$ diagonals cancel between the $(00)$ and $(11)$ halves of $\ket{0_L},\ket{1_L},\ket{2_L}$ and between the $(10)$ and $(01)$ halves of $\ket{3_L}$. The off-diagonals within the residue-$0$ block cancel term by term between the $(00)$ and $(11)$ halves, and any cross term with $\ket{3_L}$ vanishes because $Z_1$ and $Z_2$ are diagonal and the supports are disjoint. Consequently every weight-$1$ Pauli operator $E\in\{X_i,Y_i,Z_i\}$ satisfies the distance-$2$ Knill-Laflamme equations,
\[
    \bra{j_L}E\ket{k_L}=0\quad (j\neq k),
    \qquad
    \bra{j_L}E\ket{j_L}=\bra{k_L}E\ket{k_L}.
\]

Finally, the diagonal transversal operator
\[
    U=\bigotimes_{j=1}^6 Z\!\left(\tfrac{\pi}{2} w_j\right)
\]
acts on a basis state as
\[
    U\ket{x}=i^{\operatorname{res}(x)}\ket{x},
\]
so it contributes a constant phase to each occupied residue class. Since $\ket{0_L},\ket{1_L},\ket{2_L}$ lie entirely in residue value $0$ and $\ket{3_L}$ lies in residue value $1$, the induced logical action is
\[
    U_L=\mathrm{diag}(1,1,1,i).
\]
A full derivation and explicit computational-basis expansions are given in Appendix~\ref{supp:degenerate-642}. This construction lies strictly beyond the nondegenerate-residue and classical union-distance conditions used in our systematic sweeps, demonstrating that relaxing those guards yields genuinely new codes and logical gates within the SSLP framework.

\section{$7$-qubit distance-$3$ codes with transversal $T$}
\label{sec:bd16-distance3}

Having established the breadth of the distance-$2$ code space and the analytical families it contains, we now turn to the harder distance-$3$ regime, where the SSLP filters are necessary but no longer sufficient and the full nonlinear KL constraints must be resolved exactly.

At distance $3$, the SSLP filters no longer come close to settling the code-construction problem by themselves. For the distance-$2$ catalogue, residue separation together with the $Z$-marginal LP stage already reduced much of the search to combinatorics plus convex feasibility. At distance $3$, by contrast, the same subset-sum and LP stages remain useful only as filters: any surviving candidate must still satisfy all Pauli KL conditions of weight at most $2$, and the remaining $X/Y$-type constraints become coupled polynomial equations in amplitudes and phases. The small-code $((7,2,3))$ transversal-$T$ problem therefore provides a sharp test of the full workflow. It is small enough to admit an exact classification, but already rich enough that the final obstruction/realization step is genuinely nontrivial.

The $T$ gate ($\pi/8$ rotation) is of particular interest because, together with the Clifford group, it generates a universal gate set for quantum computation~\cite{Bravyi2005Universal}; transversal implementations of $T$ are therefore directly relevant to fault-tolerant architectures.

We study this problem in the binary-dihedral $\mathrm{BD}_{16}$ specialization of SSLP at modulus $m=8$, with residue pair $(S_0,S_1)=(0,7)$ and the complementary convention
\[
    \ket{1_L}=X^{\otimes 7}\ket{0_L}.
\]
In this setting the induced logical diagonal is the order-$8$ $T$-type phase appearing in the $\mathrm{BD}_{16}$ picture, so the problem is to determine exactly which small $((7,2,3))$ codes admit such a transversal realization. Under the complementary ansatz, the support of $\ket{1_L}$ is fixed by that of $\ket{0_L}$, and once a weight vector $\mathbf w$ is chosen the unknowns reduce to amplitudes on a single residue class. Ref.~\cite{Zhang2025Transversal} carried this reduction through the subset-sum and LP stages and identified twelve sorted weight vectors $\mathbf w$ that survive in this single-syndrome complementary setting. What remained open was the exact full-KL stage: for each surviving $\mathbf w$, do the resulting polynomial constraints admit a code, or do they force a contradiction?

We resolve all twelve cases exactly. The multi-agent pipeline combines large-scale search over supports and phase patterns with exactification of numerical candidates, symbolic synthesis of closed-form families, and independent Lean-based verification of all weight-$\le 2$ KL constraints and induced logical actions. The resulting constructions and no-go arguments are formalized and checked in Lean. This yields a complete classification within the complementary $\mathrm{BD}_{16}$ ansatz.

Table~\ref{tab:bd16-distance3} summarizes the complete classification within the complementary $\mathrm{BD}_{16}$ ansatz. Exact $((7,2,3))$ codes exist for ten of the twelve SS/LP-passing candidates, while the remaining two are ruled out by exact no-go arguments. The realizable cases include continuous families, finite discrete families, and isolated closed-form solutions, showing that the surviving parameter space remains structurally rich even after the subset-sum and LP reductions.

\begin{table*}[t]
    \centering
    \small
    \setlength{\tabcolsep}{3.5pt}
    \renewcommand{\arraystretch}{1.08}
    \caption{
    Classification of the 12 $\mathrm{BD}_{16}$ distance-$3$ candidates that pass the subset-sum and LP filters in the complementary setting $\ket{1_L}=X^{\otimes 7}\ket{0_L}$. Here $|C_{S_0}|$ denotes the size of the residue-$S_0$ support, and ``Nontrivial KL'' denotes the number of weight-$\le 2$ Knill-Laflamme constraints that survive the support-intersection screen and require explicit algebraic evaluation. Ten candidates admit exact full-KL solutions, all verified in Lean; the remaining two are ruled out by no-go proofs in the same complementary setting.
    }
    \label{tab:bd16-distance3}
\end{table*}

\begin{table*}[t]
    \begin{tabular}{@{}lcccl@{}}
        \toprule
        \textbf{Weight vector $\mathbf w$} & $|C_{S_0}|$ & \textbf{Nontrivial KL} & \textbf{Solution type} & \textbf{Status} \\
        \midrule
        $(1,2,2,2,2,3,3)$ & 18 & 83  & 2-parameter continuous & Lean $\checkmark$ \\
        $(1,1,2,2,2,2,5)$ & 14 & 97  & 192-element discrete & Lean $\checkmark$ \\
        $(0,1,2,2,3,3,4)$ & 16 & 68  & 3 isolated exact solutions & Lean $\checkmark$ \\
        $(1,1,1,1,3,3,5)$ & 13 & 81  & Continuous analytic & Lean $\checkmark$ \\
        $(1,1,1,1,3,4,4)$ & 12 & 85  & 128-element discrete& Lean $\checkmark$ \\
        $(1,1,1,2,2,4,4)$ & 16 & 83  & uniform-magnitude & Lean $\checkmark$ \\
        $(1,1,2,2,2,3,4)$ & 16 & 83  & phase-only $+$ families & Lean $\checkmark$ \\
        $(1,1,2,2,3,3,3)$ & 16 & 85  & continuous ($\sqrt{10}$ coeffs) & Lean $\checkmark$ \\
        $(1,1,1,2,3,3,4)$ & 16 & 85  & isolated ($\sqrt{10}$ coeffs) & Lean $\checkmark$ \\
        $(1,1,1,2,2,3,5)$ & 14 & 99  & isolated ($\sqrt{5}$ coeffs)  & Lean $\checkmark$ \\
        $(1,1,1,2,2,2,6)$ & 10 & 115 & proven infeasible & Lean $\checkmark$ \\
        $(0,1,1,2,3,3,5)$ & 14 & 78  & proven infeasible & Lean $\checkmark$ \\
        \bottomrule
    \end{tabular}
\end{table*}

Two representative cases illustrate both sides of the exact full-KL stage. As a positive example, the candidate
$\mathbf w=\tup{1,1,2,2,2,2,5}$ admits a closed-form solution on the residue-$0$ support,
\begin{equation}
\begin{aligned}
    &\ket{0_L}=
    \frac{1}{4}\ket{0000000}
    -\frac{\sqrt{3}}{4}\ket{0011110} \\
    &+\frac{i\sqrt{2}}{4}\Bigl(
    \ket{0100101}
    -\ket{0110001} -\ket{1000011}+\ket{1001001}
    \Bigr) \\
    &+\frac{i}{4}\Bigl(
    \ket{1101110}+\ket{1110110} +\ket{1111010}+\ket{1111100}
    \Bigr),
\end{aligned}
\label{eq:bd16-1122225-maintext}
\end{equation}
with amplitudes in $\mathbb{Q}(\sqrt{2},\sqrt{3},i)$. This support in fact carries a $192$-element discrete exact family, all with the same invariant $(\lambda^\ast)^2=33/16$. This example shows that, after the LP stage fixes the admissible probability profile, the remaining nonlinear constraints can still organize into a finite but nontrivial exact family rather than a single isolated point.

By contrast, the candidate $\mathbf w=\tup{1,1,1,2,2,2,6}$ passes both the subset-sum and LP filters but admits no complementary full-KL solution. Here the obstruction already appears in a $17$-equation real subsystem consisting of normalization, seven diagonal $Z$-constraints, and nine off-diagonal constraints. Using the Appendix ordering of the residue-$0$ support, Table~\ref{tab:bd16-nogo-1112226} displays this subsystem in a compact form. For the no-go instance $\mathbf w=\tup{1,1,1,2,2,2,6}$, let
$C_0=\{x_0,\dots,x_9\}$ denote the residue-$0$ support in the Appendix ordering, and write
$\ket{0_L}=\sum_{k=0}^9 c_k\ket{x_k}$ with $p_k:=|c_k|^2$. The diagonal block almost fixes the admissible probability profile, while the off-diagonal block couples only six residual amplitudes; after a global phase gauge fixing, these constraints collapse to an incompatible real bilinear system. Detailed exact codewords and full no-go derivations are given in Appendix Sections~\ref{supp:bd16-codes} and~\ref{supp:bd16-nogo}.

\begin{table*}[t]
    \centering
    \small
    \setlength{\tabcolsep}{3pt}
    \renewcommand{\arraystretch}{1.10}
    \caption{
Compact form of the $17$-equation subsystem used in the no-go proof for
$\mathbf w=(1,1,1,2,2,2,6)$ in the complementary $\mathrm{BD}_{16}$ setting.
Let $C_0=\{x_0,\dots,x_9\}$ be the Appendix ordering of the residue-$0$ support,
$\ket{0_L}=\sum_{k=0}^{9} c_k\ket{x_k}$, and $p_k:=|c_k|^2$.
The off-diagonal block involves only the six amplitude pairs that survive the support-overlap test.
}
    \label{tab:bd16-nogo-1112226}

    \begin{tabular}{@{}llll@{}}
        \toprule
        \multicolumn{2}{c}{\textbf{Diagonal subsystem}} &
        \multicolumn{2}{c}{\textbf{Off-diagonal subsystem}} \\
        \cmidrule(lr){1-2}\cmidrule(l){3-4}
        \textbf{Tag} & \textbf{Equation} & \textbf{Tag} & \textbf{Equation} \\
        \midrule

        (N) &
        \parbox[t]{0.27\textwidth}{\raggedright
        $\displaystyle \sum_{k=0}^{9} p_k = 1$
        }
        &
        (X1-0) &
        \parbox[t]{0.29\textwidth}{\raggedright
        $\Re(\overline{c}_6 c_9)+\Re(\overline{c}_7 c_8)=0$
        } \\[4pt]

        (Z1) &
        \parbox[t]{0.27\textwidth}{\raggedright
        $p_6+p_7+p_8+p_9=\tfrac12$
        }
        &
        (Z2X1) &
        \parbox[t]{0.29\textwidth}{\raggedright
        $\Im(\overline{c}_6 c_9)+\Im(\overline{c}_7 c_8)=0$
        } \\[4pt]

        (Z2) &
        \parbox[t]{0.27\textwidth}{\raggedright
        $p_4+p_5+p_8+p_9=\tfrac12$
        }
        &
        (Z4X1) &
        \parbox[t]{0.29\textwidth}{\raggedright
        $\Im(\overline{c}_6 c_9)-\Im(\overline{c}_7 c_8)=0$
        } \\[4pt]

        (Z3) &
        \parbox[t]{0.27\textwidth}{\raggedright
        $p_4+p_5+p_6+p_7=\tfrac12$
        }
        &
        (X2-0) &
        \parbox[t]{0.29\textwidth}{\raggedright
        $\Re(\overline{c}_4 c_9)+\Re(\overline{c}_5 c_8)=0$
        } \\[4pt]

        (Z4) &
        \parbox[t]{0.27\textwidth}{\raggedright
        $p_3+p_5+p_7+p_9=\tfrac12$
        }
        &
        (Z1X2) &
        \parbox[t]{0.29\textwidth}{\raggedright
        $\Im(\overline{c}_4 c_9)+\Im(\overline{c}_5 c_8)=0$
        } \\[4pt]

        (Z5) &
        \parbox[t]{0.27\textwidth}{\raggedright
        $p_2+p_5+p_7+p_9=\tfrac12$
        }
        &
        (Z4X2) &
        \parbox[t]{0.29\textwidth}{\raggedright
        $\Im(\overline{c}_4 c_9)-\Im(\overline{c}_5 c_8)=0$
        } \\[4pt]

        (Z6) &
        \parbox[t]{0.27\textwidth}{\raggedright
        $p_1+p_5+p_7+p_9=\tfrac12$
        }
        &
        (X3-0) &
        \parbox[t]{0.29\textwidth}{\raggedright
        $\Re(\overline{c}_4 c_7)+\Re(\overline{c}_5 c_6)=0$
        } \\[4pt]

        (Z7) &
        \parbox[t]{0.27\textwidth}{\raggedright
        $p_1+p_2+p_3+p_4+p_6+p_8=\tfrac12$
        }
        &
        (Z1X3) &
        \parbox[t]{0.29\textwidth}{\raggedright
        $\Im(\overline{c}_4 c_7)+\Im(\overline{c}_5 c_6)=0$
        } \\[4pt]

        &&
        (Z4X3) &
        \parbox[t]{0.29\textwidth}{\raggedright
        $\Im(\overline{c}_4 c_7)-\Im(\overline{c}_5 c_6)=0$
        } \\
        \bottomrule
    \end{tabular}
\end{table*}

The positive cases show that the $\mathrm{BD}_{16}$ parameter space supports both sparse and dense exact constructions, ranging from seven- and eight-term codewords to sixteen-term uniform-magnitude states and continuous analytic families. The two negative cases provide the complementary lesson: passing the subset-sum and LP stages does not by itself guarantee a distance-$3$ code, because the remaining $X/Y$-type KL constraints can still be inconsistent. The true bottleneck for this problem lies precisely at the exact full-KL stage.

This section therefore does more than add ten new small transversal-$T$ codes. It closes the $((7,2,3))$ complementary $\mathrm{BD}_{16}$ problem left open after the initial SSLP reduction, and it illustrates why agentic synthesis is essential in this part of the nonadditive design space. Once the easy combinatorial and LP filters are exhausted, the remaining cases are few but algebraically hard; resolving them requires iterating between search, exact ansatz discovery, no-go extraction, and formal verification until every surviving candidate is either realized or ruled out.

\section{Discussion and Outlook}
\label{sec:discuss}

We have developed a multi-agent, human-guided discovery pipeline for quantum error-correcting codes with prescribed transversal gates, implemented in TeXRA~\cite{Texra2025TeXRA} with GPT-5~\cite{Openai2025GPT}. The workflow turns the structural decomposition exposed by the SSLP framework~\cite{Zhang2025Transversal} into explicit constructions, analytical families, and no-go theorems. In the tractable distance-$2$ nondegenerate-residue regime, it produces a formally Lean-verified catalogue of $14{,}116$ new nonadditive codes for $K\in\{2,3,4\}$ on $n\le 6$ qubits, with cyclic transversal gate orders ranging from $2$ to $18$. Beyond catalogue-level enumeration, the pipeline extracts closed-form families, constructs a residue-degenerate $((6,4,2))$ code implementing $\mathrm{diag}(1,1,1,i)$, and resolves the small-code distance-$3$ transversal-$T$ problem for $((7,2,3))$ codes in the binary-dihedral $\mathrm{BD}_{16}$ setting, showing that $10$ of the $12$ candidates surviving the subset-sum and LP filters admit exact realizations while the remaining $2$ are excluded by explicit no-go proofs.

A central lesson is that the richness of the nonadditive code space is inseparable from its difficulty. SSLP makes large parts of the diagonal-transversal problem tractable by reducing it to modular support conditions and linear constraints on $Z$-marginals, but this reduction does not by itself make the full problem easy to solve. After the subset-sum and LP filters, the remaining full Knill-Laflamme equations can become highly coupled polynomial constraints on amplitudes and phases. The $((7,2,3))$ transversal-$T$ case provides a concrete example: once the easy filters are exhausted, resolving the surviving candidates requires exact algebraic construction or exact obstruction, not further screening. Agent intelligence becomes essential at precisely this stage. The agents do not merely automate a fixed proof; they iterate between conjecturing structure, generating executable tests, exactifying approximate solutions, writing formal proof code, and discarding failed ans\"atze until every case is settled. The broader significance is that SSLP opens many promising regions of code space, and agentic workflows make those regions genuinely explorable.

The multi-agent architecture was designed around a single principle: separate discovery from verification. Frontier models have finite context windows~\cite{Openai2025GPT} and can exhibit anchoring behavior, defending earlier conclusions rather than revising them~\cite{Lightman2023Lets}; isolating the Verification Agent from the search transcripts prevents these failure modes from contaminating accepted results. In practice, the Synthesis Agent derives combinatorial reformulations and proposes exact amplitude patterns; the Search Agent turns proposals into executable Python sweeps at scale; and the Verification Agent, behind a no-communication barrier, independently recomputes KL conditions, logical actions, and proof obligations in Lean~\cite{Moura2021Lean4}. This design mirrors the software-engineering principle of independent testing~\cite{Jimenez2024SWEbench} and ensures that all constructions, families, and no-go arguments reported here rest on kernel-checked proof objects rather than accumulated floating-point tolerances.
Human oversight remained essential at specific junctures. Researchers selected the scientific target, seeded the workflow with definitions and worked examples, steered search priorities when combinatorial branching was excessive, and edited the mathematical presentation for coherence and notation. Agents carried the bulk of large-scale search, exactification, family extraction, and proof scripting. This division of labor is most effective when the problem has the right structure: candidate solutions are difficult to invent, the search space branches rapidly, partial reformulations expose tractable subproblems, and final answers can be certified exactly. In such settings, agent intelligence complements rather than substitutes for human mathematical judgment.

The key technical enabler was to exploit SSLP not as a complete solution, but as an interface between tractable screening and harder exact reasoning. For distance-$2$ codes, the classical union-distance condition together with residue separation removes off-diagonal single-qubit constraints, while the remaining $Z$-marginal KL equalities reduce to convex feasibility and small LPs. This decomposition makes exhaustive search possible and allows numerical candidates to be exactified by continued fractions, lattice methods, and integer-preserving projections~\cite{Ferguson1999Analysis,Lenstra1982Factoring,Schrijver1986Theory}. The same agentic loop extends beyond this linear regime: in the distance-$3$ $\mathrm{BD}_{16}$ analysis, the subset-sum and LP stages still serve as filters, but the final step requires solving or excluding exact full-KL constraints. What changes is not the overall philosophy, but the balance between screening and synthesis. This adaptability suggests that the methodology is not confined to a single hand-tailored regime, but can move across different $((n,K,d))$ settings as long as useful intermediate structure can be exposed.

For quantum error correction, our results enlarge the known nonadditive design space~\cite{Rains1997Nonadditive,Rains2002Quantum} beyond prior constructions~\cite{Cross2009Codeword,Yu2008Nonadditive,Kubischta2023Family,Cross2025Small}. Codes with higher-order diagonal transversals and small exact transversal-$T$ realizations may inform fault-tolerant protocols such as magic-state distillation~\cite{Bravyi2005Universal} and gadget-based universality~\cite{Anderson2016Classification}. More broadly, this work contributes to the emerging picture in which AI systems assist theoretical science not only through brute-force search, but through structured reformulation, symbolic pattern extraction, and formally checked derivation~\cite{Wang2023Scientific,Romera-Paredes2024Mathematical,Brown2020Language,OpenAI2024OpenAI,Guo2025DeepSeekR1}. An important next step is to scale this pipeline to larger $n$, richer residue-degenerate regimes, and other transversal gate structures-including non-diagonal gates such as transversal $\mathrm{CZ}$ and permutation gates-while tightening the integration between search, synthesis, and proof assistants. The $((7,2,3))$ transversal-$T$ problem is only one case among many. The broader opportunity is to use agent intelligence to transform rich but difficult-to-navigate mathematical problems into domains where discovery, proof, and classification proceed systematically.

\section*{Data availability}
The catalogue of quantum codes discovered in this work, including all parameters, logical states, and verification results, is available at \url{https://github.com/LionSR/lean-qec}.

\section*{Code availability}
The Python code for the SSLP search pipeline, rational reconstruction algorithms, and verification routines is available at \url{https://github.com/LionSR/lean-qec}. The multi-agent system was implemented using the TeXRA platform~\cite{Texra2025TeXRA} with the GPT-5 API~\cite{Openai2025GPT}.

\begin{acknowledgments}
    The authors would like to express their gratitude to Alexander Frei, Ningping Cao, and Keren Li for helpful discussions.
\end{acknowledgments}

\appendix

\section{Additional background on the SSLP framework and transversal gates}
\label{sec:addi}

We briefly summarize the error-correction setting, subset-sum notation, and transversal gates that underlie the SSLP framework. These definitions were provided to the agents as seed input, partially synthesized and adapted from Ref.~\cite{Zhang2025Transversal} using the derivation-then-edit workflow described in Sec.~\ref{subsec:workspace}.

\subsection{Basic notations and definitions}
\label{subsec:notation}

We work on an $n$-qubit system with computational basis $\{\ket{x}: x\in\{0,1\}^n\}$, where $x=(x_1,\ldots,x_n)$ and $\ket{x}\equiv\ket{x_1}\otimes\cdots\otimes\ket{x_n}$. The Hamming weight is $\mathrm{wt}(x)=\sum_i x_i$, and the Hamming distance is $d_{\mathrm{H}}(x,y)=\mathrm{wt}(x\oplus y)$. The single-qubit Pauli operators are
\[
    X=\begin{bmatrix}0&1\\1&0\end{bmatrix},\quad
    Y=\begin{bmatrix}0&-i\\i&0\end{bmatrix},\quad
    Z=\begin{bmatrix}1&0\\0&-1\end{bmatrix},
\]
where $Z_i$ acts on a basis state as $Z_i\ket{x}=(-1)^{x_i}\ket{x}$. We denote code parameters by $((n,K,d))$~\cite{Nielsen2010Quantum}, for $n$ physical qubits, a code dimension of $K$, and a distance of $d$. We use the standard notation $[n]:=\{1,\ldots,n\}$ for index sets.

A quantum code with distance $d$ encodes information into a $K$-dimensional subspace spanned by orthonormal logical states $\{\ket{j_L}\}_{j=0}^{K-1}$. The KL conditions~\cite{Knill1997Theory} ensure error detectability: for a set $\mathcal{E}$ of detectable errors, $\bra{j_L}E^\dagger E'\ket{j'_L}=0$ when $j\neq j'$, and $\bra{j_L}E^\dagger E'\ket{j_L}=\lambda_{E,E'}$ is independent of $j$.

For distance $d=2$, we take the set of detectable errors to be $E_D=\{X_i,Y_i,Z_i:\ i\in[n]\}$ (all single-qubit Paulis). The KL conditions~\cite{Knill1997Theory} simplify to
\begin{align}
    \bra{j_L}P_i\ket{j'_L} &=0\ \text{for } j\neq j',\notag\\
    \bra{j_L}P_i\ket{j_L}  &=\lambda_{P_{i}} \quad \text{for all } j \in \{ 0, \dots, K-1\},
    \label{eq:KL-d2}
\end{align}
for all $P_i\in\{X_i,Y_i,Z_i\}$, where the scalars $\lambda_{P_i}$ are independent of the logical state $j$. Writing a logical state as $\ket{j_L}=\sum_x a_{j,x}\ket{x}$ with probabilities $p_{j,x}=|a_{j,x}|^2$, the diagonal constraints for $Z_i$ become
\begin{equation}
    \sum_x (1-2x_i)p_{j,x} = t_i \quad \text{for all } i\in[n], j=0,\ldots,K-1,
    \label{eq:Z-KL}
\end{equation}
for some site-wise constants $t_i\in\mathbb{R}$. The constraints for $X_i$ and $Y_i$ involve cross-terms of the form $a_{j,x}^* a_{j',x\oplus e_i}$ between Hamming-1 neighbors, which couple the amplitudes nonlinearly.

In addition, the definitions of the quantum weight enumerator and signature vector are presented as follows.

\begin{definition}[Quantum weight enumerators]\label{def:weight-enumerator}
    Let $\Pi$ be the projector onto an $((n,K,d))$ quantum code, and let $\mathcal P_n$ denote the $n$-qubit Pauli group generated by $\{I,X,Y,Z\}^{\otimes n}$ (up to overall phases). Following Rains' distance-two analysis and subsequent work~\cite{Rains2002Quantum,Du2024Characterizing,Zhang2025Transversal}, the quantum weight enumerators associated with $\Pi$ are the polynomials
    \[
        A(z) =  \sum_{j=0}^n A_j z^j,
        \qquad
        B(z) =  \sum_{j=0}^n B_j z^j,
    \]
    with coefficients
    \begin{align*}
        A_j & =  \frac{1}{K^2}
        \sum_{\substack{E\in\mathcal P_n  \\ \mathrm{wt}(E)=j}}
        \Bigl|\mathrm{Tr}(\Pi E)\Bigr|^2, \\
        B_j & =  \frac{1}{K}
        \sum_{\substack{E\in\mathcal P_n  \\ \mathrm{wt}(E)=j}}
        \mathrm{Tr}\!(\Pi E \Pi E^\dagger).
    \end{align*}
    Here $\mathrm{wt}(E)$ is the Pauli weight, i.e.\ the number of non-identity tensor factors in $E$. The sequences $A=(A_0,\dots,A_n)$ and $B=(B_0,\dots,B_n)$ are invariants of the code under local unitaries.
\end{definition}

\begin{definition}[Signature vector and signature norm $\lambda^*$]\label{def:signature-norm}
    Let $E_{\mathrm D}$ be the detectable Pauli error set used in the KL conditions for the code (for the distance-$2$ setting in this work one may take $E_{\mathrm D}=\{X_i,Y_i,Z_i : i\in[n]\}$). For each $E\in E_{\mathrm D}$, let $\lambda_E \in \mathbb C$ be the corresponding KL coefficient, characterized by $\Pi E \Pi = \lambda_E \Pi$ and equivalently $\lambda_E = \tfrac{1}{K}\mathrm{Tr}(\Pi E)$. The signature vector on $E_{\mathrm D}$ is
    \[
        \tilde\lambda :=  (\lambda_E)_{E\in E_{\mathrm D}} \in  \mathbb C^{|E_{\mathrm D}|},
    \]
    and its Euclidean norm
    \[
        \lambda^* :=  \|\tilde\lambda\|_2 =
        \left(\sum_{E\in E_{\mathrm D}} |\lambda_E|^2\right)^{1/2}
    \]
    is called the signature norm of the code~\cite{Du2024Characterizing,Zhang2025Transversal}. In particular, for the Pauli error models considered in this work, $\lambda^*$ is a scalar invariant that summarizes how the code projector overlaps with the detectable error operators.
\end{definition}

\subsection{Subset-sum classes and modular inner product}
\label{subsec:residue-classes}

Fix a modulus $m\in\mathbb{Z}_{>0}$ and a weight vector $\mathbf w=(w_1,\dots,w_n)\in(\mathbb{Z}_m)^n$. We define the modular inner product as
\[
    \langle \mathbf w,x\rangle \equiv \sum_{i=1}^n w_i x_i \pmod m,
\]
and the corresponding residue classes as
\begin{equation}
    C_{S_{j}}(\mathbf w):=\{x\in\{0,1\}^n:\ \langle \mathbf w,x\rangle\equiv S_{j} \pmod m \},
    \label{eq:res-class-pre}
\end{equation}
for $j=0,\dots,K-1$. Throughout this paper, we require each logical basis state to be supported on a single such class: $\mathrm{supp}(\ket{j_L})\subseteq C_{S_j}(\mathbf w)$ for a set of residues $\mathbf S=(S_0,\dots,S_{K-1})\in(\mathbb{Z}_m)^K$. We take $S_0=0$ without loss of generality. Let $C:=\bigcup_{j}C_{S_j}(\mathbf w)$ denote the classical union support of the code, and let $d(C):=\min_{x\neq y\in C} d_{\mathrm{H}}(x,y)$ be its minimum distance.

\subsection{Transversal diagonals and logical order}
\label{subsec:transversals}

We consider the phase gate
\[Z(\theta)=\mathrm{diag}(1,e^{i\theta}),\]
and define the transversal operator
\begin{align}
     &U(\mathbf w,m) :=\bigotimes_{i=1}^n Z\!\left(\tfrac{2\pi w_i}{m}\right), \notag\\
     &U(\mathbf w,m)\ket{x} =\omega_m^{\langle \mathbf w,x\rangle}\ket{x},
    \label{eq:U-pre}
\end{align}
where $\omega_m=e^{2\pi i/m}$. If $\mathrm{supp}(\ket{j_L})\subseteq C_{S_j}(\mathbf w)$, then $U(\mathbf w,m)$ acts diagonally on the logical basis:
\begin{equation}
    U(\mathbf w,m)\ket{j_L}=\omega_m^{S_j}\ket{j_L},\quad
    \overline U=\mathrm{diag}(\omega_m^{S_0},\dots,\omega_m^{S_{K-1}}),
    \label{eq:logical-diag-pre}
\end{equation}
The (projective) order of the induced cyclic logical group is
\begin{equation}
    \mathcal O=\frac{m}{\gcd(m,S_1,\dots,S_{K-1})}.
    \label{eq:order-pre}
\end{equation}
Using the rotation gate $R_Z(\theta)=e^{-i\theta Z/2}$ instead only introduces a global phase, leaving relative logical phases invariant; however, the absolute order of the logical cyclic group can double~\cite{Zhang2025Transversal}.

\subsection{Subset-sum linear program (SSLP)~\cite{Zhang2025Transversal} for distance-2 codes}
\label{subsec:sslp}

Given parameters $(n,K,m,\mathbf w,\mathbf S)$, where $\mathbf w=(w_1,\dots,w_n)\in(\mathbb Z_m)^n$ and $\mathbf S=\{S_0,\dots,S_{K-1}\}\subset\mathbb Z_m$ (with $S_0=0$), the SSLP framework consists of three main steps:

\entryhead{Step 1: Determining Logical-State Support Subsets.}
Compatibility with the transversal gate $U(\mathbf w,m)$ forces each logical basis state $\ket{j_L}$ to be supported exclusively on a single residue class:
\[
    \begin{aligned}
        \mathrm{supp}(\ket{j_L}) \subseteq C_{S_j}(\mathbf w)
         & :=\bigl\{x\in\{0,1\}^n:                                       \\
         & \qquad \langle \mathbf w,x\rangle \equiv S_j \pmod m \bigr\}.
    \end{aligned}
\]
If the residues $\{S_j\}$ are distinct, the supports $C_{S_j}(\mathbf w)$ are disjoint, which ensures orthogonality and dramatically simplifies the search.

\entryhead{Step 2: Linear-Programming Filter from Z-type KL Conditions.}
We introduce non-negative probabilities $p_{j,x}=|a_{j,x}|^2$ for each logical state $j$, defined on its support $C_{S_j}(\mathbf w)$ and normalized such that $\sum_{x\in C_{S_j}} p_{j,x}=1$. The single-site $Z$-marginal equalities require the existence of site parameters $t_i\in\mathbb R$ such that
\begin{equation}
    \sum_{x\in C_{S_j}(\mathbf w)} (1-2x_i)\,p_{j,x} = t_i,
\end{equation}
for $\forall\, i=1,\dots,n,\ \forall\, j=0,\dots,K-1$, along with the non-negativity constraints $p_{j,x}\ge 0$. This is a linear feasibility program; any parameter set $(\mathbf w,\mathbf S)$ for which this program is infeasible can be discarded immediately.

\entryhead{Step 3: Solving for the remaining KL conditions.}
With supports fixed (Step 1) and $Z$-type marginals feasible (Step 2), we now solve for complex amplitudes $a_{j,x}$ so that the full KL equalities hold for errors beyond the $Z$-only case. This can be done by minimizing a differentiable KL loss on the Stiefel manifold as shown in~\cite{Zhang2025Transversal}. Because $X$ (and $Y$) errors map $C_{S_j}$ into $C_{S_j\pm w_i}$, these terms couple distinct residue blocks, making the problem nonconvex and globally coupled. In practice, this stage is the computational bottleneck: solutions are typically numerical, hard to verify analytically, and difficult to scale to larger $n$.

\subsection{Distance-$3$ SSLP and the \texorpdfstring{$\mathrm{BD}_{16}$}{BD16} specialization}
\label{subsec:sslp-d3-bd16}

For distance $d=3$, the SSLP construction keeps the same subset-sum support classes as in Sec.~\ref{subsec:sslp}, but enlarges the detectable error set to all Pauli operators of weight at most $2$:
\[
    E_{\mathrm D}^{(3)}:=\{P\in\mathcal P_n:\ 1\le \mathrm{wt}(P)\le 2\}.
\]
For an $((n,2,3))$ code with logical basis $\{\ket{0_L},\ket{1_L}\}$, the Knill-Laflamme conditions become
\begin{equation}
    \bra{0_L}P\ket{1_L}=0,\
    \bra{0_L}P\ket{0_L}=\bra{1_L}P\ket{1_L},
    \ \forall\,P\in E_{\mathrm D}^{(3)}.
    \label{eq:KL-d3-appendix}
\end{equation}
Thus the subset-sum step is unchanged, while the final verification stage must now handle all weight-$1$ and weight-$2$ Pauli errors.

\paragraph{Complementary convention.}
The distance-$3$ binary-dihedral search studied in this work uses
\begin{equation}
    \ket{1_L}=X^{\otimes n}\ket{0_L}.
    \label{eq:complementary-d3}
\end{equation}
If $\ket{0_L}=\sum_x c_x\ket{x}$ and $\mathbf 1:=(1,\dots,1)$, then
\[
    \ket{1_L}=\sum_x c_x\ket{x\oplus \mathbf 1}.
\]
Hence only the amplitudes of $\ket{0_L}$ are independent. Moreover, if $\mathrm{supp}(\ket{0_L})\subseteq C_{S_0}(\mathbf w)$, then complementarity sends the support to
\begin{equation}
    \mathrm{supp}(\ket{1_L})\subseteq C_{S_1}(\mathbf w),\qquad
    S_1\equiv \sum_{i=1}^n w_i - S_0 \pmod m.
    \label{eq:complementary-residue-d3}
\end{equation}
Thus, once $S_0$ is fixed, the second residue is determined by $\mathbf w$.

\paragraph{The linear stage remains an LP.}
Let $p_x:=|c_x|^2$ on $C_{S_0}(\mathbf w)$. For any $z\in\{0,1\}^n$, write $Z^z=\bigotimes_i Z_i^{z_i}$. Because
\begin{align*}
    \bra{1_L}Z^z\ket{1_L} &= \bra{0_L}X^{\otimes n} Z^z X^{\otimes n}\ket{0_L} \\
    &=(-1)^{\mathrm{wt}(z)}\bra{0_L}Z^z\ket{0_L},
\end{align*}
the diagonal KL equalities for $Z$-type errors of weight at most $2$ collapse to the $n$ single-site balance equations
\begin{equation}
    \sum_{x\in C_{S_0}(\mathbf w)}(-1)^{x_i}p_x=0,
    \qquad i=1,\dots,n,
    \label{eq:d3-balance}
\end{equation}
while the two-body constraints for $Z_iZ_j$ are automatic. Therefore the distance-$3$ $Z$-type filter is still a linear feasibility problem in the probabilities $p_x$.

\paragraph{Full-KL finite-sum form.}
The nonlinear part comes from Pauli operators that contain bit flips. Writing
\[
    \ket{j_L}=\sum_{s\in\{0,1\}^n} c^{(j)}_s\ket{s},
    \qquad
    P=i^{N_Y(P)}X^{x(P)}Z^{z(P)},
\]
one has
\[
    P\ket{s}=i^{N_Y(P)}(-1)^{z(P)\cdot s}\ket{s\oplus x(P)},
\]
and hence
\begin{equation}
    \bra{j_L}P\ket{k_L} =    i^{N_Y(P)}
    \sum_{s\in\{0,1\}^n}
    (-1)^{z(P)\cdot s}
    \overline{c^{(j)}_{s\oplus x(P)}}\,c^{(k)}_s.
    \label{eq:full-kl-symplectic-d3}
\end{equation}
Equation~\eqref{eq:full-kl-symplectic-d3} reduces every distance-$3$ KL check to a finite sum over computational-basis amplitudes. Under Eq.~\eqref{eq:complementary-d3}, many off-diagonal constraints vanish automatically whenever the residue support contains no pairs related by the shift $\mathbf 1\oplus x(P)$; the remaining conditions are explicit phase-cancellation equations.

\paragraph{The \texorpdfstring{$\mathrm{BD}_{16}$}{BD16} case.}
For the transversal-$T$ search we specialize to $n=7$ and $m=8$ with residues
\[
    (S_0,S_1)=(0,7).
\]
Then
\begin{align}
     &U(\mathbf w,8)\ket{0_L}=\ket{0_L}, \notag\\
     &U(\mathbf w,8)\ket{1_L}=\omega_8^{7}\ket{1_L}
    =e^{-i\pi/4}\ket{1_L}, \notag
\end{align}
so the induced logical diagonal is
\[
    \overline U=\mathrm{diag}(1,\omega_8^{7})=Z(-2\pi/8)
\]
in the notation of Sec.~\ref{subsec:transversals}. Together with the complementary logical bit flip
\[
    \overline X=X^{\otimes 7},
\]
these generators satisfy
\[
    \overline X\,\overline U\,\overline X=\omega_8^{7}\overline U^{-1},
\]
so projectively the diagonal cyclic symmetry is extended by a Pauli reflection, which is exactly the binary-dihedral specialization denoted $\mathrm{BD}_{16}$ in Ref.~\cite{Zhang2025Transversal}. In this setting the SSLP pipeline consists of: (i) choosing a weight vector $\mathbf w$ with $\sum_i w_i\equiv 7 \pmod 8$, (ii) solving the linear balance equations~\eqref{eq:d3-balance} on the residue-$0$ class, and (iii) checking Eq.~\eqref{eq:KL-d3-appendix} for every Pauli error of weight at most $2$.

\section{Rational reconstruction of LP solutions}
\label{supp:rational}

In this section we describe two complementary procedures used by the Search Agent to convert floating-point LP solutions into exact rational solutions. Both exploit the integer structure of the constraint matrix and are applied after the numerical SSLP has produced a feasible point.

\subsection{Algorithm 1: Exact BFS reconstruction}

When the numerical solution is close to a basic feasible solution (BFS), at most $(K-1)n+K$ entries of $\mathbf p$ are nonzero. In that case we can identify a basis of this size and solve the corresponding subsystem exactly over $\mathbb Q$.

\begin{algorithm}[h]
    \caption{Exact BFS reconstruction}
    \KwIn{$A_{\mathrm{eq}} \in \mathbb Z^{M\times N}$, $b_{\mathrm{eq}}\in\mathbb Z^M$, $\mathbf{p}^{(\mathrm{num})}\in\mathbb R^N$.}
    \KwOut{Exact rational vector $\mathbf p\in\mathbb Q^N$ satisfying constraints, or failure.}
    \emph{Step 1}: Identify $B\subseteq\{1,\dots,N\}$ with $|B|=M=(K-1)n+K$ such that $A_B$ has full rank. \\
    \emph{Step 2}: Solve $A_B \mathbf{p}_{B}=b_{\mathrm{eq}}$ exactly over $\mathbb Q$ (e.g., Bareiss elimination). \\
    \emph{Step 3}: Set $p_i=0$ for all $i\notin B$ and assemble $\mathbf p\in\mathbb Q^N$. \\
    \emph{Step 4}: Verify $A_{\mathrm{eq}}\mathbf{p}=b_{\mathrm{eq}}$ exactly over $\mathbb Q$. \\
    \emph{Step 5}: Verify $\mathbf{p}\ge 0$ entrywise. \\
    \emph{Step 6}: Return $\mathbf{p}$ if both checks pass; otherwise report failure.
    \label{alg:exactbfs-supp}
\end{algorithm}

\subsection{Algorithm 2: Rationalization by projection}

When the numerical solution is not clearly BFS-like or when we prefer a more robust method, we first rationalize each coordinate by continued fractions and then project back onto the affine constraint space using exact rational arithmetic.

In practice, Algorithm~\ref{alg:exactbfs-supp} is used when the LP solver identifies a clear BFS with a small active set, while Algorithm~\ref{alg:ratproj-supp} provides a robust fallback when the numerical solution is more diffuse. Both are followed by exact KL and transversal checks in the Lean pipeline.

\begin{algorithm}[h]
    \caption{Rationalization by projection}
    \KwIn{$A_{\mathrm{eq}} \in \mathbb Z^{M\times N}$, $b_{\mathrm{eq}}\in\mathbb Z^M$, $\mathbf{p}^{(\mathrm{num})}\in\mathbb R^N$, $D\in\mathbb N$.}
    \KwOut{Exact rational vector $\mathbf p\in\mathbb Q^N$ satisfying constraints, or failure.}
    \emph{Step 1}: Rational approximation: for each $i=1,\dots,N$, set
    $\tilde{p}_i \gets \mathrm{CFround}\!(p^{(\mathrm{num})}_i;\ \mathrm{den}\le D)$, where $\mathrm{CFround}$ returns the nearest rational with denominator at most $D$. \\
    \emph{Step 2}: Form $\tilde{\mathbf p}\in\mathbb Q^N$. \\
    \emph{Step 3}: Exact projection: solve $A_{\mathrm{eq}}\mathbf{d}=b_{\mathrm{eq}}-A_{\mathrm{eq}}\tilde{\mathbf{p}}$ over $\mathbb Q$, and set $\mathbf{p}\gets \tilde{\mathbf{p}}+\mathbf{d}$. \\
    \emph{Step 4}: Non-negativity and re-projection: if some entries of $\mathbf p$ are slightly negative, clip to $0$ and re-solve $A_{\mathrm{eq}}\mathbf{d}=b_{\mathrm{eq}}-A_{\mathrm{eq}}\mathbf{p}$ over $\mathbb Q$. \\
    \emph{Step 5}: Block normalization: for each logical block $j$, renormalize $\{p_{j,x}\}_{x\in C_{S_j}}$ so that $\sum_{x\in C_{S_j}}p_{j,x}=1$. \\
    \emph{Step 6}: Verify $A_{\mathrm{eq}}\mathbf{p}=b_{\mathrm{eq}}$ exactly and $\mathbf{p}\ge 0$. \\
    \emph{Step 7}: Return $\mathbf{p}$ if verification passes; otherwise report failure.
    \label{alg:ratproj-supp}
\end{algorithm}


\section{Code catalogue of small diagonal distance-two codes}
\label{supp:code-catalog}

Executing the pipeline described in Section~\ref{subsec:search}, our multi-agent system performed large-scale parameter sweeps on a local high-performance computing workstation equipped with an Intel Xeon w7-3565X CPU and two Nvidia RTX 6000 Ada Generation GPUs. The search focused on distance-2 codes for $n \le 6$ qubits and logical dimensions $K \in \{2,3,4\}$. This procedure, executed by the Search Agent and verified by the Audit Agent, yielded a rich catalog of new nonadditive quantum codes. The following subsections present parameters of these codes, organized by logical dimension $K$. For $K=4$, we also present two explicit code examples.

\subsection{$K=2$ (two-dimensional logical space)}

For two-dimensional codes ($K=2$), our search on up to $n=6$ qubits revealed a rich structure, including codes with cyclic group orders as high as 18. Table~\ref{tab:K=2} summarizes the parameters of these instances, followed by their explicit logical state constructions.

\begin{table*}[t]
    \caption{Representative distance-2 diagonal-transversal codes for $K=2$.}
    \label{tab:K=2}
    \begin{tabular*}{\textwidth}{@{\extracolsep{\fill}}llllllll}
        \toprule
        $\mathcal{O}$ & $m$ & $n$ & $\mathbf w$
        & $\mathbf S$ & $(|C_0|,|C_2|)$ \\
        \midrule
        2 & 4 & 4 & (1,1,1,1) & (0,2) & (2,6) \\
        \midrule
        2   & 4   & 5 & (1, 1, 1, 1, 1)           & (0, 2) & (6, 10)  \\
        3   & 6   & 5 & (1, 1, 1, 1, 3)           & (0, 4) & (5, 5) \\
        4   & 8   & 5 & (1, 1, 1, 3, 3)           & (0, 6) & (4, 3) \\
        5   & 5   & 5 & (1, 1, 1, 1, 1)           & (0, 2) & (2, 10) \\
        6   & 6   & 5 & (2, 2, 2, 3, 3)           & (0, 1) & (4, 6) \\
        7   & 7   & 5 & (1, 1, 1, 2, 2)           & (0, 3) & (2, 7) \\
        8   & 8   & 5 & (1, 1, 2, 2, 4)           & (0, 5) & (4, 4) \\
        9   & 9   & 5 & (1, 1, 2, 2, 3)           & (0, 4) & (2, 5) \\
        \midrule
        2   & 4   & 6 & (1, 1, 1, 1, 1, 1)        & (0, 2) & (16, 16) \\
        3   & 6   & 6 & (1, 1, 1, 1, 1, 1)        & (0, 2) & (2, 15) \\
        4   & 8   & 6 & (1, 1, 1, 1, 3, 3)        & (0, 6) & (7, 9) \\
        5   & 5   & 6 & (1, 1, 1, 1, 1, 1)        & (0, 2) & (7, 15) \\
        6   & 6   & 6 & (2, 2, 2, 2, 3, 3)        & (0, 1) & (10, 12) \\
        7   & 7   & 6 & (1, 1, 1, 1, 1, 2)        & (0, 3) & (2, 15) \\
        8   & 8   & 6 & (1, 1, 1, 1, 1, 4)        & (0, 5) & (6, 6) \\
        9   & 9   & 6 & (1, 1, 1, 1, 2, 3)        & (0, 4) & (2, 11) \\
        10  & 10  & 6 & (1, 1, 1, 1, 4, 6)        & (0, 7) & (3, 8) \\
        11  & 11  & 6 & (1, 1, 1, 1, 4, 4)        & (0, 8) & (5, 3) \\
        12  & 12  & 6 & (1, 1, 1, 2, 3, 4)        & (0, 5) & (2, 8) \\
        13  & 13  & 6 & (1, 1, 1, 2, 5, 5)        & (0, 10) & (5, 3)\\
        14  & 14  & 6 & (1, 1, 1, 3, 3, 6)        & (0, 9)  & (4, 4)\\
        15  & 15  & 6 & (1, 1, 2, 2, 5, 6)        & (0, 11) & (4, 4)\\
        16  & 16  & 6 & (1, 1, 2, 3, 4, 5)        & (0, 7) & (2, 6) \\
        17  & 17  & 6 & (1, 1, 2, 4, 4, 6)        & (0, 8) & (3, 5)\\
        18  & 18  & 6 & (1, 2, 3, 4, 5, 6)        & (0, 11) & (3, 5)\\
        \bottomrule
    \end{tabular*}
\end{table*}

\subsection{$K=3$ (qutrit logical space)}

For $K=3$ on $n=6$ qubits, the search yielded codes with orders up to 16 as in Table~\ref{tab:K=3}. The attainable orders appear less continuous in $m$ compared to the $K=2$ case. Key instances are detailed below.

\begin{table*}[t]
    \caption{Representative distance-2 diagonal-transversal codes for $K=3$.}
    \label{tab:K=3}
    \begin{tabular*}{\textwidth}{@{\extracolsep{\fill}}llllllll}
        \toprule
        $\mathcal{O}$ & $m$ & $n$ & $\mathbf w$
        & $\mathbf S$ & $(|C_{S_{0}}|,|C_{S_{1}}|, |C_{S_{2}}|)$ \\
        \midrule
        3   & 6   & 6 & (1, 1, 1, 1, 3, 3)        & (0, 2, 4)  & (10, 12, 10)   \\
        4   & 8   & 6 & (1, 1, 1, 3, 3, 3)        & (0, 2, 4)  & (10, 6, 10) \\
        6   & 12  & 6 & (1, 1, 1, 5, 5, 7)        & (0, 6, 10) & (6, 9, 2) \\
        8   & 16  & 6 & (1, 1, 4, 4, 7, 7)        & (0, 2, 8)  & (6, 3, 6)  \\
        10  & 10  & 6 & (1, 1, 1, 4, 4, 4)        & (0, 2, 5)  & (10, 4, 10) \\
        12  & 12  & 6 & (2, 2, 3, 3, 4, 4)        & (0, 6, 7)  & (6, 6, 6) \\
        14  & 14  & 6 & (1, 1, 3, 4, 6, 6)        & (0, 2, 7)  & (6, 4, 6) \\
        15  & 15  & 6 & (1, 1, 4, 4, 6, 9)        & (0, 2, 10) & (6, 3, 6) \\
        16  & 16  & 6 & (1, 2, 4, 4, 6, 7)        & (0, 8, 11) & (4, 4, 5) \\
        \bottomrule
    \end{tabular*}
\end{table*}

\subsection{$K=4$ (four-dimensional logical space)}

For $K=4$ on $n=6$ qubits, our search identified fewer instances due to the increasing number of constraints as presented in Table~\ref{tab:K=4}. We report two notable codes with orders 4 and 6.

\begin{table*}[t]
    \caption{Representative distance-2 diagonal-transversal codes for $K=4$.}
    \label{tab:K=4}
    \begin{tabular*}{\textwidth}{@{\extracolsep{\fill}}llllll}
        \toprule
        $\mathcal{O}$ & $m$ & $n$ & $\mathbf w$
        & $\mathbf S$ & $(|C_{S_{0}}|,|C_{S_{1}}|, |C_{S_{2}}|, |C_{S_{3}}|)$ \\
        \midrule
        4   & 8   & 6 & (1, 1, 1, 3, 3, 3)        & (0, 2, 4, 6)  & (10, 6, 10, 6) \\
        6   & 12  & 6 & (1, 1, 3, 3, 5, 5)        & (0, 2, 6, 10) & (6, 5, 6, 5)\\
        \bottomrule
    \end{tabular*}
\end{table*}

\entryhead{Order 4:}
\noindent\textbf{Parameters:} $n=6,  m=8,  K=4;\quad \mathbf w=(1, 1, 1, 3, 3, 3);\quad \mathbf S=\{0, 2, 4, 6\};\quad \text{sizes}=\left(10, 6, 10, 6\right)$.
$ \text{order}=\frac{m}{\gcd(m,S_1,\dots,S_{K-1})}=\frac{8}{2}=4$.

\noindent\textbf{Transversal gates:}
\[
    U=Z \left(\tfrac{2\pi}{8}\right)^{\otimes3} \otimes Z \left(\tfrac{6\pi}{8}\right)^{\otimes3},\quad
    \overline U=\mathrm{diag}(1,\omega_8^{2},\omega_8^{4},\omega_8^{6}).
\]

\noindent\textbf{Logical states:}
\begin{align*}
     & \ket{0_L} = \sqrt{\tfrac{1}{4}} \ket{000000} + \sqrt{\tfrac{1}{4}} \ket{011011} + \sqrt{\tfrac{1}{4}} \ket{101101} \\
     & \quad + \sqrt{\tfrac{1}{4}} \ket{110110},                                                                          \\
     & \ket{1_L} = \sqrt{\tfrac{1}{2}} \ket{001111} + \sqrt{\tfrac{1}{2}} \ket{110000},                                   \\
     & \ket{2_L} = \sqrt{\tfrac{1}{4}} \ket{001100} + \sqrt{\tfrac{1}{4}} \ket{010010} + \sqrt{\tfrac{1}{4}} \ket{100001} \\
     & \quad + \sqrt{\tfrac{1}{4}} \ket{111111},                                                                          \\
     & \ket{3_L} = \sqrt{\tfrac{1}{2}} \ket{000011} + \sqrt{\tfrac{1}{2}} \ket{111100}.
\end{align*}

\noindent\textbf{Weight enumerators:}
\begin{align*}
     & A(z) = 1 + \tfrac{7}{4}\,z^{2} + \tfrac{1}{2}\,z^{3} + \tfrac{7}{2}\,z^{4} + \tfrac{5}{2}\,z^{5} + \tfrac{27}{4}\,z^{6}, \\
     & B(z) = 1 + \tfrac{31}{2}\,z^{2} + 28\,z^{3} + 76\,z^{4} + 80\,z^{5} + \tfrac{111}{2}\,z^{6}.
\end{align*}

\entryhead{Order 6:}

\noindent\textbf{Parameters:} $n=6,  m=12,  K=4;\quad \mathbf w=(1, 1, 3, 3, 5, 5);\quad \mathbf S=\{0, 2, 6, 10\};\quad \text{sizes}=\left(6, 5, 6, 5\right)$. $ \text{order}=\frac{m}{\gcd(m,S_1,\dots,S_{K-1})}=\frac{12}{2}=6$.

\noindent\textbf{Transversal gates:}
\[
    \begin{aligned}
        U
         & = Z \left(\tfrac{2\pi}{12}\right)^{\otimes2}
        \otimes Z \left(\tfrac{6\pi}{12}\right)^{\otimes2}                      \\
         & \quad \otimes Z \left(\tfrac{10\pi}{12}\right)^{\otimes2},           \\
        \overline U
         & = \mathrm{diag}(1,\omega_{12}^{2},\omega_{12}^{6},\omega_{12}^{10}).
    \end{aligned}
\]

\noindent\textbf{Logical states:}
\begin{align*}
     & \ket{0_L} = \sqrt{\tfrac{1}{6}} \ket{000000} + \sqrt{\tfrac{1}{6}} \ket{011101} + \sqrt{\tfrac{1}{6}} \ket{101110}  \\
     & \quad + \sqrt{\tfrac{1}{2}} \ket{110011},                                                                           \\
     & \ket{1_L} = \sqrt{\tfrac{1}{3}} \ket{010111} + \sqrt{\tfrac{1}{3}} \ket{101011} + \sqrt{\tfrac{1}{3}} \ket{110000}, \\
     & \ket{2_L} = \sqrt{\tfrac{1}{3}} \ket{010001} + \sqrt{\tfrac{1}{3}} \ket{100010} + \sqrt{\tfrac{1}{3}} \ket{111111}, \\
     & \ket{3_L} = \sqrt{\tfrac{1}{3}} \ket{000011} + \sqrt{\tfrac{1}{3}} \ket{110101} + \sqrt{\tfrac{1}{3}} \ket{111010}.
\end{align*}

\noindent\textbf{Weight enumerators:}
\begin{align*}
     & A(z) = 1 + \tfrac{2}{3}\,z + \tfrac{2}{3}\,z^{2}                  \\
     & \quad + \tfrac{2}{3}\,z^{3} + 4\,z^{4} + \tfrac{14}{3}\,z^{5}     \\
     & \quad + \tfrac{13}{3}\,z^{6},                                     \\
     & B(z) = 1 + \tfrac{2}{3}\,z + \tfrac{40}{3}\,z^{2}                 \\
     & \quad + 40\,z^{3} + \tfrac{247}{3}\,z^{4} + \tfrac{238}{3}\,z^{5} \\
     & \quad + \tfrac{118}{3}\,z^{6}.
\end{align*}

This systematic search produced a large number of new quantum codes for $n \in \{4, 5, 6\}$. The verified instances in this dataset reveal recurring patterns that can be elevated to analytical families.

\section{Catalogue examples for analytical families}
\label{supp:families-examples}

In this section, we list explicit small-$n$ instances for the analytical families introduced in the main text. These examples were first discovered by the Search Agent and subsequently recognized and generalized by the Synthesis Agent into the closed-form constructions described in the Results and Appendix sections.

\subsection{Family I: $C_0=\{0^n,1^n\}$}
\label{subsubsec:supp:family1}

We recall the general setting.
Fix $n\ge 2$, a modulus $m\ge n$, and a weight vector
\[
    \mathbf w=(1,1,\dots,1,m-(n-1))\in(\mathbb Z_m)^n.
\]
For $s\in\{0,\dots,m-1\}$ define the subset-sum residue classes
\[
    C_s=\{x\in\{0,1\}^n : \langle \mathbf w,x\rangle \equiv s \pmod m\}.
\]

For $m\ge n$, the residue class $C_0$ reduces to the two extremal strings,
$C_0=\{0^n,1^n\}$, and for each $s$ in the window $m-(n-1)\le s\le n-1$ the class $C_s$ decomposes into two Hamming-weight slices $A_s$ and $B_t$ as described in the main text.
We choose $\ket{0_L}$ supported on $C_0$ and $\ket{1_L}$ supported on $C_s$ and obtain distance-$2$ $((n,2,2))$ codes with logical diagonal action $\overline U=\mathrm{diag}(1,\omega_m^s)$ when the residue-shift screen is satisfied.

Below we list the concrete instances for $n=5$ and $n=6$ that motivated this family.

\subsubsection{Instances with $n=5$}

For $n=5$ we have
\[
    \mathbf w=(1,1,1,1,m-4),\qquad
    C_0=\{00000,11111\}.
\]
The two-slice window for $s$ is
$m-4\le s\le 4$. We focus on a choice that satisfies the residue-shift screen and yields a distance-2 code.

\entryhead{Example 1 ($n=5$, $m=5$, $s=2$).}
Here
\begin{equation}
    m = 5,\quad \mathbf w=(1,1,1,1,1), \quad s = 2,\quad t=1.
\end{equation}
The residue classes are determined by Hamming weight modulo $5$:
\[
    \begin{aligned}
        C_0 & = \{x:\mathrm{wt}(x)\equiv 0\ (\mathrm{mod}\ 5)\}, \\
        C_2 & = \{x:\mathrm{wt}(x)\equiv 2\ (\mathrm{mod}\ 5)\}.
    \end{aligned}
\]
In particular
\[
    C_0=\{00000,11111\},\qquad
    C_2=A_2\cup B_1,
\]
where
\[
    \begin{aligned}
        A_2 & = \{y\in\{0,1\}^5 : \mathrm{wt}(y)=2\}, \\
        B_1 & = \{y\in\{0,1\}^5 : \mathrm{wt}(y)=1\}.
    \end{aligned}
\]
We take $\ket{0_L}$ supported on $C_0$ and $\ket{1_L}$ supported on $C_2$, with probabilities
\[
    p(0^5)=1-\frac{s}{m}=\frac{3}{5},\qquad
    p(1^5)=\frac{s}{m}=\frac{2}{5},
\]
and
\[
    q(y)=
    \begin{cases}
        \dfrac{m-s}{m}\dfrac{1}{\binom{4}{2}}=\dfrac{3}{5}\cdot \dfrac{1}{6}, & y\in A_2, \\
        \dfrac{s}{m}\dfrac{1}{\binom{4}{1}}=\dfrac{2}{5}\cdot \dfrac{1}{4},   & y\in B_1.
    \end{cases}
\]
The normalized logical states are
\begin{align}
    \ket{0_L}
     &=\sqrt{\tfrac{3}{5}}\ket{00000}+\sqrt{\tfrac{2}{5}}\ket{11111}, \notag\\
    \ket{1_L}
     &=\frac{\sqrt{3/5}}{\sqrt{6}}\Bigl(\ket{11000}+\ket{10100} +\ket{10010}+\ket{01100} \notag\\
     &\qquad\qquad +\ket{01010}+\ket{00110}
    \Bigr) \notag\\
     &\quad +\frac{\sqrt{2/5}}{2}\Bigl(\ket{10001}+\ket{01001} +\ket{00101}+\ket{00011}
    \Bigr). \notag
\end{align}

The transversal gate
\[
    U(\mathbf w,5)=Z(2\pi/5)^{\otimes 5}
\]
acts as
\[
    \overline U=\mathrm{diag}(1,\omega_5^2),
\]
with order $5/\gcd(5,2)=5$. The residue-shift screen
$s\not\equiv \pm1\ (\mathrm{mod}\ 5)$ and
$s\not\equiv \pm(m-(n-1))\ (\mathrm{mod}\ 5)$ is satisfied for $s=2$, so all weight-1 $X/Y$ off-diagonal KL terms vanish combinatorially, and the code has distance~2.

\subsubsection{\texorpdfstring{Instances with $n=6$}{Instances with n=6}}

For $n=6$ we have
\[
    \mathbf w=(1,1,1,1,1,m-5),\qquad
    C_0=\{000000,111111\}.
\]
The two-slice window for $s$ is
$m-5\le s\le 5$.

\entryhead{Example 2 ($n=6$, $m=7$, $s=3$).}
Here
\begin{equation}
    m = 7,\ \mathbf w=(1,1,1,1,1,2), \ s = 3,\ t=n-1+s-m=1.
\end{equation}

The residue classes satisfy
\[
    C_0=\{000000,111111\},\qquad C_3=A_3\cup B_1,
\]
with
\[
    \begin{aligned}
        A_3 & = \{(u,0):\mathrm{wt}(u)=3\},                       \\
        B_1 & = \{(u,1):\mathrm{wt}(u)=1\},\quad u\in\{0,1\}^{5}.
    \end{aligned}
\]
We again choose $\ket{0_L}$ supported on $C_0$ and $\ket{1_L}$ supported on $C_3$, with probabilities
\[
    p(0^6)=1-\frac{3}{7}=\frac{4}{7},\qquad
    p(1^6)=\frac{3}{7},
\]
and
\[
    q(y)=
    \begin{cases}
        \dfrac{m-s}{m}\dfrac{1}{\binom{5}{3}}=\dfrac{4}{7}\cdot\dfrac{1}{10}, & y\in A_3, \\
        \dfrac{s}{m}\dfrac{1}{\binom{5}{1}}=\dfrac{3}{7}\cdot\dfrac{1}{5},    & y\in B_1.
    \end{cases}
\]
The logical states are
\begin{align}
    \ket{0_L}
     &=\sqrt{\tfrac{4}{7}}\ \ket{000000} +\sqrt{\tfrac{3}{7}}\ \ket{111111}, \notag\\
    \ket{1_L}
     &=\frac{\sqrt{4/7}}{\sqrt{\binom{5}{3}}}
    \sum_{\mathrm{wt}(u)=3} \ket{u\,0} +\frac{\sqrt{3/7}}{\sqrt{\binom{5}{1}}}
    \sum_{\mathrm{wt}(u)=1} \ket{u\,1}. \notag
\end{align}
One convenient explicit expansion is
\begin{align*}
    \ket{1_L}
     & =\frac{\sqrt{4/7}}{\sqrt{10}}\Bigl(\ket{111000}+\ket{110100}+\ket{110010}     \\
     & \qquad +\ket{101100}+\ket{101010}+\ket{100110}                                \\
     & \qquad +\ket{011100}+\ket{011010}+\ket{010110}+\ket{001110}
    \Bigr)                                                                           \\
     & \quad+\frac{\sqrt{3/7}}{\sqrt{5}}\Bigl(\ket{100001}+\ket{010001}+\ket{001001} \\
     & \qquad\qquad +\ket{000101}+\ket{000011}
    \Bigr).
\end{align*}
The $Z$-equalities give
\[
    \langle Z_i\rangle_{\ket{0_L}}=\langle Z_i\rangle_{\ket{1_L}}
    =1-\frac{2s}{m}=\frac{1}{7}
\]
for all sites $i$. The transversal gate
\[
    U(\mathbf w,7)
    =Z(2\pi/7)^{\otimes 5}\otimes Z(4\pi/7)
\]
acts as
\[
    \overline U=\mathrm{diag}(1,\omega_7^{3}),
\]
with order $7/\gcd(7,3)=7$. The residue-shift screen
$s\not\equiv \pm1,\pm2\ (\mathrm{mod}\ 7)$ holds for $s=3$, so the code has distance~2.

\subsection{Family II: even-parity subset-sum codes}
\label{subsubsec:supp:family2}

We recall the general construction.
Let $n$ be even and
\[
    \mathsf E=\{\sigma\in\{0,1\}^n:\mathrm{wt}(\sigma)\equiv 0\ (\mathrm{mod}\ 2)\}
\]
denote the even-parity subcode.
For a modulus $m$, weights $\mathbf w\in(\mathbb Z_m)^n$, and residues
$\mathbf S=\{S_0,\dots,S_{K-1}\}\subset\mathbb Z_m$ with $S_0=0$, define
\[
    C^{(+)}_{S_k}(\mathbf w)
    =\left\{\sigma\in\mathsf E:\textstyle\sum_{j=1}^n w_j\sigma_j\equiv S_k\pmod m\right\},
\]
and logical states
\[
    \ket{k_L}=\frac{1}{\sqrt{|C^{(+)}_{S_k}|}}
    \sum_{\sigma\in C^{(+)}_{S_k}}\ket{\sigma},\quad k=0,\dots,K-1.
\]

Because each support is contained in $\mathsf E$, single-qubit $X_i$ and $Y_i$ errors map basis states to odd-parity strings outside the support, and therefore all $X/Y$ KL matrix elements vanish. If, in addition, each support $C^{(+)}_{S_k}$ is column-balanced (exactly half of the strings in the set have a $1$ in each coordinate), then all $Z$-type KL constraints are satisfied and the code has distance~2. The transversal diagonal $U(\mathbf w,m)$ acts as
\[
    \overline U=\mathrm{diag}(\omega_m^{S_0},\dots,\omega_m^{S_{K-1}}),
\]
with logical order
\[
    \mathcal O=\frac{m}{\gcd(m,S_1,\dots,S_{K-1})}.
\]

Below we list the catalogue examples that instantiate this family.

\subsubsection{\texorpdfstring{$K=2$ examples}{K=2 examples}}

\entryhead{Example 3 ($n=4,K=2$; order $2$).}
Take
\[
    m=6,\quad \mathbf w=(1,2,4,5),\quad \mathbf S=\{0,3\}.
\]
The supports are
\[
    \begin{aligned}
        C^{(+)}_{0} & =\{0000,0110,1001,1111\}, \\
        C^{(+)}_{3} & =\{0011,1100\}.
    \end{aligned}
\]
The logical states are
\[
    \begin{aligned}
        \ket{0_L} & = \tfrac12(\ket{0000}+\ket{0110}+\ket{1001}+\ket{1111}), \\
        \ket{1_L} & = \tfrac{1}{\sqrt2}(\ket{0011}+\ket{1100}).
    \end{aligned}
\]

Both supports are column-balanced: for each $i\in\{1,2,3,4\}$ exactly half of the strings in $C^{(+)}_0$ and in $C^{(+)}_3$ have a $1$ at position $i$. Thus $\langle Z_i\rangle=0$ for both logical states, and all single-qubit KL conditions hold. The transversal diagonal
\[
    U(\mathbf w,6)=\bigotimes_{j=1}^4 Z\!\left(\tfrac{2\pi w_j}{6}\right)
\]
acts as
\[
    \overline U=\mathrm{diag}(1,e^{i\pi})
\]
with order $2$.

\entryhead{Example 4 ($n=6,K=2$; order $2$).}
Take
\[
    m=8,\quad \mathbf w=(1,2,3,5,6,7),\quad \mathbf S=\{0,4\}.
\]
The supports are
\[
    \begin{aligned}
        C^{(+)}_{0} & =\{000000,001100,010010,011110,      \\
                    & \quad 100001,101101,110011,111111\}, \\
        C^{(+)}_{4} & =\{000101,010111,101000,111010\}.
    \end{aligned}
\]
The logical states are uniform superpositions:
\[
    \ket{0_L}=\tfrac{1}{\sqrt8} \sum_{s\in C^{(+)}_{0}}\ket{s},\qquad
    \ket{1_L}=\tfrac{1}{2} \sum_{s\in C^{(+)}_{4}}\ket{s}.
\]
Both supports are column-balanced in all six coordinates, so $\langle Z_i\rangle=0$ for both logical states and each $i$. The logical action is
\[
    \overline U=\mathrm{diag}(1,\omega_8^{4})=\mathrm{diag}(1,-1),
\]
again of order $2$.

\entryhead{Example 5 ($n=6,K=2$; order $4$).}
Take
\[
    m=8,\quad \mathbf w=(6,4,0,2,7,5),\quad \mathbf S=\{0,2\}.
\]
The supports are
\[
    \begin{aligned}
        C^{(+)}_{0} & =\{000000,011011,100100,111111\}, \\
        C^{(+)}_{2} & =\{001100,010111,101011,110000\}.
    \end{aligned}
\]
The logical states are
\[
    \ket{0_L}=\tfrac{1}{2}(\ket{000000}+\ket{011011}+\ket{100100}+\ket{111111}),
\]
\[
    \ket{1_L}=\tfrac{1}{2}(\ket{001100}+\ket{010111}+\ket{101011}+\ket{110000}).
\]

On both supports, the one-counts in each column are
\[
    [2,2,2,2,2,2],
\]
so each set is column-balanced and $\langle Z_i\rangle=0$ for both logical states and all $i$. The transversal action is
\[
    \overline U=\mathrm{diag}(1,\omega_8^{2}),
\]
with order $8/\gcd(8,2)=4$.

\subsubsection{\texorpdfstring{$K>2$ example}{K>2 example}}

\entryhead{Example 6 ($n=6,K=3$; order $3$).}
Take
\[
    m=9,\quad \mathbf w=(1,2,5,5,7,1),\quad \mathbf S=\{0,3,6\}.
\]
The even-parity supports are
\[
    \begin{aligned}
        C^{(+)}_{0} & =\{000000,001111,010010,             \\
                    & \quad 101110,110101,111001\},        \\
        C^{(+)}_{3} & =\{000110,001010,010001,             \\
                    & \quad 101101,110000,111111\},        \\
        C^{(+)}_{6} & =\{000101,001001,010111,011011,      \\
                    & \quad 100100,101000,110110,111010\}.
    \end{aligned}
\]
We define logical states as uniform superpositions:
\[
    \begin{aligned}
        \ket{0_L} & =\tfrac{1}{\sqrt6}(\ket{000000}+\ket{001111}+\ket{010010}
        +\ket{101110}                                                            \\
                  & \quad +\ket{110101}+\ket{111001}),                           \\
        \ket{1_L} & =\tfrac{1}{\sqrt6}(\ket{000110}+\ket{001010}+\ket{010001}
        +\ket{101101}                                                            \\
                  & \quad +\ket{110000}+\ket{111111}),                           \\
        \ket{2_L} & =\tfrac{1}{\sqrt8}(\ket{000101}+\ket{001001}+\ket{010111}
        +\ket{011011}                                                            \\
                  & \quad +\ket{100100}+\ket{101000}+\ket{110110}+\ket{111010}).
    \end{aligned}
\]
The one-counts in each column are
\[
    \begin{aligned}
         & [3,3,3,3,3,3]\ \text{on } C^{(+)}_{0}, \\
         & [3,3,3,3,3,3]\ \text{on } C^{(+)}_{3}, \\
         & [4,4,4,4,4,4]\ \text{on } C^{(+)}_{6},
    \end{aligned}
\]
so each support is column-balanced and all $Z$-type KL conditions hold. The transversal action is
\[
    \overline U=\mathrm{diag}(1,\omega_9^{3},\omega_9^{6}),
\]
with order
\[
    \mathcal O=\frac{9}{\gcd(9,3,6)}=3.
\]

These catalogue instances illustrate how the even-parity SSLP construction supports different logical dimensions $K$ and a range of logical orders $\mathcal O$ at small $n$.

\section{Residue-degenerate \texorpdfstring{$((6,4,2))$}{((6,4,2))} controlled-phase code}
\label{supp:degenerate-642}

In this section, we give a detailed construction of the residue-degenerate $((6,4,2))$ code with logical controlled-phase action described in the main text. This example lies beyond the strictly nondegenerate residue regime used in our systematic sweeps, and also beyond the classical union-distance guard: the union support contains Hamming-$1$ neighbours, but the remaining Knill-Laflamme constraints are enforced by structured sign cancellation.

\subsection{Residues and parity structure}

We take modulus $m=4$ and weight vector
\[
    \mathbf w=(1,3,2,2,2,2)\in\mathbb Z_4^6.
\]
For a bit string $x=(x_1,\dots,x_6)$ the residue is
\begin{align}
    \operatorname{res}(x) &\equiv \mathbf w\cdot x \pmod 4 \notag\\
                          &= x_1 - x_2 + 2\,(x_3\oplus x_4\oplus x_5\oplus x_6)\pmod 4. \notag
\end{align}
Thus the residue class is determined by $(x_1,x_2)$ together with the parity of the last four bits.

We organize the last four qubits into even- and odd-parity subsets indexed by $t=(t_1,t_2,t_3)\in\mathbb F_2^3$ via
\begin{align}
     &\phi(t)=(t_1,t_2,t_3,t_1\oplus t_2\oplus t_3) \ \ \quad \qquad\text{(even parity)}, \notag\\
     &\psi(t)=(t_1,t_2,t_3,1\oplus t_1\oplus t_2\oplus t_3)\qquad\text{(odd parity)}. \notag
\end{align}
With this notation,
\begin{align}
     &\operatorname{res}(00\,\phi(t))=\operatorname{res}(11\,\phi(t))=0, \notag\\
     &\operatorname{res}(10\,\phi(t))=\operatorname{res}(01\,\psi(t))=1, \notag
\end{align}
while
\begin{align}
     &\operatorname{res}(01\,\phi(t))=\operatorname{res}(10\,\psi(t))=3, \notag\\
     &\operatorname{res}(00\,\psi(t))=\operatorname{res}(11\,\psi(t))=2. \notag
\end{align}

We also introduce the $\{\pm1\}$-valued characters
\[
    \chi_3(t)=(-1)^{t_1},\qquad
    \chi_4(t)=(-1)^{t_2},\qquad
    \chi_5(t)=(-1)^{t_3}.
\]
For later use, define
\[
    \eta_3:=\chi_3,\qquad
    \eta_4:=\chi_4,\qquad
    \eta_5:=\chi_5,\qquad
    \eta_6:=\chi_3\chi_4\chi_5.
\]
These are the sign functions corresponding to $(-1)^{x_i}$ on the even-parity strings $\phi(t)$ for qubits $i=3,4,5,6$.

\subsection{Logical states and sign patterns}

We define four orthonormal logical states. Three states, $\ket{0_L},\ket{1_L},\ket{2_L}$, are supported on residue-$0$ strings, while $\ket{3_L}$ is supported on residue-$1$ strings. All coefficients have magnitude $\tfrac14$.

For the residue-$0$ block we introduce sign patterns
\[
    s_0(t)=1,\qquad
    s_1(t)=\chi_3(t)\chi_4(t),\qquad
    s_2(t)=\chi_3(t)\chi_5(t),
\]
and set
\begin{equation}
    \ket{j_L}
    = \frac{1}{4}\sum_{t\in\mathbb F_2^3} s_j(t)
    (\ket{00\,\phi(t)}+\ket{11\,\phi(t)}),
    \qquad j=0,1,2.
    \label{eq:res0-block}
\end{equation}
For the residue-$1$ state we use
\begin{equation}
    \ket{3_L} = \frac{1}{4}\sum_{t\in\mathbb F_2^3}\chi_5(t)
    (\ket{10\,\phi(t)}+\ket{01\,\psi(t)}).
    \label{eq:res1-state}
\end{equation}

\subsection{Explicit expansions in the computational basis}

For completeness, we expand the four logical states explicitly in the computational basis $\ket{x_1x_2x_3x_4x_5x_6}$.

\entryhead{State $\ket{0_L}$.}
\begin{align*}
    \ket{0_L}
     & =\tfrac{1}{4}\Big(\ket{000000}+\ket{001001}+\ket{000101}+\ket{000011} \\
     & \qquad +\ket{001100}+\ket{001010}+\ket{000110}+\ket{001111}           \\
     & \qquad +\ket{110000}+\ket{111001}+\ket{110101}+\ket{110011}           \\
     & \qquad +\ket{111100}+\ket{111010}+\ket{110110}+\ket{111111}
    \Big).
\end{align*}

\entryhead{State $\ket{1_L}$.}
\begin{align*}
    \ket{1_L}
     & =\tfrac{1}{4}\Big(\ket{000000}-\ket{001001}-\ket{000101}+\ket{000011} \\
     & \qquad +\ket{001100}-\ket{001010}-\ket{000110}+\ket{001111}           \\
     & \qquad +\ket{110000}-\ket{111001}-\ket{110101}+\ket{110011}           \\
     & \qquad +\ket{111100}-\ket{111010}-\ket{110110}+\ket{111111}
    \Big).
\end{align*}

\entryhead{State $\ket{2_L}$.}
\begin{align*}
    \ket{2_L}
     & =\tfrac{1}{4}\Big(\ket{000000}-\ket{001001}+\ket{000101}-\ket{000011} \\
     & \qquad -\ket{001100}+\ket{001010}-\ket{000110}+\ket{001111}           \\
     & \qquad +\ket{110000}-\ket{111001}+\ket{110101}-\ket{110011}           \\
     & \qquad -\ket{111100}+\ket{111010}-\ket{110110}+\ket{111111}
    \Big).
\end{align*}

\entryhead{State $\ket{3_L}$.}
\begin{align*}
    \ket{3_L}
     & =\tfrac{1}{4}\Big(\ket{100000}+\ket{010001} +\ket{101001}+\ket{011000} \\
     & \qquad +\ket{100101}+\ket{010100}
    -\ket{100011}-\ket{010010}                                                \\
     & \qquad +\ket{101100}+\ket{011101}
    -\ket{101010}-\ket{011011}                                                \\
     & \qquad -\ket{100110}-\ket{010111}
    -\ket{101111}-\ket{011110}
    \Big).
\end{align*}

Each state contains $16$ basis strings with amplitudes $\pm 1/4$, so $\|\ket{j_L}\|^2=1$ for all $j$.

\subsection{Transversal controlled-phase action}

Define $Z(\theta)=\mathrm{diag}(1,e^{i\theta})$ and
\[
    U=\bigotimes_{j=1}^6 Z\!\left(\tfrac{\pi}{2}w_j\right).
\]
On a computational basis state $\ket{x}$ this yields
\[
    U\ket{x}=e^{i\frac{\pi}{2}\mathbf w\cdot x}\ket{x}
    = i^{\operatorname{res}(x)}\ket{x}.
\]
Thus $U$ acts by a constant phase on each residue class.

Since $\ket{0_L},\ket{1_L},\ket{2_L}$ lie entirely in residue value $0$ and $\ket{3_L}$ lies entirely in residue value $1$, the induced logical action is
\[
    U_L=\mathrm{diag}(1,1,1,i).
\]
This explicit example lies beyond the nondegenerate-residue filter of the strict subset-sum pipeline: three logical states share residue value $0$, and the union support contains Hamming-$1$ neighbours. Nevertheless, structured sign cancellation enforces the remaining distance-$2$ KL constraints and yields a $((6,4,2))$ code with a nontrivial diagonal transversal gate.


\section{Distance-3 \texorpdfstring{$((7,2,3))$}{((7,2,3))} BD16 codes}
\label{supp:bd16-codes}

Throughout this section we work at modulus $m=8$ with residues $(S_0,S_1)=(0,7)$ and the complementary convention
\[
    \ket{1_L}=X^{\otimes 7}\ket{0_L}.
\]
For each sorted weight vector $\mathbf w$ listed below, the displayed $\ket{0_L}$ is supported on the residue-$0$ subset-sum class of $\mathbf w$, and the pair $\{\ket{0_L},\ket{1_L}\}$ satisfies the full Knill-Laflamme conditions for every Pauli error of weight at most $2$.

\paragraph{$\mathbf w=\tup{1,2,2,2,2,3,3}$.}
This is the two-parameter family already identified in Ref.~\cite{Zhang2025Transversal}. An exact form is
\begin{align}
    4\ket{0_L}= &e^{i\theta_1}\ket{0000000}+\sqrt{3}\ket{0111100}+\sqrt{3}\ket{1001110} \notag\\
                &+\sqrt{2}\ket{1010101}+\sqrt{2}\ket{1011001} +e^{i\theta_2}\ket{1110010} \notag\\
                &+2i\,\ket{0100011}, \notag
\end{align}
with arbitrary $\theta_1,\theta_2\in\mathbb{R}$. Every member of the family has $(\lambda^{\ast})^2=21/8$.

\paragraph{$\mathbf w=\tup{1,1,2,2,2,2,5}$.}
A representative exact solution is
\[
    \begin{aligned}
        \ket{0_L}=
         & \ \frac{1}{4}\ket{0000000}
        -\frac{\sqrt{3}}{4}\ket{0011110} +\frac{i\sqrt{2}}{4}\Bigl(\ket{0100101} \\
        &-\ket{0110001}-\ket{1000011}+\ket{1001001}
        \Bigr) \\
        &+\frac{i}{4}\Bigl(\ket{1101110}+\ket{1110110}+\ket{1111010} \\
        &+\ket{1111100}
        \Bigr).
    \end{aligned}
\]
Within the same residue support there is a $192$-element discrete family with amplitudes in $\mathbb{Q}(\sqrt{2},\sqrt{3},i)$; all members share $(\lambda^{\ast})^2=33/16$.

\paragraph{$\mathbf w=\tup{0,1,2,2,3,3,4}$.}
There are three closed-form exact solutions. One convenient representative is
\[
    \begin{aligned}
        \ket{0_L}=
         & \frac{1}{4}\ket{0000000}
        +\frac{i\sqrt{2}}{4}\ket{0010110}
        -\frac{i\sqrt{3}}{4}\ket{0100011} \notag \\
         & -\frac{\sqrt{2}}{4}\ket{0111100}
        +\frac{\sqrt{2}}{4}\ket{1001110}
        -\frac{\sqrt{3}}{4}\ket{1011001} \notag  \\
         & +\frac{\sqrt{2}}{4}\ket{1100101}
        +\frac{i}{4}\ket{1111010}. \notag
    \end{aligned}
\]
The three exact solutions have $(\lambda^{\ast})^2\in\{51/16,23/16,15/8\}$.

\paragraph{$\mathbf w=\tup{1,1,2,2,3,3,3}$.}
An exact one-parameter family is
\[
    \begin{aligned}
        \ket{0_L(\theta)}=
         & \frac{1}{4}\ket{0000000}
        +\frac{\sqrt{2}}{4}e^{i\theta}\ket{0001011} \notag  \\
         & -\frac{\sqrt{2}}{4}e^{i\theta}\ket{0010011}
        +i\frac{\sqrt{3}}{4}e^{i\theta}\ket{0111100} \notag \\
         & +i\frac{\sqrt{6}}{8}e^{i\theta}\ket{1011001}
        -i\frac{\sqrt{6}}{8}e^{i\theta}\ket{1011010} \notag \\
         & -i\frac{\sqrt{10}}{8}e^{i\theta}\ket{1100101}
        -i\frac{\sqrt{10}}{8}e^{i\theta}\ket{1100110},
    \end{aligned}
\]
with arbitrary $\theta\in\mathbb{R}$. Every family member has $\lambda^{\ast}=9\sqrt{2}/8$.

\paragraph{$\mathbf w=\tup{1,1,2,2,2,3,4}$.}
A simple uniform-magnitude exact solution is
\[
    \begin{aligned}
        \ket{0_L}=
         & \frac{1}{4}\ket{0000000}
        -\frac{i}{4}\ket{0001101}
        +\frac{i}{4}\ket{0010101} \notag     \\
         & -\frac{i}{4}\ket{0011001}
        -\frac{i}{4}\ket{0100011}
        -\frac{1}{4}\ket{0101110} \notag     \\
         & +\frac{1}{4}\ket{0110110}
        +\frac{1}{4}\ket{0111010}
        -\frac{i}{4}\ket{1000011} \notag     \\
         & +\frac{1}{4}\ket{1001110}
        +\frac{1}{4}\ket{1010110}
        -\frac{1}{4}\ket{1011010} \notag     \\
         & -\frac{1}{4}\ket{1100101}
        -\frac{1}{4}\ket{1101001}
        -\frac{1}{4}\ket{1110001} \notag     \\
         & -\frac{i}{4}\ket{1111100}. \notag
    \end{aligned}
\]
For this weight vector there are three uniform-magnitude exact solutions with $(\lambda^{\ast})^2\in\{7/16,11/16,19/16\}$. Allowing $\sqrt{2}$-reweightings yields additional exact families with
\[
    (\lambda^{\ast})^2\in\left\{\frac{7}{16},\frac{11}{16},\frac{17}{16},\frac{19}{16},\frac{21}{16},\frac{25}{16},\frac{17}{8},\frac{7}{4},2,\frac{35}{16},\frac{9}{4}\right\}.
\]

\paragraph{$\mathbf w=\tup{1,1,1,2,3,3,4}$.}
An explicit self-complementary solution is
\[
    \begin{aligned}
        \ket{0_L}=
         & \frac{1}{4}\ket{0000000}
        +\frac{i\sqrt{2}}{4}\ket{0001110}
        +\frac{\sqrt{2}}{4}\ket{0110110} \notag   \\
         & +\frac{\sqrt{3}}{4}\ket{0111001}
        +\frac{i\sqrt{10}}{8}\ket{1000011} \notag \\
         & +\frac{i\sqrt{10}}{8}\ket{1000101}
        +\frac{i\sqrt{6}}{8}\ket{1111010} \notag  \\
         & -\frac{i\sqrt{6}}{8}\ket{1111100}.
    \end{aligned}
\]
Here $(\lambda^{\ast})^2=75/32$. Keeping the same support and magnitudes yields $64$ exact phase variants, and further solutions follow from permuting qubits $(1,2,3)$ and $(5,6)$.

\paragraph{$\mathbf w=\tup{1,1,1,2,2,4,4}$.}
One representative uniform-magnitude solution is
\[
    \begin{aligned}
        \ket{0_L}= \frac{1}{4}\Big(
        &\ket{0000000}-\ket{0000011}\\
        &+\ket{0001101}+\ket{0001110}\\
        &+i\ket{0110101}-i\ket{0110110}\\
        &-i\ket{0111001}+i\ket{0111010}\\ &+i\ket{1010101}+i\ket{1010110}\\
        &-i\ket{1011001}-i\ket{1011010}\\
        &+i\ket{1100101}-i\ket{1100110}\\
        &+i\ket{1101001}-i\ket{1101010}
        \Big).
    \end{aligned}
\]
For this weight vector there are three closed-form uniform-magnitude solutions with $(\lambda^{\ast})^2\in\{1,5/4,11/8\}$.

\paragraph{$\mathbf w=\tup{1,1,1,2,2,3,5}$.}
An exact self-complementary codeword is
\[
    \begin{aligned}
        \ket{0_L}=
         & \frac14\ket{0000000}+\frac14\ket{0000011}\\
         &-\frac{i\sqrt3}{8}\Bigl(\ket{0010101}+\ket{0011001} +\ket{0100101} \\
        &\quad \quad \ \ +\ket{0101001}+\ket{1000101}+\ket{1001001}
        \Bigr) \\
        &-\frac{\sqrt6}{8}\Bigl(\ket{0011110}+\ket{0101110}+\ket{1001110}
        \Bigr) \\ 
        &-\frac{\sqrt{10}}{8}\ket{1110001}
        +\frac{i\sqrt5}{8}\left(\ket{1110110}+\ket{1111010}\right).
    \end{aligned}
\]
Its invariant is $(\lambda^{\ast})^2=369/128$.

\paragraph{$\mathbf w=\tup{1,1,1,1,3,4,4}$.}
A paste-ready representative of the $128$-element analytic family is
\[
    \begin{aligned}
        \ket{0_L}=
         & \ \frac{1}{4}\ket{0000000}
        +\frac{1}{4}\ket{0000011}
        +\frac{i}{4}\ket{0001101} \\
        &-\frac{i}{4}\ket{0001110} 
         +\frac{i}{4}\ket{0010101}
        -\frac{i}{4}\ket{0010110} \\
        &+\frac{i}{4}\ket{0100101}
        -\frac{i}{4}\ket{0100110}
        +\frac{i}{4}\ket{1000101} \\
        &-\frac{i}{4}\ket{1000110}
        +\frac{i\sqrt{3}}{4}\ket{1111001}
        +\frac{i\sqrt{3}}{4}\ket{1111010}.
    \end{aligned}
\]
Varying six sign choices and one parity bit gives $128$ exact family members. All share $(\lambda^{\ast})^2=31/8$.

\paragraph{$\mathbf w=\tup{1,1,1,1,3,3,5}$.}
A closed-form representative is
\[
    \begin{aligned}
        \ket{0_L(\theta)}=
         & \frac14\left(\ket{0000000}-\ket{0000011}-\ket{0000101}\right)           \\
         & +e^{i\theta}\Biggl[
            \frac14\left(\ket{0011110}+\ket{1100110}\right) \\
        &\qquad+\frac18\Bigl(\ket{0101110} +\ket{1010110}\Bigr) \\
        &\qquad+\frac38\left(\ket{0110110}+\ket{1001110}\right) \\
        &\qquad+\frac{\sqrt6}{8}\Bigl(\ket{1011001}+\ket{1101001} \\
        &\qquad \qquad \ -\ket{0111001} -\ket{1110001}\Bigr)
            \Biggr],
    \end{aligned}
\]
with arbitrary $\theta\in\mathbb{R}$. More generally there is a continuous family parameterized by $x,y>0$ satisfying $x^2+y^2+xy=7$ and $z=x+y$. The invariant is $(\lambda^{\ast})^2=177/64$.


\section{No-go cases for \texorpdfstring{$((7,2,3))$}{((7,2,3))} BD16 candidates}
\label{supp:bd16-nogo}

We again work in the BD16 setting $m=8$ with residues $(S_0,S_1)=(0,7)$ and the complementary ansatz
\[
    \ket{1_L}=X^{\otimes 7}\ket{0_L}.
\]
For each of the two remaining SS/LP-passing weight vectors, we write down an explicit real-coordinate subsystem of the weight-$\le 2$ Knill-Laflamme equations and show that this subsystem is inconsistent. Since every complementary full-KL solution would satisfy the corresponding subsystem, this rules out both candidates.

\subsection{\texorpdfstring{$\mathbf w=\tup{1,1,1,2,2,2,6}$}{w=(1,1,1,2,2,2,6)}}

Write
\[
    C_0:=C_{S_0}(\mathbf w),\qquad C_1:=C_{S_1}(\mathbf w),
    \qquad (S_0,S_1)=(0,7).
\]
For $\mathbf w=\tup{1,1,1,2,2,2,6}$, the residue-$0$ class is
\[
    \begin{aligned}
        C_0=\{ &
        0000000,\ 0000011,\ 0000101,\ 0001001,\ 0110001,            \\
               & 0111110,\ 1010001,\ 1011110,\ 1100001,\ 1101110\}.
    \end{aligned}
\]
Since $\sum_{i=1}^7 w_i=15\equiv 7 \pmod 8$, bitwise complement sends $C_0$ to $C_1$.

\begin{proposition}
    For $\mathbf w=\tup{1,1,1,2,2,2,6}$, there is no choice of amplitudes on $C_0$ such that
    \[
        \ket{0_L}=\sum_{s\in C_0} c_s\ket{s},
        \qquad
        \ket{1_L}=X^{\otimes 7}\ket{0_L},
    \]
    satisfies the Knill-Laflamme conditions \eqref{eq:KL-d3-appendix} for every Pauli operator of weight at most $2$.
\end{proposition}

\begin{proof}
    Fix the order
    \[
        \begin{aligned}
             & x_0=0000000,\ x_1=0000011,\ x_2=0000101,\notag \\ &x_3=0001001,\ x_4=0110001,\ x_5=0111110, \notag \\ &x_6=1010001,\ x_7=1011110,\ x_8=1100001, \notag \\ &x_9=1101110,
        \end{aligned}
    \]
    and write
    \[
        \ket{0_L}=\sum_{k=0}^9 c_k\ket{x_k},
        \qquad
        \ket{1_L}=X^{\otimes 7}\ket{0_L}=\sum_{k=0}^9 c_k\ket{\bar x_k},
    \]
    where $\bar x_k:=x_k\oplus \mathbf 1$ and $\mathbf 1=(1,\dots,1)$.

    For $x,z\in\{0,1\}^7$, let
    \[
        P(x,z):=i^{x\cdot z}X^x Z^z,
        \qquad
        \mathrm{wt}(x,z):=\bigl|\operatorname{supp}(x)\cup \operatorname{supp}(z)\bigr|.
    \]
    Then
    \[
        P(x,z)\ket{s}=i^{x\cdot z}(-1)^{z\cdot s}\ket{s\oplus x},
    \]
    and the full distance-$3$ KL system is
    \begin{align}
        \label{eq:bd16-1112226-fullKL}
         &\forall\,x,z\in\{0,1\}^7 \text{ with }\mathrm{wt}(x,z)\le 2: \notag\\
         &\bra{0_L}P(x,z)\ket{1_L}=0, \notag\\
         &\bra{0_L}P(x,z)\ket{0_L}=\bra{1_L}P(x,z)\ket{1_L}.
    \end{align}

    We use only the following $17$ scalar equations: normalization, the seven diagonal equations for $Z_1,\dots,Z_7$, and the nine off-diagonal equations for
    \[
        X_1,\ Z_2X_1,\ Z_4X_1,\quad
        X_2,\ Z_1X_2,\ Z_4X_2,\quad
        X_3,\ Z_1X_3,\ Z_4X_3.
    \]
    For clarity, the provenance is:
    \begin{align}
         &\text{(N)} P(0,0)=I \notag\\
         &\ \quad \bra{0_L}I\ket{0_L}=1; \notag\\
         &\text{(Z1)} P(0,e_1)=Z_1 \notag\\
         &\ \quad \bra{0_L}Z_1\ket{0_L}=\bra{1_L}Z_1\ket{1_L}\iff \bra{0_L}Z_1\ket{0_L}=0; \notag\\
         &\text{(Z2)} P(0,e_2)=Z_2 \notag\\
         &\ \quad \bra{0_L}Z_2\ket{0_L}=\bra{1_L}Z_2\ket{1_L}\iff \bra{0_L}Z_2\ket{0_L}=0; \notag\\
         &\text{(Z3)} P(0,e_3)=Z_3 \notag\\
         &\ \quad \bra{0_L}Z_3\ket{0_L}=\bra{1_L}Z_3\ket{1_L}\iff \bra{0_L}Z_3\ket{0_L}=0; \notag\\
         &\text{(Z4)} P(0,e_4)=Z_4 \notag\\
         &\ \quad \bra{0_L}Z_4\ket{0_L}=\bra{1_L}Z_4\ket{1_L}\iff \bra{0_L}Z_4\ket{0_L}=0; \notag\\
         &\text{(Z5)} P(0,e_5)=Z_5 \notag\\
         &\ \quad \bra{0_L}Z_5\ket{0_L}=\bra{1_L}Z_5\ket{1_L}\iff \bra{0_L}Z_5\ket{0_L}=0; \notag\\
         &\text{(Z6)} P(0,e_6)=Z_6 \notag\\
         &\ \quad \bra{0_L}Z_6\ket{0_L}=\bra{1_L}Z_6\ket{1_L}\iff \bra{0_L}Z_6\ket{0_L}=0; \notag\\
         &\text{(Z7)} P(0,e_7)=Z_7 \notag\\
         &\ \quad \bra{0_L}Z_7\ket{0_L}=\bra{1_L}Z_7\ket{1_L}\iff \bra{0_L}Z_7\ket{0_L}=0; \notag\\
         &\text{(X1-0)} P(e_1,0)=X_1 \notag\\
         &\ \quad \bra{0_L}X_1\ket{1_L}=0; \notag\\
         &\text{(Z2X1)} P(e_1,e_2)=X_1Z_2=Z_2X_1 \notag\\
         &\ \quad \bra{0_L}(Z_2X_1)\ket{1_L}=0; \notag\\
         &\text{(Z4X1)} P(e_1,e_4)=X_1Z_4=Z_4X_1 \notag\\
         &\ \quad \bra{0_L}(Z_4X_1)\ket{1_L}=0; \notag\\
         &\text{(X2-0)} P(e_2,0)=X_2 \notag\\
         &\ \quad \bra{0_L}X_2\ket{1_L}=0; \notag\\
         &\text{(Z1X2)} P(e_2,e_1)=X_2Z_1=Z_1X_2 \notag\\
         &\ \quad \bra{0_L}(Z_1X_2)\ket{1_L}=0; \notag\\
         &\text{(Z4X2)} P(e_2,e_4)=X_2Z_4=Z_4X_2 \notag\\
         &\ \quad \bra{0_L}(Z_4X_2)\ket{1_L}=0; \notag\\
         &\text{(X3-0)} P(e_3,0)=X_3 \notag\\
         &\ \quad \bra{0_L}X_3\ket{1_L}=0; \notag\\
         &\text{(Z1X3)} P(e_3,e_1)=X_3Z_1=Z_1X_3 \notag\\
         &\ \quad \bra{0_L}(Z_1X_3)\ket{1_L}=0; \notag\\
         &\text{(Z4X3)} P(e_3,e_4)=X_3Z_4=Z_4X_3 \notag\\
         &\ \quad \bra{0_L}(Z_4X_3)\ket{1_L}=0. \notag
    \end{align}

    (For the nine Pauli operators involving $X_i$, only the off-diagonal clause is nontrivial here; for $i=1,2,3$, one has $x_k\oplus e_i\notin C_0$, so $\bra{0_L}X_i\ket{0_L}=0$ automatically.)

    \paragraph{Real variables and the explicit $17$-equation subsystem.}
    Set
    \[
        c_k=v_{2k+1}+i v_{2k+2}\qquad (k=0,\dots,9),
    \]
    and
    \[
        p_k:=|c_k|^2=v_{2k+1}^2+v_{2k+2}^2\qquad (k=0,\dots,9).
    \]

    Normalization becomes
    \begin{equation}
        \label{eq:bd16-1112226-N}
        (v_1^2+v_2^2)+(v_3^2+v_4^2)+\cdots+(v_{19}^2+v_{20}^2)=1.
        \tag{N}
    \end{equation}

    The seven diagonal $Z_i$ equations become
    \begin{align}
        \label{eq:bd16-1112226-Z1}
         &(v_{13}^2+v_{14}^2)+(v_{15}^2+v_{16}^2)+(v_{17}^2+v_{18}^2)+(v_{19}^2+v_{20}^2)=\tfrac12,
        \tag{Z1}\\
        \label{eq:bd16-1112226-Z2}
         &(v_{9}^2+v_{10}^2)+(v_{11}^2+v_{12}^2)+(v_{17}^2+v_{18}^2)+(v_{19}^2+v_{20}^2)=\tfrac12,
        \tag{Z2}\\
        \label{eq:bd16-1112226-Z3}
         &(v_{9}^2+v_{10}^2)+(v_{11}^2+v_{12}^2)+(v_{13}^2+v_{14}^2)+(v_{15}^2+v_{16}^2)=\tfrac12,
        \tag{Z3}\\
        \label{eq:bd16-1112226-Z4}
         &(v_{7}^2+v_{8}^2)+(v_{11}^2+v_{12}^2)+(v_{15}^2+v_{16}^2)+(v_{19}^2+v_{20}^2)=\tfrac12,
        \tag{Z4}\\
        \label{eq:bd16-1112226-Z5}
         &(v_{5}^2+v_{6}^2)+(v_{11}^2+v_{12}^2)+(v_{15}^2+v_{16}^2)+(v_{19}^2+v_{20}^2)=\tfrac12,
        \tag{Z5}\\
        \label{eq:bd16-1112226-Z6}
         &(v_{3}^2+v_{4}^2)+(v_{11}^2+v_{12}^2)+(v_{15}^2+v_{16}^2)+(v_{19}^2+v_{20}^2)=\tfrac12,
        \tag{Z6}\\
        \label{eq:bd16-1112226-Z7}
         &(v_{3}^2+v_{4}^2)+(v_{5}^2+v_{6}^2)+(v_{7}^2+v_{8}^2)+(v_{9}^2+v_{10}^2)
        +(v_{13}^2 \notag\\
         &+v_{14}^2)+(v_{17}^2+v_{18}^2)=\tfrac12.
        \tag{Z7}
    \end{align}

    A finite check of the support shows
    \[
        \bar x_6\oplus e_1=x_9,\quad \bar x_9\oplus e_1=x_6,\quad
        \bar x_7\oplus e_1=x_8,\quad \bar x_8\oplus e_1=x_7,
    \]
    \[
        \bar x_4\oplus e_2=x_9,\quad \bar x_9\oplus e_2=x_4,\quad
        \bar x_5\oplus e_2=x_8,\quad \bar x_8\oplus e_2=x_5,
    \]
    \[
        \bar x_4\oplus e_3=x_7,\quad \bar x_7\oplus e_3=x_4,\quad
        \bar x_5\oplus e_3=x_6,\quad \bar x_6\oplus e_3=x_5.
    \]
    Thus only the pairs $(6,9)$, $(7,8)$, $(4,9)$, $(5,8)$, $(4,7)$, and $(5,6)$ contribute to the chosen off-diagonal KL equations.

    For $x=e_1$, one finds
    \begin{align*}
        \bra{0_L}X_1\ket{1_L}
         & =\overline{c_9}c_6+\overline{c_6}c_9+\overline{c_8}c_7+\overline{c_7}c_8  \\
         & =2\Re(\overline{c_6}c_9)+2\Re(\overline{c_7}c_8),                         \\
        \bra{0_L}(Z_2X_1)\ket{1_L}
         & =-\overline{c_9}c_6+\overline{c_6}c_9-\overline{c_8}c_7+\overline{c_7}c_8 \\
         & =2i(\Im(\overline{c_6}c_9)+\Im(\overline{c_7}c_8)),                       \\
        \bra{0_L}(Z_4X_1)\ket{1_L}
         & =-\overline{c_9}c_6+\overline{c_6}c_9+\overline{c_8}c_7-\overline{c_7}c_8 \\
         & =2i(\Im(\overline{c_6}c_9)-\Im(\overline{c_7}c_8)).
    \end{align*}
    For $x=e_2$,
    \begin{align*}
        \bra{0_L}X_2\ket{1_L}
         & =\overline{c_9}c_4+\overline{c_4}c_9+\overline{c_8}c_5+\overline{c_5}c_8  \\
         & =2\Re(\overline{c_4}c_9)+2\Re(\overline{c_5}c_8),                         \\
        \bra{0_L}(Z_1X_2)\ket{1_L}
         & =-\overline{c_9}c_4+\overline{c_4}c_9-\overline{c_8}c_5+\overline{c_5}c_8 \\
         & =2i(\Im(\overline{c_4}c_9)+\Im(\overline{c_5}c_8)),                       \\
        \bra{0_L}(Z_4X_2)\ket{1_L}
         & =-\overline{c_9}c_4+\overline{c_4}c_9+\overline{c_8}c_5-\overline{c_5}c_8 \\
         & =2i(\Im(\overline{c_4}c_9)-\Im(\overline{c_5}c_8)).
    \end{align*}
    For $x=e_3$,
    \begin{align*}
        \bra{0_L}X_3\ket{1_L}
         & =\overline{c_7}c_4+\overline{c_4}c_7+\overline{c_6}c_5+\overline{c_5}c_6  \\
         & =2\Re(\overline{c_4}c_7)+2\Re(\overline{c_5}c_6),                         \\
        \bra{0_L}(Z_1X_3)\ket{1_L}
         & =-\overline{c_7}c_4+\overline{c_4}c_7-\overline{c_6}c_5+\overline{c_5}c_6 \\
         & =2i(\Im(\overline{c_4}c_7)+\Im(\overline{c_5}c_6)),                       \\
        \bra{0_L}(Z_4X_3)\ket{1_L}
         & =-\overline{c_7}c_4+\overline{c_4}c_7+\overline{c_6}c_5-\overline{c_5}c_6 \\
         & =2i(\Im(\overline{c_4}c_7)-\Im(\overline{c_5}c_6)).
    \end{align*}
    Expanding these in the $v$-coordinates gives the remaining nine equations:
    \begin{align}
        \label{eq:bd16-1112226-X1-0}
         &v_{13}v_{19}+v_{14}v_{20}+v_{15}v_{17}+v_{16}v_{18}=0,
        \tag{X1-0}\\
        \label{eq:bd16-1112226-Z2X1}
         &v_{13}v_{20}-v_{14}v_{19}+v_{15}v_{18}-v_{16}v_{17}=0,
        \tag{Z2X1}\\
        \label{eq:bd16-1112226-Z4X1}
         &v_{13}v_{20}-v_{14}v_{19}-v_{15}v_{18}+v_{16}v_{17}=0,
        \tag{Z4X1}\\
        \label{eq:bd16-1112226-X2-0}
         &v_{11}v_{17}+v_{12}v_{18}+v_{19}v_{9}+v_{20}v_{10}=0,
        \tag{X2-0}\\
        \label{eq:bd16-1112226-Z1X2}
         &v_{11}v_{18}-v_{12}v_{17}-v_{19}v_{10}+v_{20}v_{9}=0,
        \tag{Z1X2}\\
        \label{eq:bd16-1112226-Z4X2}
         &-v_{11}v_{18}+v_{12}v_{17}-v_{19}v_{10}+v_{20}v_{9}=0,
        \tag{Z4X2}\\
        \label{eq:bd16-1112226-X3-0}
         &v_{11}v_{13}+v_{12}v_{14}+v_{15}v_{9}+v_{16}v_{10}=0,
        \tag{X3-0}\\
        \label{eq:bd16-1112226-Z1X3}
         &v_{11}v_{14}-v_{12}v_{13}-v_{15}v_{10}+v_{16}v_{9}=0,
        \tag{Z1X3}\\
        \label{eq:bd16-1112226-Z4X3}
         &-v_{11}v_{14}+v_{12}v_{13}-v_{15}v_{10}+v_{16}v_{9}=0.
        \tag{Z4X3}
    \end{align}

    Assume, for contradiction, that \eqref{eq:bd16-1112226-N}-\eqref{eq:bd16-1112226-Z4X3} admit a real solution.

    \paragraph{Step 1: solve the diagonal subsystem.}
    Let $p_k=v_{2k+1}^2+v_{2k+2}^2\ge 0$.
    From \eqref{eq:bd16-1112226-Z4}-\eqref{eq:bd16-1112226-Z6} we get
    \[
        p_3=p_2,\qquad p_2=p_1,
    \]
    so
    \begin{equation}
        \label{eq:bd16-1112226-t}
        p_1=p_2=p_3=:t.
    \end{equation}
    Subtracting \eqref{eq:bd16-1112226-Z3} from \eqref{eq:bd16-1112226-Z2} yields $p_8+p_9=p_6+p_7$, and together with \eqref{eq:bd16-1112226-Z1} this gives
    \begin{equation}
        \label{eq:bd16-1112226-halves}
        p_6+p_7=\frac14,\qquad p_8+p_9=\frac14.
    \end{equation}
    Plugging $p_8+p_9=\tfrac14$ into \eqref{eq:bd16-1112226-Z2} gives
    \begin{equation}
        \label{eq:bd16-1112226-p45}
        p_4+p_5=\frac14.
    \end{equation}
    Normalization \eqref{eq:bd16-1112226-N} now reads
    \[
        1=p_0+3t+\frac14+\frac14+\frac14,
    \]
    so
    \begin{equation}
        \label{eq:bd16-1112226-p0t}
        p_0+3t=\frac14.
    \end{equation}
    Set $S:=p_4+p_6+p_8$. Rewriting \eqref{eq:bd16-1112226-Z6} with \eqref{eq:bd16-1112226-p45} and \eqref{eq:bd16-1112226-halves} yields $t=S-\tfrac14$, while \eqref{eq:bd16-1112226-Z7} gives $3t+S=\tfrac12$. Hence
    \[
        S=\frac{5}{16},\qquad t=\frac{1}{16}.
    \]
    Therefore
    \begin{equation}
        \label{eq:bd16-1112226-fourfixed}
        p_0=p_1=p_2=p_3=\frac{1}{16}.
    \end{equation}
    Finally, set
    \[
        U:=p_7,\qquad V:=p_9.
    \]
    Then \eqref{eq:bd16-1112226-halves} and \eqref{eq:bd16-1112226-p45} give
    \begin{equation}
        \label{eq:bd16-1112226-param}
        p_6=\frac14-U,\
        p_8=\frac14-V,\
        p_4=U+V-\frac{3}{16},\
        p_5=\frac{7}{16}-U-V,
    \end{equation}
    with
    \begin{equation}
        \label{eq:bd16-1112226-UV}
        0\le U\le \frac14,\quad
        0\le V\le \frac14,\quad
        \frac{3}{16}\le U+V\le \frac{7}{16}.
    \end{equation}

    \paragraph{Step 2: extract the cross-relations.}
    Adding and subtracting the paired off-diagonal equations gives
    \begin{align}
        \label{eq:bd16-1112226-cross1}
         &v_{13}v_{20}-v_{14}v_{19}=0,\qquad v_{15}v_{18}-v_{16}v_{17}=0,\\
        \label{eq:bd16-1112226-cross2}
         &v_{20}v_{9}-v_{19}v_{10}=0,\qquad v_{11}v_{18}-v_{12}v_{17}=0,\\
        \label{eq:bd16-1112226-cross3}
         &v_{16}v_{9}-v_{15}v_{10}=0,\qquad v_{11}v_{14}-v_{12}v_{13}=0.
    \end{align}

    \paragraph{Step 3: six amplitudes must be nonzero.}
    Define
    \[
        a:=c_6=v_{13}+iv_{14},\
        b:=c_9=v_{19}+iv_{20},\
        c:=c_7=v_{15}+iv_{16},
    \]
    \[
        d:=c_8=v_{17}+iv_{18},\
        e:=c_4=v_{9}+iv_{10},\
        f:=c_5=v_{11}+iv_{12}.
    \]
    We claim that
    \begin{equation}
        \label{eq:bd16-1112226-nonzero}
        a\neq 0,\ b\neq 0,\ c\neq 0,\ d\neq 0,\ e\neq 0,\ f\neq 0.
    \end{equation}
    Indeed:

    (i) If $b=0$, then $V=p_9=0$, so $p_8=\tfrac14$ and $d\neq 0$. Equations \eqref{eq:bd16-1112226-X1-0} and \eqref{eq:bd16-1112226-cross1} reduce to
    \[
        v_{15}v_{17}+v_{16}v_{18}=0,\qquad v_{15}v_{18}-v_{16}v_{17}=0,
    \]
    forcing $c=0$ and hence $U=p_7=0$. Then \eqref{eq:bd16-1112226-param} gives $p_4=-\tfrac{3}{16}<0$, contradiction.

    (ii) If $e=0$, then $p_4=0$, so $U+V=\tfrac{3}{16}$ and $p_5=\tfrac14$, hence $f\neq 0$. Equations \eqref{eq:bd16-1112226-X2-0} and \eqref{eq:bd16-1112226-cross2} reduce to
    \[
        v_{11}v_{17}+v_{12}v_{18}=0,\qquad v_{11}v_{18}-v_{12}v_{17}=0,
    \]
    forcing $d=0$ and $p_8=0$. Then $V=\tfrac14$, hence $U=\tfrac{3}{16}-\tfrac14<0$, contradiction.

    (iii) If $c=0$, then $U=p_7=0$, so $p_6=\tfrac14$ and $a\neq 0$. Since $b\neq 0$ by (i), equations \eqref{eq:bd16-1112226-X1-0} and \eqref{eq:bd16-1112226-cross1} force $a=0$, contradiction.

    (iv) If $d=0$, then $p_8=0$, so $V=\tfrac14$. Equations \eqref{eq:bd16-1112226-X2-0} and \eqref{eq:bd16-1112226-cross2} then force $e=0$, contradicting (ii).

    (v) If $a=0$, then $p_6=0$, so $U=\tfrac14$ and $c\neq 0$. Equations \eqref{eq:bd16-1112226-X3-0} and \eqref{eq:bd16-1112226-cross3} force $e=0$, contradicting (ii).

    (vi) If $f=0$, then $p_5=0$, so $U+V=\tfrac{7}{16}$ and $p_4=\tfrac14$, hence $e\neq 0$. But then \eqref{eq:bd16-1112226-X2-0} and \eqref{eq:bd16-1112226-cross2} force $e=0$, contradiction.

    So \eqref{eq:bd16-1112226-nonzero} holds.

    \paragraph{Step 4: global phase gauge forces all six amplitudes to be real.}
    The full system is invariant under multiplying every $c_k$ by the same phase $e^{i\theta}$. Since $b\neq 0$, we may choose $\theta$ so that
    \begin{equation}
        \label{eq:bd16-1112226-gauge}
        v_{20}=0,\qquad v_{19}>0.
    \end{equation}
    Now \eqref{eq:bd16-1112226-cross1}-\eqref{eq:bd16-1112226-cross3} imply successively:
    \begin{align*}
         & v_{13}v_{20}-v_{14}v_{19}=0\ \Rightarrow\ v_{14}=0, \\
         & v_{20}v_{9}-v_{19}v_{10}=0\ \Rightarrow\ v_{10}=0,
    \end{align*}
    and since $e\neq 0$, we have $v_9\neq 0$, so
    \[
        v_{16}v_{9}-v_{15}v_{10}=0\ \Rightarrow\ v_{16}=0.
    \]
    Since $c\neq 0$, we have $v_{15}\neq 0$, hence
    \[
        v_{15}v_{18}-v_{16}v_{17}=0\ \Rightarrow\ v_{18}=0.
    \]
    Since $d\neq 0$, we have $v_{17}\neq 0$, hence
    \[
        v_{11}v_{18}-v_{12}v_{17}=0\ \Rightarrow\ v_{12}=0.
    \]
    Therefore
    \begin{equation}
        \label{eq:bd16-1112226-allreal}
        a,b,c,d,e,f\in \mathbb R\setminus\{0\}.
    \end{equation}

    \paragraph{Step 5: contradiction from the remaining dot equations.}
    Under \eqref{eq:bd16-1112226-allreal}, equations \eqref{eq:bd16-1112226-X1-0}, \eqref{eq:bd16-1112226-X2-0}, and \eqref{eq:bd16-1112226-X3-0} simplify to
    \[
        ab+cd=0,\qquad fd+be=0,\qquad fa+ce=0.
    \]
    Equivalently,
    \begin{equation}
        \label{eq:bd16-1112226-realrel}
        ba=-dc,\qquad be=-df,\qquad ce=-af.
    \end{equation}
    Multiply $ba=-dc$ by $f$ and $be=-df$ by $c$:
    \[
        baf=-dcf,\qquad bce=-dfc=-dcf.
    \]
    Thus $baf=bce$. Since $b\neq 0$, we obtain $af=ce$. But \eqref{eq:bd16-1112226-realrel} also gives $ce=-af$, so $af=-af$, hence $af=0$, contradicting $a\neq 0$ and $f\neq 0$.

    Therefore the $17$-equation real subsystem is inconsistent. Since every complementary full-KL solution would satisfy this subsystem, no such solution exists.
\end{proof}

\subsection{\texorpdfstring{$\mathbf w=\tup{0,1,1,2,3,3,5}$}{w=(0,1,1,2,3,3,5)}}

Write again
\[
    C_0:=C_{S_0}(\mathbf w),\qquad C_1:=C_{S_1}(\mathbf w),
    \qquad (S_0,S_1)=(0,7).
\]
For $\mathbf w=\tup{0,1,1,2,3,3,5}$, the residue-$0$ class is
\[
    \begin{aligned}
        C_0=\{ &
        0000000,\ 0000011,\ 0000101,\ 0001110,\ 0011001, \\ & 0101001,\ 0110110, 1000000,\ 1000011,\ 1000101,\\ & 1001110,\ 1011001,\ 1101001,\ 1110110\}.
    \end{aligned}
\]
Since $\sum_{i=1}^7 w_i=15\equiv 7\pmod 8$, bitwise complement sends $C_0$ to $C_1$. Note also that $w_1=0$, so $C_0$ splits into seven pairs $(0u,1u)$ with the same last six bits.

\begin{proposition}
    For $\mathbf w=\tup{0,1,1,2,3,3,5}$, there is no choice of amplitudes on $C_0$ such that
    \[
        \ket{0_L}=\sum_{s\in C_0} c_s\ket{s},
        \qquad
        \ket{1_L}=X^{\otimes 7}\ket{0_L},
    \]
    satisfies the Knill-Laflamme conditions \eqref{eq:KL-d3-appendix} for every Pauli operator of weight at most $2$.
\end{proposition}

\begin{proof}
    Let
    \[
        \mathbf w\cdot s:=\sum_{i=1}^7 w_i s_i \pmod 8,
        \qquad
        \bar s:=s\oplus \mathbf 1.
    \]
    For $x,z\in\{0,1\}^7$, use the Hermitian Pauli convention
    \[
        P(x,z):=i^{x\cdot z}X^x Z^z,
        \qquad
        \mathrm{wt}(x,z):=\bigl|\operatorname{supp}(x)\cup \operatorname{supp}(z)\bigr|.
    \]
    Thus $P(1,1)=Y$ on one qubit, and
    \[
        P(e_1+e_2,e_1)=Y_1X_2,\qquad P(e_1+e_3,e_1)=Y_1X_3.
    \]

    We use the following $19$ KL equations: normalization; the seven diagonal equations for $Z_1,\dots,Z_7$; the diagonal equation for $Y_1$; the off-diagonal equations for $X_2,X_2Z_1,X_2Z_3,X_3,X_3Z_1,X_3Z_2,X_1X_2,Y_1X_2,X_1X_3,Y_1X_3$. Explicitly:
    \begin{align*}
         & \text{(E1)} P(0,0)=I \qquad \qquad \qquad \ \bra{0_L}I\ket{0_L}=1;                              \\
         & \text{(E2)} P(0,e_1)=Z_1                                                 \\
         & \bra{0_L}Z_1\ket{0_L}=\bra{1_L}Z_1\ket{1_L}\iff \bra{0_L}Z_1\ket{0_L}=0; \\
         & \text{(E3)} P(0,e_2)=Z_2                                                 \\
         & \bra{0_L}Z_2\ket{0_L}=\bra{1_L}Z_2\ket{1_L}\iff \bra{0_L}Z_2\ket{0_L}=0; \\
         & \text{(E4)} P(0,e_3)=Z_3                                                 \\
         & \bra{0_L}Z_3\ket{0_L}=\bra{1_L}Z_3\ket{1_L}\iff \bra{0_L}Z_3\ket{0_L}=0; \\
         & \text{(E5)} P(0,e_4)=Z_4                                                 \\
         & \bra{0_L}Z_4\ket{0_L}=\bra{1_L}Z_4\ket{1_L}\iff \bra{0_L}Z_4\ket{0_L}=0; \\
         & \text{(E6)} P(0,e_5)=Z_5                                                 \\
         & \bra{0_L}Z_5\ket{0_L}=\bra{1_L}Z_5\ket{1_L}\iff \bra{0_L}Z_5\ket{0_L}=0; \\
         & \text{(E7)} P(0,e_6)=Z_6                                                 \\
         & \bra{0_L}Z_6\ket{0_L}=\bra{1_L}Z_6\ket{1_L}\iff \bra{0_L}Z_6\ket{0_L}=0; \\
         & \text{(E8)} P(0,e_7)=Z_7                                                 \\
         & \bra{0_L}Z_7\ket{0_L}=\bra{1_L}Z_7\ket{1_L}\iff \bra{0_L}Z_7\ket{0_L}=0; \\
         & \text{(E9)} P(e_1,e_1)=Y_1                                               \\
         & \bra{0_L}Y_1\ket{0_L}=\bra{1_L}Y_1\ket{1_L}\iff \bra{0_L}Y_1\ket{0_L}=0; \\
         & \text{(E10)} P(e_2,0)=X_2 \qquad \quad \quad \ \bra{0_L}X_2\ket{1_L}=0;                       \\
         & \text{(E11)} P(e_2,e_1)=X_2Z_1 \qquad \quad \bra{0_L}X_2Z_1\ket{1_L}=0;               \\
         & \text{(E12)} P(e_2,e_3)=X_2Z_3 \qquad \quad  \bra{0_L}X_2Z_3\ket{1_L}=0;               \\
         & \text{(E13)} P(e_3,0)=X_3 \qquad \qquad \ \bra{0_L}X_3\ket{1_L}=0;                       \\
         & \text{(E14)} P(e_3,e_1)=X_3Z_1 \qquad \quad \bra{0_L}X_3Z_1\ket{1_L}=0;               \\
         & \text{(E15)} P(e_3,e_2)=X_3Z_2 \qquad \quad  \bra{0_L}X_3Z_2\ket{1_L}=0;               \\
         & \text{(E16)} P(e_1+e_2,0)=X_1X_2 \quad \bra{0_L}X_1X_2\ket{1_L}=0;             \\
         & \text{(E17)} P(e_1+e_2,e_1)=Y_1X_2 \quad \bra{0_L}Y_1X_2\ket{1_L}=0;           \\
         & \text{(E18)} P(e_1+e_3,0)=X_1X_3 \quad \bra{0_L}X_1X_3\ket{1_L}=0;             \\
         & \text{(E19)} P(e_1+e_3,e_1)=Y_1X_3 \quad \bra{0_L}Y_1X_3\ket{1_L}=0.
    \end{align*}

    \paragraph{Real variables and the $19$ explicit quadratic equations.}
    Introduce real variables $v_1,\dots,v_{28}$ by
    \begin{align*}
        c_{0000000} & :=v_1+i v_2,       &
        c_{0000011} & :=v_3+i v_4,         \\
        c_{0000101} & :=v_5+i v_6,       &
        c_{0001110} & :=v_7+i v_8,         \\
        c_{0011001} & :=v_9+i v_{10},    &
        c_{0101001} & :=v_{11}+i v_{12},   \\
        c_{0110110} & :=v_{13}+i v_{14}, &
        c_{1000000} & :=v_{15}+i v_{16},   \\
        c_{1000011} & :=v_{17}+i v_{18}, &
        c_{1000101} & :=v_{19}+i v_{20},   \\
        c_{1001110} & :=v_{21}+i v_{22}, &
        c_{1011001} & :=v_{23}+i v_{24},   \\
        c_{1101001} & :=v_{25}+i v_{26}, &
        c_{1110110} & :=v_{27}+i v_{28}.
    \end{align*}
    Let $p_s:=|c_s|^2$. Then normalization is
    \begin{equation}
        \label{eq:bd16-0112335-E1}
        v_1^2+v_2^2+\cdots+v_{27}^2+v_{28}^2=1.
        \tag{E1}
    \end{equation}

    The seven diagonal $Z_i$ equations become
    \begin{align}
        \label{eq:bd16-0112335-E2}
         &2\Big((v_{15}^2+v_{16}^2)+(v_{17}^2+v_{18}^2)+(v_{19}^2+v_{20}^2)+(v_{21}^2+v_{22}^2) \notag\\
         &+(v_{23}^2+v_{24}^2)+(v_{25}^2+v_{26}^2)+(v_{27}^2+v_{28}^2)\Big)=1,
        \tag{E2}\\
        \label{eq:bd16-0112335-E3}
         &2\big((v_{11}^2+v_{12}^2)+(v_{13}^2+v_{14}^2)+(v_{25}^2+v_{26}^2) \notag \\
         &+(v_{27}^2+v_{28}^2)\big)=1,
        \tag{E3}\\
        \label{eq:bd16-0112335-E4}
         &2\big((v_{9}^2+v_{10}^2)+(v_{13}^2+v_{14}^2)+(v_{23}^2+v_{24}^2) \notag \\
         &+(v_{27}^2+v_{28}^2)\big)=1,
        \tag{E4}\\
        \label{eq:bd16-0112335-E5}
         &2\Big((v_{7}^2+v_{8}^2)+(v_{9}^2+v_{10}^2)+(v_{11}^2+v_{12}^2)+(v_{21}^2+v_{22}^2) \notag\\
         &+(v_{23}^2+v_{24}^2)+(v_{25}^2+v_{26}^2)\Big)=1,
        \tag{E5}\\
        \label{eq:bd16-0112335-E6}
         &2\Big((v_{5}^2+v_{6}^2)+(v_{7}^2+v_{8}^2)+(v_{13}^2+v_{14}^2)+(v_{19}^2+v_{20}^2) \notag\\
         &+(v_{21}^2+v_{22}^2)+(v_{27}^2+v_{28}^2)\Big)=1,
        \tag{E6}\\
        \label{eq:bd16-0112335-E7}
         &2\Big((v_{3}^2+v_{4}^2)+(v_{7}^2+v_{8}^2)+(v_{13}^2+v_{14}^2)+(v_{17}^2+v_{18}^2) \notag\\
         &+(v_{21}^2+v_{22}^2)+(v_{27}^2+v_{28}^2)\Big)=1,
        \tag{E7}\\
        \label{eq:bd16-0112335-E8}
         &2\Big((v_{3}^2+v_{4}^2)+(v_{5}^2+v_{6}^2)+(v_{9}^2+v_{10}^2)+(v_{11}^2+v_{12}^2)+(v_{17}^2 \notag \\
         &+v_{18}^2)+(v_{19}^2+v_{20}^2)+(v_{23}^2+v_{24}^2)+(v_{25}^2+v_{26}^2)\Big)=1.
        \tag{E8}
    \end{align}

    The diagonal $Y_1$ equation becomes
    \begin{align}
        \label{eq:bd16-0112335-E9}
         &(v_1v_{16}-v_2v_{15})+(v_3v_{18}-v_4v_{17})+(v_5v_{20}-v_6v_{19}) \notag\\
         &+(v_7v_{22}-v_8v_{21})
        +(v_9v_{24}-v_{10}v_{23})+(v_{11}v_{26}-v_{12}v_{25}) \notag\\
         &+(v_{13}v_{28}-v_{14}v_{27})=0.
        \tag{E9}
    \end{align}

    For the off-diagonal equations, the only contributing pairs are:
    \begin{align*}
        &0101001\leftrightarrow 1110110,\ 0110110\leftrightarrow 1101001 \ \text{for }x=e_2, \\
        &0011001\leftrightarrow 1110110,\ 0110110\leftrightarrow 1011001 \ \text{for }x=e_3, \\
        &0101001\leftrightarrow 0110110,\ 1101001\leftrightarrow 1110110 \ \text{for }x=e_1+e_2, \\
        &0011001\leftrightarrow 0110110,\ 1011001\leftrightarrow 1110110 \ \text{for }x=e_1+e_3.
    \end{align*}
    Expanding the corresponding matrix elements gives
    \begin{align}
        \label{eq:bd16-0112335-E10}
         &(v_{11}v_{27}+v_{12}v_{28})+(v_{13}v_{25}+v_{14}v_{26})=0,
        \tag{E10}\\
        \label{eq:bd16-0112335-E11}
         &(v_{11}v_{28}-v_{12}v_{27})+(v_{13}v_{26}-v_{14}v_{25})=0,
        \tag{E11}\\
        \label{eq:bd16-0112335-E12}
         &(v_{11}v_{28}-v_{12}v_{27})-(v_{13}v_{26}-v_{14}v_{25})=0,
        \tag{E12}\\
        \label{eq:bd16-0112335-E13}
         &(v_{9}v_{27}+v_{10}v_{28})+(v_{13}v_{23}+v_{14}v_{24})=0,
        \tag{E13}\\
        \label{eq:bd16-0112335-E14}
         &(v_{9}v_{28}-v_{10}v_{27})+(v_{13}v_{24}-v_{14}v_{23})=0,
        \tag{E14}\\
        \label{eq:bd16-0112335-E15}
         &(v_{9}v_{28}-v_{10}v_{27})-(v_{13}v_{24}-v_{14}v_{23})=0,
        \tag{E15}\\
        \label{eq:bd16-0112335-E16}
         &(v_{25}v_{27}+v_{26}v_{28})+(v_{11}v_{13}+v_{12}v_{14})=0,
        \tag{E16}\\
        \label{eq:bd16-0112335-E17}
         &(v_{25}v_{27}+v_{26}v_{28})-(v_{11}v_{13}+v_{12}v_{14})=0,
        \tag{E17}\\
        \label{eq:bd16-0112335-E18}
         &(v_{23}v_{27}+v_{24}v_{28})+(v_{9}v_{13}+v_{10}v_{14})=0,
        \tag{E18}\\
        \label{eq:bd16-0112335-E19}
         &(v_{23}v_{27}+v_{24}v_{28})-(v_{9}v_{13}+v_{10}v_{14})=0.
        \tag{E19}
    \end{align}

    Assume, for contradiction, that \eqref{eq:bd16-0112335-E1}-\eqref{eq:bd16-0112335-E19} admit a real solution.

    \paragraph{Step 1: eliminate the $Z$-system.}
    Let
    \[
        p_0:=p_{0000000},\ p_1:=p_{0000011},\ \dots,\ p_{13}:=p_{1110110}.
    \]
    Linear elimination of \eqref{eq:bd16-0112335-E1}-\eqref{eq:bd16-0112335-E8} gives the pair-sum identities
    \begin{align}
        \label{eq:bd16-0112335-pairsums}
         &p_0+p_7=\frac{1}{16},
        p_1+p_8=\frac{1}{16},
        p_2+p_9=\frac{1}{16},
        p_3+p_{10}=\frac{1}{8}, \notag\\
         &p_4+p_{11}=\frac{3}{16},
        p_5+p_{12}=\frac{3}{16},
        p_6+p_{13}=\frac{5}{16}.
    \end{align}

    \paragraph{Step 2: multiplicative identities from the $x=e_2$ and $x=e_3$ blocks.}
    From \eqref{eq:bd16-0112335-E11} and \eqref{eq:bd16-0112335-E12}, adding and subtracting gives
    \[
        v_{11}v_{28}-v_{12}v_{27}=0,\qquad v_{13}v_{26}-v_{14}v_{25}=0.
    \]
    Together with \eqref{eq:bd16-0112335-E10}, the identity
    \[
        (ac+bd)^2+(ad-bc)^2=(a^2+b^2)(c^2+d^2)
    \]
    implies
    \begin{equation}
        \label{eq:bd16-0112335-mult1}
        p_5\,p_{13}=p_6\,p_{12}.
    \end{equation}
    Exactly the same argument applied to \eqref{eq:bd16-0112335-E13}-\eqref{eq:bd16-0112335-E15} gives
    \begin{equation}
        \label{eq:bd16-0112335-mult2}
        p_4\,p_{13}=p_6\,p_{11}.
    \end{equation}

    \paragraph{Step 3: linearize these identities.}
    From \eqref{eq:bd16-0112335-pairsums}, we have $p_5=\tfrac{3}{16}-p_{12}$ and $p_6=\tfrac{5}{16}-p_{13}$. Substituting into \eqref{eq:bd16-0112335-mult1} yields
    \[
        \left(\frac{3}{16}-p_{12}\right)p_{13}
        =
        \left(\frac{5}{16}-p_{13}\right)p_{12},
    \]
    hence
    \[
        p_{12}=\frac35\,p_{13}.
    \]
    Similarly, \eqref{eq:bd16-0112335-mult2} and $p_4=\tfrac{3}{16}-p_{11}$ imply
    \[
        p_{11}=\frac35\,p_{13}.
    \]
    Therefore
    \begin{equation}
        \label{eq:bd16-0112335-linearized}
        p_{11}=p_{12}=\frac35\,p_{13},
        \qquad
        p_4=p_5=\frac35\,p_6.
    \end{equation}

    \paragraph{Step 4: lower bounds on $p_{13}$ and $p_6$.}
    Let $t:=p_{13}\ge 0$. Then \eqref{eq:bd16-0112335-linearized} gives
    \[
        p_{11}+p_{12}+p_{13}=\frac{11}{5}t.
    \]
    Using $p_7\ge 0$ and the linear system, one gets $\frac{11}{5}t\le \frac12$, hence $t\le \frac{5}{22}$. Since $p_6+p_{13}=\frac{5}{16}$, this implies
    \[
        p_6=\frac{5}{16}-t\ge \frac{15}{176}.
    \]
    For the lower bound on $t$, use $p_0\ge 0$ together with
    \[
        p_0=(p_8+p_9+p_{10}+p_{11}+p_{12}+p_{13})-\frac{7}{16}.
    \]
    From \eqref{eq:bd16-0112335-pairsums}, $p_8\le \frac{1}{16}$, $p_9\le \frac{1}{16}$, and $p_{10}\le \frac18$, so $p_8+p_9+p_{10}\le \frac14$. Hence
    \[
        0\le \frac14+\frac{11}{5}t-\frac{7}{16},
    \]
    which gives $t\ge \frac{15}{176}$. Therefore
    \begin{equation}
        \label{eq:bd16-0112335-bounds}
        p_{13}\ge \frac{15}{176}>0,
        \qquad
        p_6\ge \frac{15}{176}>0.
    \end{equation}

    \paragraph{Step 5: real-part vanishings from the $X_1X_2$ and $X_1X_3$ blocks.}
    Adding and subtracting \eqref{eq:bd16-0112335-E16} and \eqref{eq:bd16-0112335-E17} gives
    \begin{equation}
        \label{eq:bd16-0112335-re1}
        v_{25}v_{27}+v_{26}v_{28}=0,
        \qquad
        v_{11}v_{13}+v_{12}v_{14}=0.
    \end{equation}
    Similarly, from \eqref{eq:bd16-0112335-E18} and \eqref{eq:bd16-0112335-E19} we get
    \begin{equation}
        \label{eq:bd16-0112335-re2}
        v_{23}v_{27}+v_{24}v_{28}=0,
        \qquad
        v_{9}v_{13}+v_{10}v_{14}=0.
    \end{equation}

    \paragraph{Step 6: fix a global phase.}
    Because $p_{13}=v_{27}^2+v_{28}^2>0$ by \eqref{eq:bd16-0112335-bounds}, we may rotate all amplitudes by a common phase so that
    \begin{equation}
        \label{eq:bd16-0112335-gauge}
        v_{28}=0,\qquad v_{27}>0.
    \end{equation}
    Then \eqref{eq:bd16-0112335-E11}, \eqref{eq:bd16-0112335-E14}, \eqref{eq:bd16-0112335-re1}, and \eqref{eq:bd16-0112335-re2} imply
    \[
        v_{12}=0,\quad v_{10}=0,\quad v_{25}=0,\quad v_{23}=0,\quad v_{13}=0.
    \]
    Since $p_6=v_{13}^2+v_{14}^2>0$, we have
    \begin{equation}
        \label{eq:bd16-0112335-v14}
        v_{14}\neq 0.
    \end{equation}
    Also, \eqref{eq:bd16-0112335-linearized} gives $p_4=\frac35 p_6>0$ and $p_5=\frac35 p_6>0$, so under \eqref{eq:bd16-0112335-gauge},
    \[
        v_9\neq 0,\qquad v_{11}\neq 0.
    \]

    \paragraph{Step 7: split the $Y_1$ equation into a small part and a large part.}
    Write \eqref{eq:bd16-0112335-E9} as
    \[
        U+T=0,
    \]
    where
    \begin{align*}
        U:= & (v_1v_{16}-v_2v_{15})+(v_3v_{18}-v_4v_{17})+(v_5v_{20}-v_6v_{19}) \\
            & +(v_7v_{22}-v_8v_{21}),
    \end{align*}
    \[
        T:=(v_9v_{24}-v_{10}v_{23})+(v_{11}v_{26}-v_{12}v_{25})+(v_{13}v_{28}-v_{14}v_{27}).
    \]
    Under the consequences of \eqref{eq:bd16-0112335-gauge}, this simplifies to
    \[
        T=v_9v_{24}+v_{11}v_{26}-v_{14}v_{27}.
    \]
    Now \eqref{eq:bd16-0112335-E10} and \eqref{eq:bd16-0112335-E13} become
    \begin{align*}
        v_{11}v_{27}+v_{14}v_{26}=0
        &\quad\Longrightarrow\quad
        v_{26}=-\frac{v_{11}v_{27}}{v_{14}}, \\
        v_{9}v_{27}+v_{14}v_{24}=0
        &\quad\Longrightarrow\quad
        v_{24}=-\frac{v_{9}v_{27}}{v_{14}},
    \end{align*}
where division is valid by \eqref{eq:bd16-0112335-v14}. Substituting gives
    \[
        T=-\frac{v_{27}}{v_{14}}\,(v_9^2+v_{11}^2+v_{14}^2),
    \]
    hence
    \[
        |T|=\frac{|v_{27}|}{|v_{14}|}\,(v_9^2+v_{11}^2+v_{14}^2).
    \]
    In terms of the probabilities, this is
    \[
        |T|=\frac{\sqrt{p_{13}}}{\sqrt{p_6}}\,(p_4+p_5+p_6).
    \]
    Using \eqref{eq:bd16-0112335-linearized}, $p_4=p_5=\frac35 p_6$, so
    \begin{equation}
        \label{eq:bd16-0112335-T}
        |T|=\frac{11}{5}\sqrt{p_6p_{13}}.
    \end{equation}
    With \eqref{eq:bd16-0112335-bounds}, this yields
    \begin{equation}
        \label{eq:bd16-0112335-Tlb}
        |T|\ge \frac{11}{5}\cdot \frac{15}{176}=\frac{3}{16}.
    \end{equation}

    \paragraph{Step 8: uniform upper bound for $U$.}
    Each summand of $U$ has the form $ad-bc$, so
    \[
        |ad-bc|\le \sqrt{(a^2+b^2)(c^2+d^2)}
        \le \frac{(a^2+b^2)+(c^2+d^2)}{2}.
    \]
    Applying this to the four pairs $(v_1,v_2)$ with $(v_{15},v_{16})$, $(v_3,v_4)$ with $(v_{17},v_{18})$, etc., and using \eqref{eq:bd16-0112335-pairsums}, we obtain
    \begin{align*}
        &|v_1v_{16}-v_2v_{15}|\le \frac{1}{32},\qquad
        |v_3v_{18}-v_4v_{17}|\le \frac{1}{32}, \\
        &|v_5v_{20}-v_6v_{19}|\le \frac{1}{32},\qquad
        |v_7v_{22}-v_8v_{21}|\le \frac{1}{16}.
    \end{align*}
    Therefore
    \begin{equation}
        \label{eq:bd16-0112335-Uub}
        |U|\le \frac{1}{32}+\frac{1}{32}+\frac{1}{32}+\frac{1}{16}=\frac{5}{32}.
    \end{equation}

    \paragraph{Step 9: contradiction.}
    From $U+T=0$, we have $|U|=|T|$. But \eqref{eq:bd16-0112335-Tlb} gives $|T|\ge \frac{3}{16}=\frac{6}{32}$, while \eqref{eq:bd16-0112335-Uub} gives $|U|\le \frac{5}{32}$. This is impossible.

    Hence the $19$-equation real subsystem is inconsistent. Since every complementary full-KL solution would satisfy this subsystem, no such solution exists.
\end{proof}

\nocite{*}

\bibliography{refs}%

\end{document}